\documentclass[11pt,a4paper,english,oneside]{article}

\usepackage{authblk} % <-- For managing authors and affiliations

\usepackage[utf8]{inputenc}
\usepackage[left=2.5cm,top=2.5cm,right=2.5cm,bottom=2cm]{geometry}

\usepackage{amsmath, amssymb, amsthm}

\usepackage{makeidx}
\makeindex

\usepackage{fancyhdr}
\usepackage{graphicx}
\usepackage{subcaption}
\usepackage{epsfig}

\usepackage{bm}

\usepackage{multirow}
\usepackage{cancel}
\usepackage{adjustbox}
\usepackage{float}
\usepackage{caption}
\usepackage{hyperref}
\usepackage{quiver}
\usepackage{enumitem}
\usepackage[most]{tcolorbox}
\usepackage{xcolor}
\usepackage{setspace}
\usepackage{parskip}
\usepackage{afterpage}
\usepackage{csquotes}
\usepackage[backend=biber, sorting=none, style=numeric, maxbibnames=99]{biblatex}
\def \punt {\leaders \hbox to 4mm {\hfil . \hfil } \hfill}

\theoremstyle{definition}

\newcommand{\N}{\mathbb N}

\makeatletter
\newcommand{\skipitems}[1]{%
  \addtocounter{\@enumctr}{#1}%
}
\makeatother

\title{Benchmark of Pauli Correlation Encoding for different optimisation problems}

\author[1,2]{Fernando Alonso}
\author[1]{Colomán Samprón}
\author[1]{Jacobo Veiga}
\author[1]{Mariamo Mussa Juane}
\author[1,*]{Andrés Gómez}

\affil[1]{Galicia Supercomputing Center (CESGA), Spain}
\affil[2]{Centro de Investigación TIC (CITIC), Spain}
\affil[*]{Corresponding author: agomez@cesga.es}
\date{\today}

\begin{document}

\maketitle

% REQUIRED
\begin{abstract}
The continuous progress of quantum technologies has spurred the exploration of their potential applications across diverse fields, particularly in combinatorial optimisation. In this work, we study a quantum-classical optimisation framework based on Pauli Correlation Encoding, an encoding scheme that can represent $m$ binary variables using a polynomial number of qubits. To evaluate the performance of the method, we use three classical optimisation problems against the instances of the \texttt{QOPTLib}~\cite{qoptlib} benchmark. The study includes an analysis of the impact of the compression order of the encoding scheme, the problem structure, and hyperparameter selection on solution quality, as well as the role of post-processing in improving performance. Additionally, we study the effect of shot-based execution and hardware noise, showing how these factors influence both the accuracy of expected value estimation and the overall dynamics of the optimisation process. The results indicate that the proposed PCE-based framework achieves competitive performance against the benchmark and, in several cases, obtains equivalent or even superior solutions, highlighting its potential as an efficient encoding strategy for quantum optimisation in the NISQ and near fault-tolerant era.

\end{abstract}

% % REQUIRED
% \begin{keywords}
% Quantum Machine Learning, Fourier Series approximation, Parametrised Quantum Circuits, Distribution estimation, Option pricing
% \end{keywords}

% % REQUIRED
% \begin{MSCcodes}
% 65C05, 65R20, 42A10, 81P68
% \end{MSCcodes}

\section{Introduction}

Quantum computing is a field that studies the processing of information by exploiting the principles of quantum mechanics. Since its inception, significant advances have been made both in algorithm design and in the development of quantum hardware, which has driven rapid growth in quantum computational solutions and a constant search for applications across very diverse fields, such as operational research, material science, or drug discovery

% Nevertheless, in the current \textit{Noisy Intermediate-Scale Quantum} (NISQ)~\cite{Preskill_2018} era, uncertainty persists as to whether quantum computers will be able to offer a practical advantage in solving these problems before the arrival of fully or partially fault-tolerant devices~\cite{Shor_tolerant_1997}. In this context, variational quantum algorithms~\cite{Bharti_2022,Cerezo_2021} are an heuristic approach that emerge as a promising alternative for solving diverse computationally costly problems using classical computing, such as the calculation of ground-state energy in chemistry or physics, or the resolution of combinatorial optimisation problems.

Nevertheless, in the current \textit{Noisy Intermediate-Scale Quantum} (NISQ)~\cite{Preskill_2018} era, in which the size of quantum computers is limited to a number of qubits that ranges from 50 to a few hundreds and the quality of those qubits is limited by noise and decoherence, uncertainty persists as to whether quantum computers will be able to offer a practical advantage before the arrival of fully or partially fault-tolerant devices~\cite{Shor_tolerant_1997}. In this context, \textit{Variational Quantum Algorithms} (VQAs)~\cite{Bharti_2022,Cerezo_2021}, a family of heuristic randomised search algorithms, emerge as the leading strategy for reaching quantum advantage by addressing the limitations of NISQ devices using an optimisation-based approach that keeps the circuits depth shallow.

These algorithms are hybrid quantum-classical approaches whose main purpose is to minimize a given cost or loss function by combining quantum subroutines, which typically consist in the evaluation of the cost function through a \textit{Parametrised Quantum Circuit} (PQC), with classical subroutines, namely the optimisation of those parameters. Due to its general framework, VQAs have been proposed in diverse areas, ranging from fundamental physics to combinatorial optimisation.

In particular, even though the most immediate application of quantum algorithms relates to problems whose nature is inherently quantum, VQAs have also been explored for solving classical discrete optimisation problems. Within this framework, the \textit{Quantum Approximate optimisation Algorithm} (QAOA) constitutes the predominant approach. The method maps the combinatorial optimisation problem onto a spin Hamiltonian such that the ground state of the resulting Hamiltonian encodes the optimal solution of the original problem \cite{farhi2014quantum}.

However, they also present certain practical limitations and difficulties related to device noise ~\cite{Stilck_2021, Zhang_2022, Bornens_2023}. The presence of nearly flat regions in the cost-function landscape, known as \textit{barren plateaus}~\cite{Larocca_2025}, together with the risk of becoming trapped in local minima~\cite{Anschuetz_2022}, can significantly hinder the optimisation process and prevent convergence to high-quality solutions. Moreover, in the particular case of QAOA, the number of qubits required grows linearly with the number of classical variables, which constitutes an additional scalability constraint~\cite{Weidenfeller_2022}.

As a response to this last difficulty, new and more efficient encoding schemes have been investigated with the aim of reducing the cost in terms of the number of qubits (and therefore quantum resources). Among others, linear compression methods~\cite{Patti_2022, Fuller_2024} have been proposed, as well as exponential compression strategies~\cite{Perelshtein_2023, Huber_2024}. The latter are typically applied to relaxed versions of the problems, which may affect the quality of the solutions obtained. Moreover, in~\cite{Tene_Cohen_2026, Tan_2021}  is shown that such exponential compressions render the scheme classically simulable and significantly limit the expressivity of the models.

\textit{Pauli Correlation Encoding} (PCE)~\cite{Sciorilli_2025} is a recently proposed encoding that drastically reduces the number of qubits required in binary optimisation problems. Using Pauli operator strings, it encodes $m$ binary variables with a polynomial number $n$ of qubits, where $n \ll m$. This encoding is used within a variational scheme that classically optimises a PQC ~\cite{Benedetti_2019}. The whole framework is designed to minimise a loss function adapted to the problem at hand. In the original paper describing the algorithm, solutions are obtained for \textit{Maximum Cut Problem} instances with several thousand variables, which is highly promising. Indeed, \cite{soloviev2025, Padin_2026} have demonstrated its utility in portfolio optimisation problems, of great interest to financial and economic sectors.

PCE high-variable-compression capacity, together with the promising results obtained, motivates the investigation of whether the algorithm can be applied effectively to other combinatorial optimisation problems. To answer this question, in this work its performance is evaluated on three classical combinatorial optimisation problems, the \textit{Maximum Cut Problem} (MCP), the \textit{Bin Packing Problem} (BPP) and the \textit{Travelling Salesman Problem} (TSP), using datasets from the \texttt{QOPTLib}~\cite{qoptlib} benchmark. This library contains 40 instances spanning four optimisation problems and covering a wide range of sizes, from computationally easy cases to the largest instances for which there remains a non-negligible probability of obtaining high-quality solutions. The fourth problem included in the benchmark is the \textit{Vehicle Routing Problem} (VRP), which is not considered in the present study, as its analysis falls outside the scope of this work.

Additionally, this benchmark includes a complete resolution carried out with two solvers employing a D-WAVE system (\textit{Advantage 6.1})~\cite{Catherine_2020}: the first is based on \textit{Quantum Annealing} operating exclusively on QPU, and the second is a quantum-classical hybrid \textit{LeapHybridBQMSampler}. The D-WAVE's hybrid solver service integrates a portfolio of heuristic solvers that
leverage both quantum and classical resources to read and solve optimisation problems and improve their performance. The solutions obtained with the \textit{LeapHybridBQMSampler} will serve as a reference benchmark for the PCE results, as the quantum solver does not consistently provide solutions for all problem instances.

The manuscript is organised as follows. Section~\ref{sec_2} introduces the fundamental concepts that serve as the basis of the encoding scheme employed throughout this work. Section~\ref{sec_3} describes the methodology used for the execution and evaluation of the benchmark, specifying the computational components employed, such as the ansatz, the optimiser, the encoding and the software environment. Section~\ref{sec_4} is structured into several subsections, each dedicated to the formulation of a specific optimisation problem and to the description and analysis of the results obtained, in comparison with those of the benchmark. Section~\ref{subsec_41} addresses the MCP, Section~\ref{subsec_42} the BPP, and Section~\ref{subsec_43} the TSP. Section~\ref{sec_5} studies the influence of real hardware noise on the optimisation process through shot-based executions. Section~\ref{subsec_51} establishes a reference for precision in expected value computation for a real QPU backend, whilst Section~\ref{subsec_52} analyses the potential benefits of noise in the iterative process and provides an estimate of shot-based execution times under the same optimisation conditions. Finally, the main conclusions of the work are summarised in Section~\ref{sec_6}, and Section~\ref{sec_7} concludes with a discussion of the principal results and future needs for a deeper understanding of the capabilities of the algorithm, both in its quantum and classical components.

\section{Preliminaries}\label{sec_2}

Let a combinatorial optimisation problem of a function $F$ defined in terms of $m$  binary variables
\begin{equation}\label{eq:optimisation}
\begin{gathered}
   \min_{\vec{z}}(F(\vec{z})) \\
   \vec{z} := \Big\{ (z_1, \ldots, z_m) \;\Big|\; z_i \in \{-1,1\} \;\forall i \in \mathcal{I}\Big\}\\
   \mathcal{I} =\{1,\ldots,m\} ,
\end{gathered}
\end{equation}
and let
\begin{equation*}
    \Pi := \Big\{ \Pi_i \;\Big|\; i \in \mathcal{I} \Big\},
\end{equation*}
be a specific subset of $m \leq 4^n - 1$ traceless Pauli strings $\Pi_i$ for a certain number of qubits $n$, that is, a subset of operators constructed as tensor products of combinations of $n$ Pauli matrices $\{\mathbb{I},X,Y,Z\}$, excluding $\mathbb{I}^{\otimes n}$.

The PCE of the variables $\vec{x}$ relative to $\Pi$ is defined as the encoding of those variables in the expected values of the Pauli string operators contained in the subset $\Pi$, that is,
\begin{equation}
z_{i} := \operatorname{sgn} \left( \langle \Pi_i \rangle_{\Psi} \right) = \operatorname{sgn} \left( \langle \Psi | \Pi_i | \Psi \rangle \right), \quad \forall i \in \mathcal{I},
\label{cod_pce}
\end{equation}
where $\operatorname{sgn}$ denotes the sign function and $|\Psi\rangle$ is an arbitrary quantum state.

% According to the standard formulation known in the literature, given an index $k$, known as the compression order, we select the subset
% \begin{equation*}
%     \Pi^{(k)} := \{\Pi^{(k)}_1, \ldots, \Pi^{(k)}_m\},
% \end{equation*}
% where each $\Pi^{(k)}_i$ is a permutation of the Pauli strings
% \begin{equation*}
%     X^{\otimes k} \otimes \mathbb{I}^{\otimes (n-k)}, \quad
%     Y^{\otimes k} \otimes \mathbb{I}^{\otimes (n-k)}, \quad
%     Z^{\otimes k} \otimes \mathbb{I}^{\otimes (n-k)},
% \end{equation*}

According to the standard formulation known in the literature, given an index $k\in \N$, known as the compression order, the subset
\begin{equation*}
    \Pi^{(k)} := \Big\{\Pi_X^{(k)}, ~\Pi_Y^{(k)}, ~\Pi_Z^{(k)}\Big\},
\end{equation*}
is selected, where
\begin{align*}
    \Pi_X^{(k)} = \{ \sigma(X^{\otimes k} \otimes \mathbb{I}^{\otimes (n-k)})~|~ \sigma:\mathcal{U}\left(\mathcal{H}^n\right) \to \mathcal{U}\left(\mathcal{H}^n\right)\}, \\
    \Pi_Y^{(k)} = \{ \sigma(Y^{\otimes k} \otimes \mathbb{I}^{\otimes (n-k)})~|~ \sigma:\mathcal{U}\left(\mathcal{H}^n\right) \to \mathcal{U}\left(\mathcal{H}^n\right)\}, \\
    \Pi_Z^{(k)} = \{ \sigma(Z^{\otimes k} \otimes \mathbb{I}^{\otimes (n-k)})~|~ \sigma:\mathcal{U}\left(\mathcal{H}^n\right) \to \mathcal{U}\left(\mathcal{H}^n\right)\},
\end{align*}
where $\sigma$ is the map that generates permutations of the Pauli strings in the operators space $\mathcal{U}\left(\mathcal{H}^n\right)$. Using all possible permutations for the encoding gives rise to a total of
\begin{equation*}
    m = 3 \binom{n}{k}
\end{equation*}
variables for a certain number of qubits $n$. For example, if the encoding is performed with a compression order $k=2$ or $k=3$, one may have up to
\begin{equation*}
    m = \frac{3}{2} n (n-1) \quad \text{or} \quad m = \frac{1}{2} n (n-1)(n-2)
\end{equation*}
variables, respectively. Moreover, this choice requires only three measurement configurations throughout the entire procedure for the computation of expected values.

Once the encoding of the variables has been performed, the optimisation problem~\eqref{eq:optimisation} is transformed into finding a state $| \Psi \rangle$ that minimises the function. In general, this state is unknown and is therefore encoded as a PQC of $n$ qubits, where quantum operations (known as gates) depending on free parameters are applied. Those parameters are defined as:
\begin{equation*}
\begin{gathered}
   \vec{\Theta} := \Big\{ (\theta_1, \ldots, \theta_p) \;\Big|\; \theta_j \in [0, 2\pi) \;, \forall j \in \mathcal{P}\Big\}, \\
   \mathcal{P} =\{1,\ldots,p\} ,
\end{gathered}
\end{equation*}
% \begin{equation*}
%     \vec\Theta=\{\theta_j ~\big|~ j \in P =\{1, ..., p\}\},
% \end{equation*}
where $p$ is the number of free parameters in the circuit. That is, problem~\eqref{eq:optimisation} is rewritten as:
\begin{equation}\label{eq:optimisation2}
\begin{gathered}
   \min_{\vec{\Theta}} \left(F \left(\vec{z} \left( \vec{\Theta} \right) \right) \right) \\
   \vec{z}(\vec{\Theta}) := \Big\{ (z_1, \ldots, z_m) \;\Big|\; z_i := \operatorname{sgn} \langle \Psi(\vec\Theta) | \Pi_i^{(k)} | \Psi(\vec\Theta)\rangle\;\forall i \in \mathcal{I} \Big\},
\end{gathered}
\end{equation}
% \begin{equation}\label{eq:optimisation2}
% \begin{gathered}
%    \min_{\vec{\Theta}} \left(F \left(\vec{x} \left( \vec{\Theta} \right) \right) \right) \\
%    \vec{x} := \Big\{ (x_1, \ldots, x_m) \;\Big|\; x_i := \operatorname{sgn} \langle \Psi(\vec\Theta) | \Pi_i^{(k)} | \Psi(\vec\Theta)\rangle / \vec\Theta \in [0, 2\pi)^{\otimes p}   \;\forall i \in \mathcal{I} \Big\},
% \end{gathered}
% \end{equation}
for a given compression order $k$ to be selected.

% \textcolor{red}{[METODOLOGÍA:]} This hybrid classical--quantum scheme may be affected by the rigidity of the $\operatorname{sgn}$ function used in~\eqref{cod_pce}. In some cases, it may be convenient to relax the binary problem by replacing this function with the $\tanh(\cdot)$\footnote{The specific choice of the hyperbolic tangent is not particularly relevant, as any non-linear sigmoid function could be used instead.}, which is also more suitable for use with gradient-based methods. Consequently, throughout this work the following is employed in general:
% \begin{equation}
%   x_i := \tanh\big(\alpha \langle \Pi_i \rangle\big),
% \label{cod_pce_tanh}
% \end{equation}
% where the hyperparameter $\alpha > 1$ is as a tuneable rescaling factor that prevents the function from operating in the linear regime when the expected values are too small, given that $\tanh(z) \approx z$ for $|z| \ll 1$.

% % Once training is complete, a bit string $x$ is obtained.
% It is also common for a classical post-processing step, specific to each problem, to be applied in order to determine whether a better solution can be found.

% \textcolor{red}{[METHODOLOGY] Finally, starting from the bit string obtained through optimisation, an additional stage of classical post-processing based on local bit-swaps is applied, with the aim of refining the solution and improving its quality.} 

\section{Methodology}\label{sec_3}

This section describes the main technical aspects that characterise the experimental framework and specify the conditions under which the simulations and experiments considered have been conducted.

\begin{figure}[H]
\centering
\includegraphics[width=0.75\textwidth]{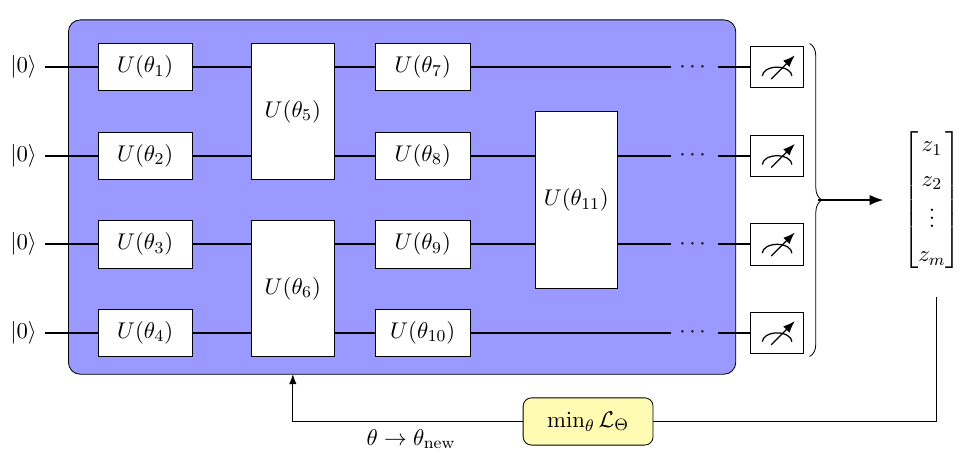}
\caption{Schematic of the quantum circuit used to construct the PQC.}
\label{fig:esquema_ansatz}
\end{figure}

Firstly, following the original proposal, the employed PQCs are constructed using a problem-agnostic and hardware-efficient ansatz based on a brick-like structure, as shown schematically in Figure~\ref{fig:esquema_ansatz}. These type of circuits are extensively used because they are sufficiently expressive when the available information about the problem at hand cannot be readily leveraged.

In particular, the circuits are built by alternating single- and two-qubit gate layers. The former, corresponding to even layers, consist of parameterised single-qubit rotation gates in which every layer corresponds to a particular type of gate selected ciclically from the set $\{ \texttt{RX}, \texttt{RY}, \texttt{RZ} \}$. The latter, corresponding to odd layers, consist of parameterised two-qubit rotation gates -\texttt{RXX}, \texttt{RYY} and \texttt{RZZ}- among pairs of neighbouring qubits. Again, for each layer, the rotation axis is selected ciclically from the set $\{ \texttt{RXX}, \texttt{RYY}, \texttt{RZZ} \}$, but, in this case, the ordering of the neighbouring qubit pairs alternates between $(\text{even}, \text{odd})$ and $(\text{odd}, \text{even})$ from one layer to the next. Thus, each rotation introduces a single variational parameter, guaranteeing controlled growth of the number of variational degrees of freedom throughout the circuit.

Furthermore, this type of design has the advantage of being adjustable in the circuit depth, which allows its expressive capacity and, consequently, the complexity of the model to be increased in a controlled manner~\cite{Schuld_2021}. However, in accordance with the original proposal, the circuit depth, in terms of number of layers, is considered to depend on the number of variables $m$ and the compression order $k$ as $\mathcal{O}\left(m^{1-\frac{1}{k}}\right)$.

Regarding optimisation, it has been shown in~\cite{Failde2023} that the local minima challenge for VQAs, can be avoided using a gradient-free optimisation strategy called \textit{Differential Evolution} (DE). This strategy is based on an evolutionary algorithm in which, for each iteration, a population of individuals goes through a mutation and recombination scheme that allows the optimisation to continue evolving even in the cases in which some individuals reach a local minima. Therefore, DE has been chosen as optimiser in this work. Additionally, it should be noted that, due to the stochastic nature of DE, the circuit parameters were initialised randomly within the interval $[0, 2\pi)$.

% \textcolor{red}{[¿a discusiones?]} Furthermore, no warm-starting strategy was employed, although such approaches have been proposed in the existing literature~\cite{doCarmo2025WarmStartingPCE}.

The specific parameters used for the optimiser are indicated in Table~\ref{tab:DEparameters}.

\begin{table}[H]
\centering
\begin{tabular}{|c|c|}
\hline
\textbf{Parameter} & \textbf{Value} \\
\hline
strategy & best1exp \\
maxiter & 2500 \\
popsize & 3 \\
tol & $1\times10^{-7}$ \\
mutation & $(0.5,\,1)$ \\
recombination & 0.7 \\
disp & True \\
polish & True \\
\hline
\end{tabular}
\caption{\textit{Differential Evolution} configuration parameters.}\label{tab:DEparameters}
\end{table}

Regarding the encoding, the resulting binary variables $z_{i}$ in~\eqref{cod_pce} are relaxed by replacing the sign function with the hyperbolic tangent function:
\begin{equation}
  z_{i} := \tanh\big(\alpha \langle \Pi_i \rangle\big),
\label{cod_pce_tanh}
\end{equation}
where $\alpha > 1$ is a tunable hyperparameter that prevents the function from operating in the linear regime when the expected values are too small, given that $\tanh(z) \approx z$ for $|z| \ll 1$. It should be noted that this relaxation was introduced in the original proposal to facilitate gradient-based optimisation methods and is therefore not strictly required when using DE. Nevertheless, we retain it in our implementation to investigate its influence on the optimisation process. Although the original sign function is asymptotically recovered as $\alpha$ increases, different values of $\alpha$ may induce different search dynamics, allowing us to assess the advantages and limitations associated with varying degrees of relaxation.

A regularisation term of the form
\begin{equation*}
\beta ~ \lambda_{\text{reg}}
\left[ \frac{1}{m} \sum_{i \in \mathcal{I}} \big( \tanh(\alpha \langle \Pi_i \rangle) \big)^2 \right]^2,
\end{equation*}
was also introduced in the original proposal, where $\lambda_{\text{reg}}$ is a problem-dependent constant that fixes the magnitude of this term within the cost function, and $\beta$ is a tunable hyperparameter. According to the original authors, this term improves the optimisation performance by concentrating the distribution of expected values around zero with lighter tails. In this work, we retain this regularisation term to assess its actual impact on the optimisation process and determine whether the reported benefits are consistently observed in our setting.

Finally, the software used is indicated in Table~\ref{tab:software_versions}.

\begin{table}[H]
\centering
\begin{tabular}{|c|c|}
\hline
\textbf{Software} & \textbf{Version} \\
\hline
Python & 3.11.9 \\
NumPy & 1.26.4 \\
SciPy & 1.13.0 \\
Pandas & 2.2.2 \\
NetworkX & 3.3 \\
Qiskit & 1.2.4 \\
Qiskit Aer & 0.15.1 \\
CUNQA & 2.2.0 \\
\hline
\end{tabular}
\caption{Software environment and versions used in the experiments.}
\label{tab:software_versions}
\end{table}

Exact simulations were carried out using \textit{QiskitAer}, whilst the shot-based experiments were conducted with CUNQA~\cite{CUNQA_25} with the \textit{AerSimulator} emulator \footnote{For more details about CUNQA configuration see \ref{subsec_51}.}; in both cases the \textit{Statevector} and \textit{Matrix Product State} default, i.e. exact, methods were employed.

\section{Benchmarked problems}\label{sec_4}

This section studies three classical combinatorial optimisation problems included in the \texttt{QOPTLib} benchmark: MCP, BPP and TSP. For each problem, both the original formulation and its adaptations to the PCE encoding are presented, enabling an evaluation of how this approach can capture different combinatorial structures and problem-specific constraints. Subsequently, the model's performance on each instance is analysed by comparing the results obtained with the reference solutions provided by the benchmark, which allows not only assessing the expressive capacity and effectiveness of PCE, but also identifying the limitations and advantages of its application to problems of different scales and complexities.

\subsection{Maximum Cut Problem}\label{subsec_41}

The MCP is a combinatorial optimisation problem whose main objective is to divide the nodes of a graph into two disjoint subsets  to maximise the sum of the weights of the edges connecting both subsets.

% The need to consider all possible partitions introduces into the MCP a complexity that grows exponentially as the number of nodes increases and consequently the edge density. This combinatorial nature makes exact resolution of the problem impractical for large instances, which is why it is classified as NP-hard~\cite{Du2022,Bodlaender_1994}. Nevertheless, due to its relevance in very diverse fields, the MCP has been one of the first problems on which quantum-computing-based solvers such as QAOA~\cite{Blekos_2024,Wang_2018,Crooks_2018} or \textit{Quantum Annealers}~\cite{Sharma_2026,Pecci_2024,Vodeb_2024} have been used.

Formally, the MCP can be defined on an undirected graph $G = (V,E)$, where $V = \{v_1, v_2, \ldots, v_{\mathcal{N}}\}$ is the set of nodes and $E \subset \{(v_i, v_j) ~|~ v_i, v_j \in V,\; i \neq j\}$ is the set of edges between these nodes \footnote{Recall that the graph considered in this problem is not necessarily complete.}, each with an associated weight $w_{ij} = w_{ji}$. A cut $(S,S')$ is defined as a partition of $V$ into two subsets $S$ and $S' = V \setminus S$, and its value is calculated as the sum of the weights $w_{ij}$ of the edges connecting both subsets.

Thus, the problem consists of solving
\begin{equation}
\max_{\vec{z}} C(\vec{z}) = \max_{\vec{z}} \left( \sum_{(v_i,v_j) \in E} w_{ij} \, (1 - z_i z_j)\right),
\label{obj_maxcut_1}
\end{equation}
where $z_i \in \{-1,1\}$ takes the value 1 if node $v_i$ is in partition $S$ and $-1$ otherwise. It should be noted that, since $\left(\sum_{v_i,v_j \in E} w_{ij}\right)$ is constant with respect to $z$, the problem can be reinterpreted by passing to its \textit{Quadratic Unconstrained Binary Optimisation} (QUBO)~\cite{Glover_2019} form as
\begin{equation}
\min_{\vec{z}}~\vec{z}^\top W \vec{z} = \min_{\vec{z}} \left(\sum_{(v_i,v_j) \in E} w_{ij} \, z_i z_j\right).
\label{obj_maxcut_2}
\end{equation}
Now it is necessary to define an cost function that properly adapts to the PCE encoding. However, since the QUBO formulation in~\eqref{obj_maxcut_2} is already sufficiently accurate, it can be directly employed,
\begin{equation*}
\min_{\vec{z}} F(\vec{z}) =
\min_{\vec{z}} \left(\sum_{(v_i,v_j) \in E} w_{ij} \, z_i z_j\right).
\label{obj_maxcut_pce}
\end{equation*}
Once the optimisation of $F(\vec{z})$ has been carried out, the corresponding solution is constructed based on the values $z_{i}$, assigning each node to a partition according to whether its value lies on one side or the other of zero. Subsequently, a classical post-processing step is performed in which a bit-by-bit swap of the resulting string is carried out to check whether is found a solution that provides a larger cut.

\subsubsection{Experiments and results}

\begin{figure}[H]
    \centering
     \subfloat[$k\in\{1,2,3,4,5\}$]{\includegraphics[width = 0.5\textwidth]{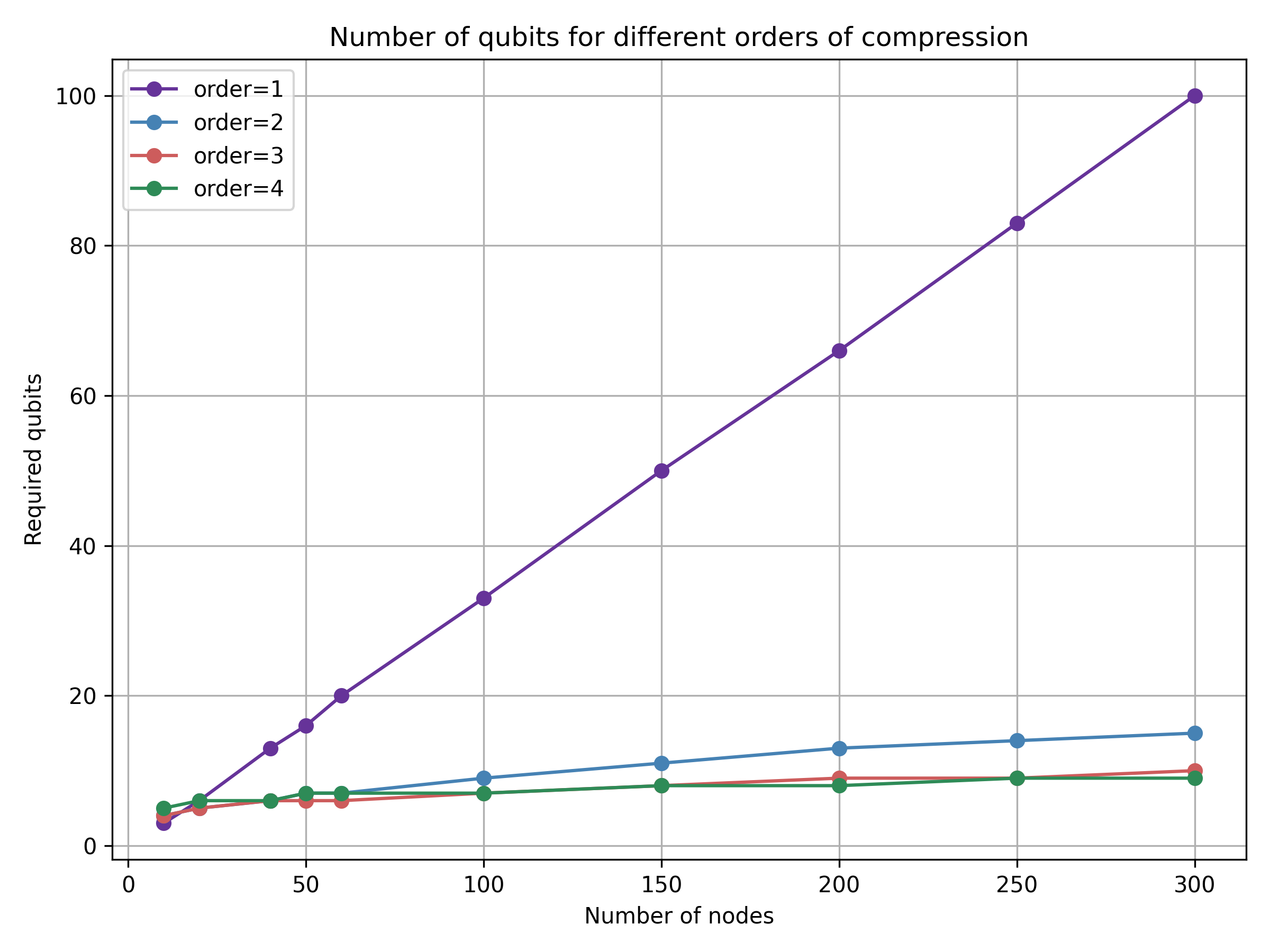}}
     \subfloat[$k\in\{2,3,4,5\}$]{\includegraphics[width = 0.5\textwidth]{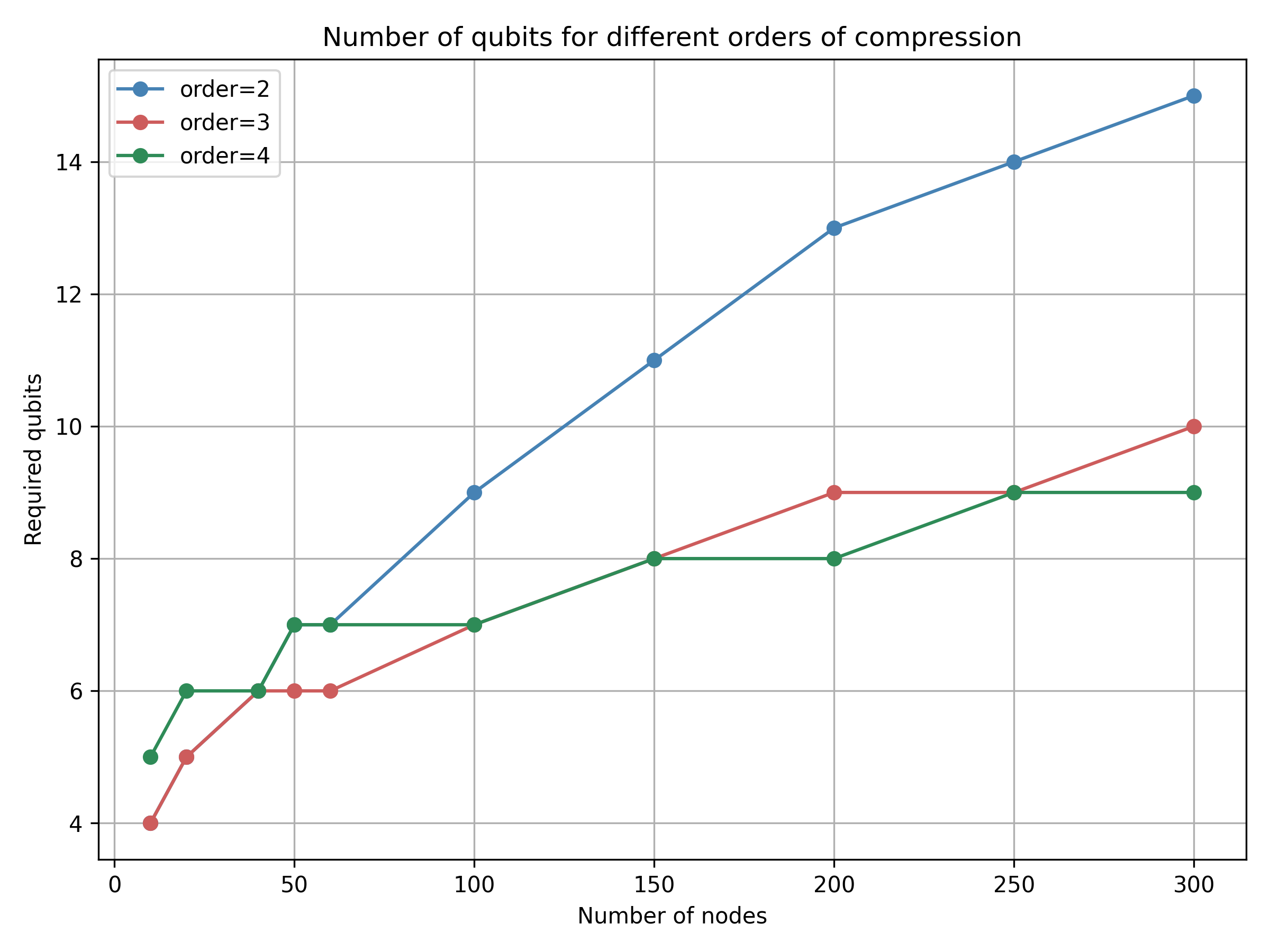}}
    \caption{Scaling of the number of qubits as a function of the number of variables for MCP.}
    \label{fig: MC_num_qubits}
\end{figure}

The \texttt{QOPTLib} dataset considers 10 instances of the MCP, characterised by a number of nodes ${\mathcal{N}}\in \{10,20,40,50,60,100,150,200,250,300\}$. Since the model requires encoding $m={\mathcal{N}}$ binary variables, the number of qubits required scales according to Figure~\ref{fig: MC_num_qubits}.

\begin{figure}[H]
    \centering
    \subfloat[Solution approximation ratio for $k=2$.]{%
        \includegraphics[width=0.8\textwidth]{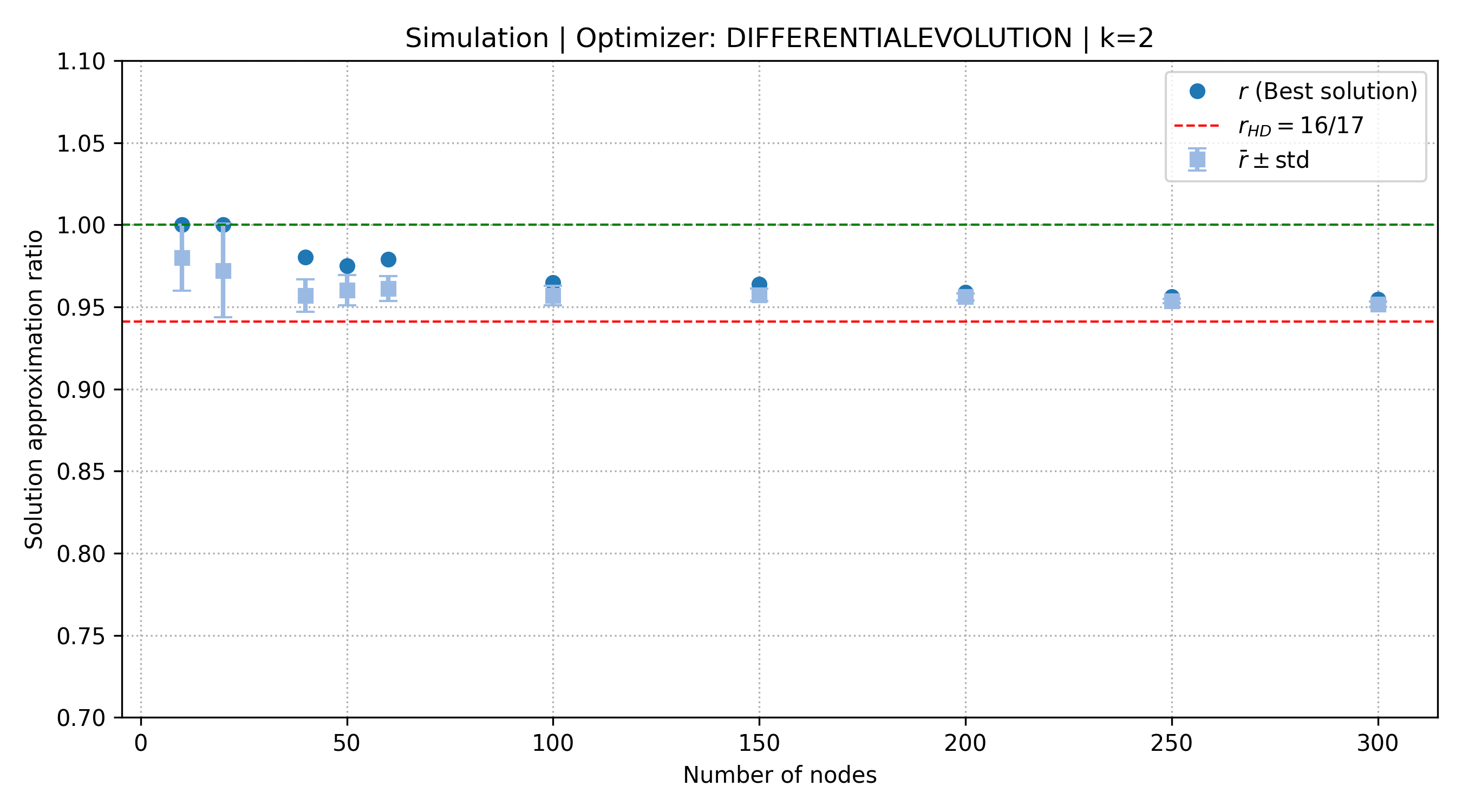}%
    }\\[1ex]
    \subfloat[Solution approximation ratio for $k=3$.]{%
        \includegraphics[width=0.8\textwidth]{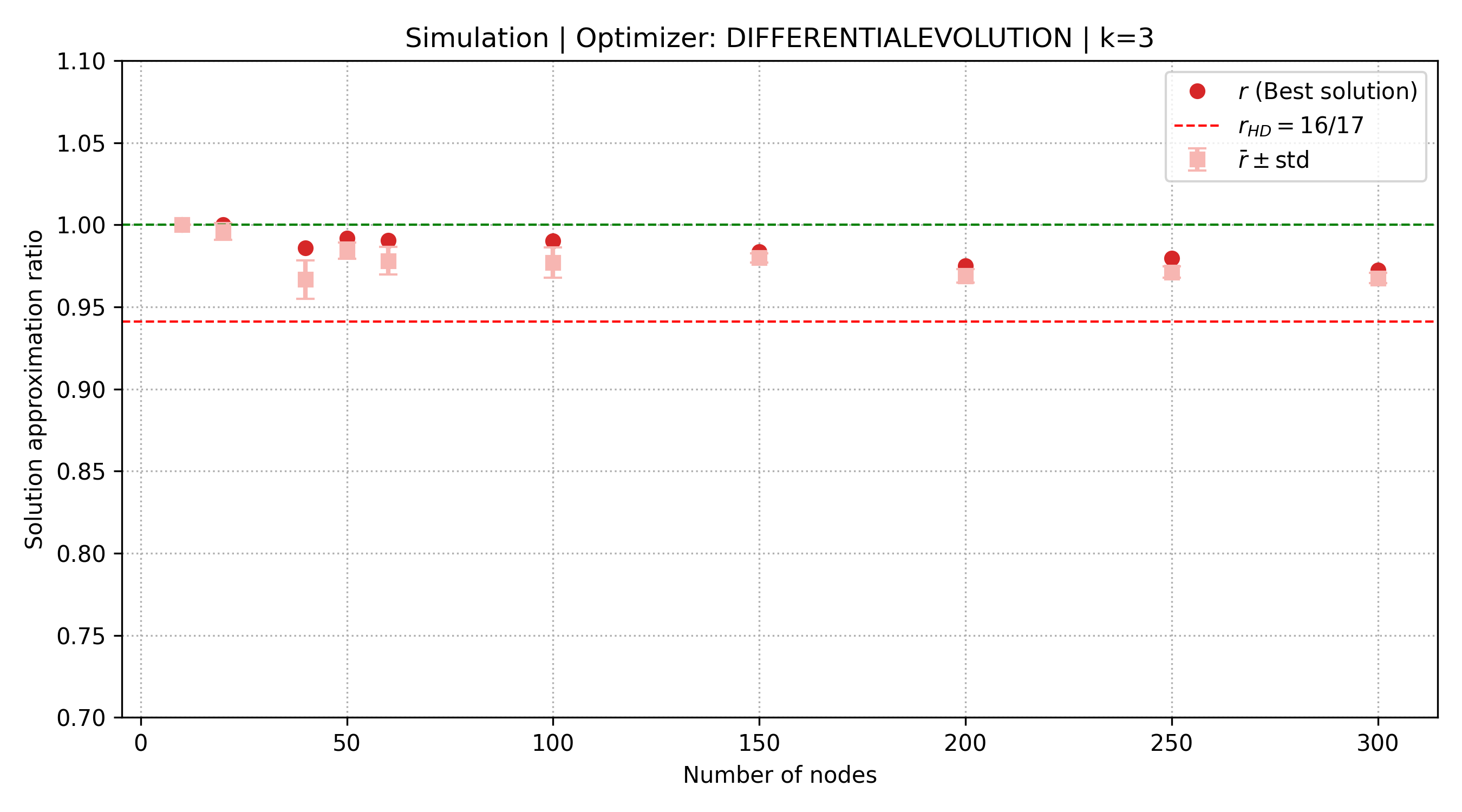}%
    }\\[1ex]
    \subfloat[Solution approximation ratio for $k=4$.]{%
        \includegraphics[width=0.8\textwidth]{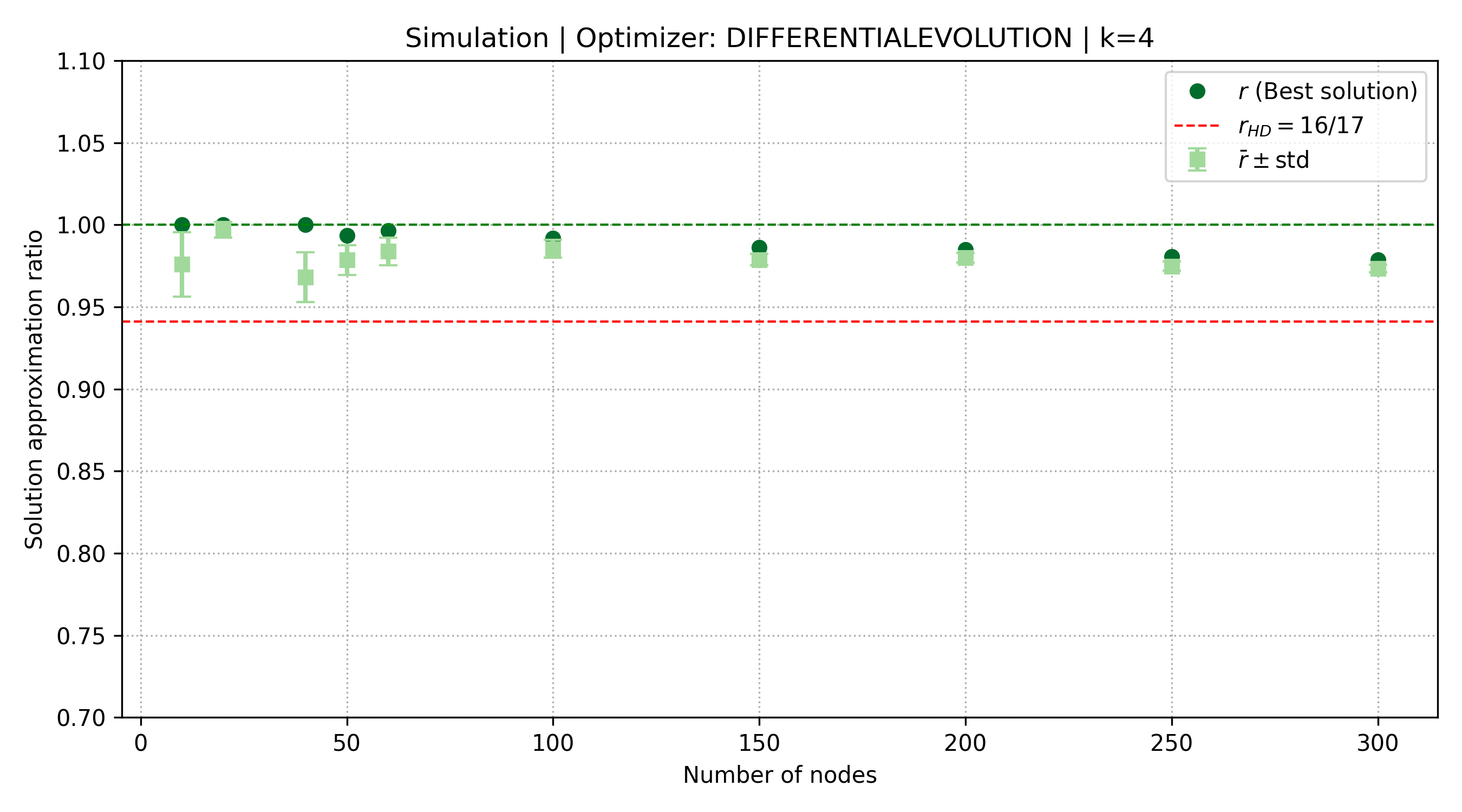}%
    }
    \caption{Results obtained using regularisation for the 10 MCP instances.}
    \label{fig:MC_results}
\end{figure}

The analysis presented in~\cite{Sciorilli_2025} for the MCP problem has been reproduced and
validated, confirming the consistency of its results and adopting the configurations proposed therein. In particular, a hyperparameter scaling given by
\begin{equation*}
\alpha \approx n^{\lfloor k/2 \rfloor}, \quad \beta = 0.5
\end{equation*}
is employed. Finally, all instances have been executed for compression orders $k \in \{2,3,4\}$ using their optimal hyperparameter configurations $\alpha$ and $\beta$. Results are shown in Figure~\ref{fig:MC_results}. To evaluate them adequately, it is necessary to have a metric for graphs whose exact solution $C_{\text{max}} := \max_{\vec{z}}C(\vec{z})$ is generally unknown. We denote the exact approximation ratio~\cite{farhi2014quantum} as
\begin{equation*}
r_{\text{exact}} := \frac{C(z^\ast)}{C_{\text{max}}}
\end{equation*}
Since $C_{\text{max}}$ is, in general, unknown, $r_{\text{exact}}$ can be approximated by the estimated approximation ratio
\begin{equation*}
r := \frac{C(z^\ast)}{C_{\text{best}}},
\end{equation*}
which is computed based on the best known solution $C_{\text{best}}$ available. The latter will be used to compare directly against the benchmark solution. This value $r$ will be less than 1 unless the algorithm finds a better solution to the problem than the known solution.

\begin{figure}[H]
    \centering
    \subfloat[Depth as a function of variables.]{%
        \includegraphics[width=0.5\textwidth]{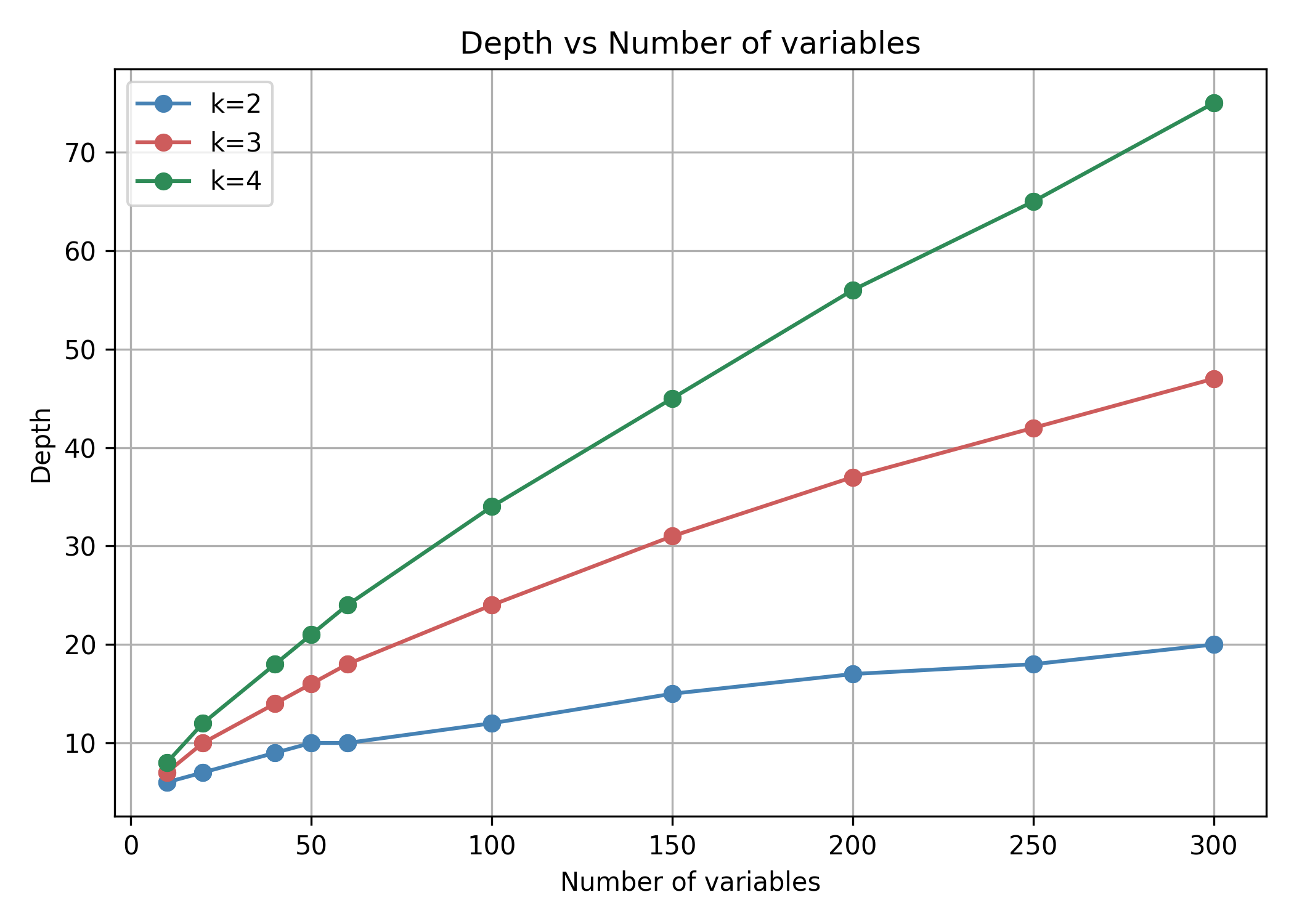}%
    }
    \hfill
    \subfloat[Parameters as a function of variables.]{%
        \includegraphics[width=0.5\textwidth]{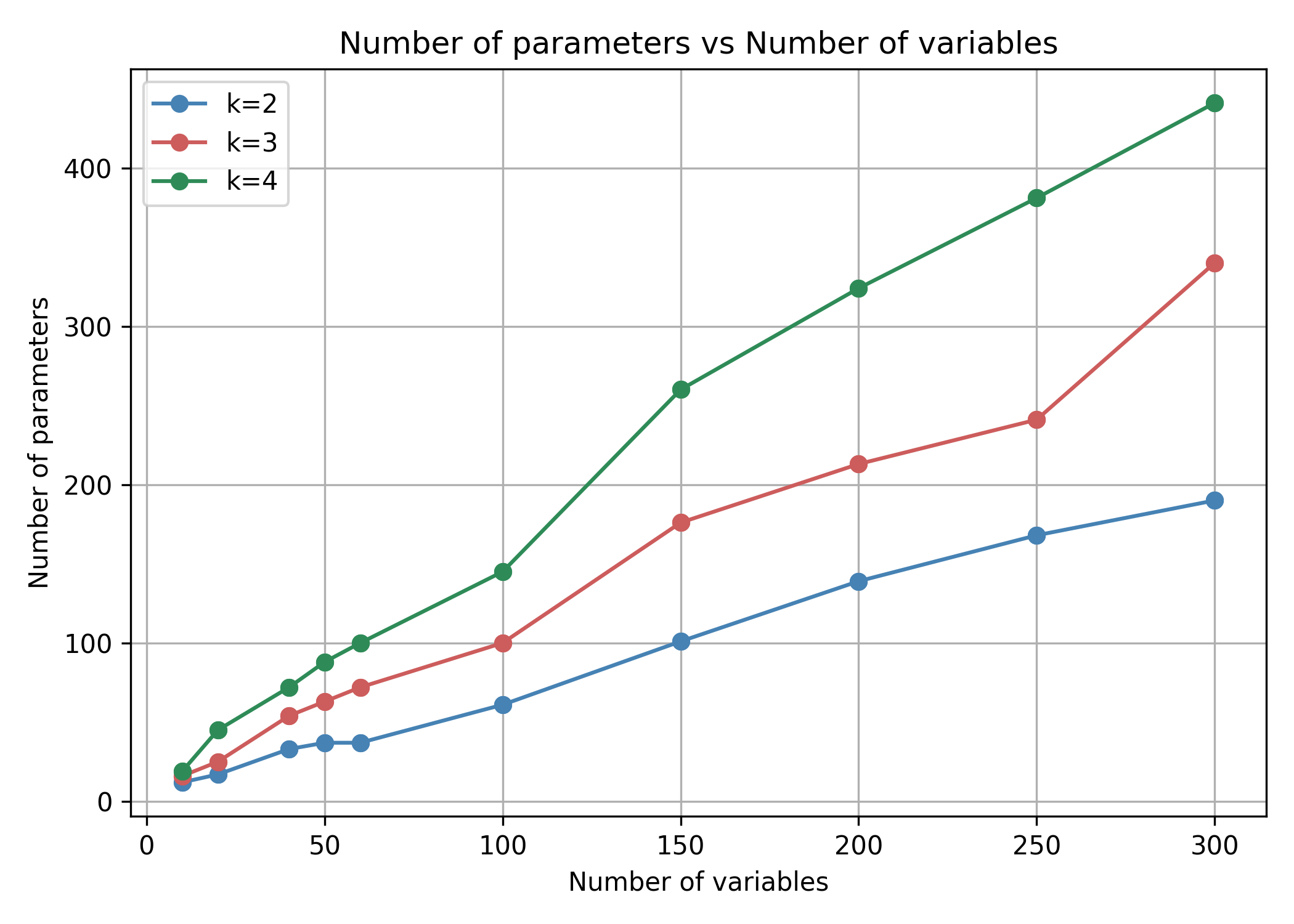}%
    }\\[1ex]
    % \subfloat[Parameters as a function of depth.]{%
    %     \includegraphics[width=0.5\textwidth]{images/Noisy/params_vs_depth.png}%
    % }
    \caption{Scaling of the number of layers and parameters under the assumed configuration.}
    \label{fig:Params_Depth_Nodes}
\end{figure}

Figure~\ref{fig:MC_results} shows the results for 50 different initialisations, indicating that for the majority of compression orders and evaluated instances, the mean and standard deviation of the ratio are close to unity. Similarly, these values also lie within the known theoretical bounds for the problem, whose hardness threshold establishes that it is not possible to exceed $r_{HD}=16/17$ in polynomial time~\cite{Khot_2004}. On the other hand, it is observed that a higher compression order leads to improved solution quality. This behaviour can be explained by the fact that, in the implementation of the algorithm used, the reduction in the number of qubits is compensated by an increase in the depth of the quantum circuits, which in turn enlarges the number of variational parameters, as illustrated in Figure \ref{fig:Params_Depth_Nodes}, consequently enhancing the expressive capacity of the model. Thus, an inherent trade-off is established between expressivity and trainability prior to reaching computational overhead: the combinatorial problem relies on the optimisation of a larger set of parameters, which increases the computational cost but can ultimately yield improved results.

\subsection{Bin Packing Problem}\label{subsec_42}

The BPP is an optimisation problem whose objective is to distribute a series of objects or items, each with a given cost, into the fewest possible bins without exceeding their capacity. Although the BPP may appear simple when framed as a straightforward distribution task, the need to consider all possible assignment combinations introduces a complexity that grows exponentially as the number of elements increases. This combinatorial nature, combined with strict capacity constraints, makes exact resolution of the problem impractical for large instances, which is also why it is considered an NP-hard problem~\cite{Garey_1979}.

Despite this, given its great utility in the field of operational research and industry, this problem has been the subject of notable interest in recent years, and depending on the characteristics of both the objects and the available bins, different variants of it can be formulated~\cite{Martello_1998, Lodi_2010, Liu_2021}. This has led to multiple versions of the problem being addressed using quantum or hybrid methods, as for example in~\cite{Andoin_2022, Romero_2023, Bozhedarov_2024}.

In this case, we address what is known as the one-dimensional BPP~\cite{Munien_2021}, which, mathematically, can be formulated through a set of bins $B = \{b_1, \ldots, b_s\}$ with a maximum capacity $C$ for each bin (identical for all of them) and a set of objects $O = \{o_1, \ldots, o_{\mathcal{N}}\}$ , each of which has an associated weight $w_j$ for each $j\in\{1,\ldots,{\mathcal{N}}\}$, satisfying $0 \leq w_j < C$. The objective is to determine the smallest number of bins ($s_{\text{min}}\leq s$) such that there exists a partition of $O$ into $s_{\text{min}}$ disjoint subsets in such a way that each subset does not exceed capacity $C$.
% \begin{equation*}
% \begin{aligned}
% O &= \bigcup_{i=1}^{s_{\text{min}}} b_i, \\ 
% b_i \cap b_\ell &= \emptyset, \quad \forall i \neq \ell, \\
% \sum_{o_j \in b_i} w_j &\leq C, \quad \forall i \in \{1, \ldots, s_{\text{min}}\}.
% \end{aligned}
% \end{equation*}
Translating this into binary variables, the problem consists of solving
\begin{equation}
\min_{y_j \in \{0,1\}} \sum_{j=1}^{s} y_j
\label{obj_bpp}
\end{equation}
subject to
\begin{align}
&\sum_{j=1}^{s} x_{ij} = 1,
\qquad 1 \le i \le {\mathcal{N}}, \label{rest_bpp_1}\\
&\sum_{i=1}^{\mathcal{N}} w_i \, x_{ij} \;\le\; C \, y_j,
\qquad 1 \le j \le s, \label{rest_bpp_2}
\end{align}
where the variable $y_j$ takes the value 1 if there is any object inside bin $b_j$, and the variable $x_{ij}$ takes the value 1 if object $o_i$ is inside bin $b_j$, which necessarily implies $x_{ij}\leq y_j$. That is, apart from the cost function~\eqref{obj_bpp}, the problem comprises constraints~\eqref{rest_bpp_1} and~\eqref{rest_bpp_2}, which enforce that an object cannot be in multiple bins and that the capacity of the bins used is never exceeded, respectively.

% It is important to note that here we are dealing with a binary problem where variables take values $\{0,1\}$, whilst recall that the original encoding in the PCE algorithm, as defined in~\eqref{cod_pce}, takes values $\{-1,1\}$. Therefore, to guarantee a correct problem construction, a change of variable must be performed to pass from one domain to the other.
To define an cost function that properly adapts to the PCE encoding, with a formulation that resembles a standard QUBO model, it requires encoding $\mathcal{N}^2$ variables corresponding to all possible combinations of $x_{ij}$, together with the worst-case scenario in which all $\mathcal{N}$ bins are used. Consequently, the formulation involves $m = \mathcal{N}^2$ variables, while the problem constraints are incorporated as penalty terms. Specifically, the cost function is defined to include these variables together with the necessary penalties that enforce the restrictions of the problem.

\begin{equation*}
\min_{\vec{y},\vec{x}} F(\vec{y},\vec{x}) =
\min_{\vec{y},\vec{x}}
\lambda_1 Q_1(\vec{y}) +
\lambda_2 Q_2(\vec{x}) +
\lambda_3 Q_3(\vec{y},\vec{x})
\label{obj_bpp_pce}
\end{equation*}
where the different terms are given by
\begin{equation}
Q_1(\vec{y}) = \sum_{j=1}^{s} y_j,
\label{penal_bpp_1}
\end{equation}
\begin{equation}
Q_2(\vec{x}) = \sum_{i=1}^{\mathcal{N}}
\left( \sum_{j=1}^{s} x_{ij} - 1 \right)^2,
\label{penal_bpp_2}
\end{equation}
\begin{equation}
Q_3(\vec{y},\vec{x}) = \sum_{j=1}^{s}
\left[
\max\!\left(
0,\,
\sum_{i=1}^{\mathcal{N}} w_i x_{ij} - C y_j
\right)
\right]^2,
\label{penal_bpp_3}
\end{equation}
Terms~\eqref{penal_bpp_1} and~\eqref{penal_bpp_2} constitute the QUBO-equivalent transformations of the original cost function~\eqref{obj_bpp} and constraint~\eqref{rest_bpp_1}, respectively. %In particular, term~\eqref{penal_bpp_1} penalises the number of active bins, whilst~\eqref{penal_bpp_2} ensures that each object is assigned to exactly one bin.
Term~\eqref{penal_bpp_3} corresponds to a modification of the capacity constraint~\eqref{rest_bpp_2}, with the aim of preventing the quadratic penalty from acting in those cases where the bin capacity is not exceeded.

On the other hand, it is important to note that here we are dealing with a binary problem where variables take values $\{0,1\}$, whilst recall that the original encoding in the PCE algorithm, as defined in~\eqref{cod_pce}, takes values $\{-1,1\}$. Therefore, to guarantee a correct problem construction, the  following change of variables
\begin{equation}
    x_{ij} = \frac{1 + z_{ij}}{2},
\label{ising_to_binary}
\end{equation}
needs to be performed to pass from one domain to the other.

Once the optimisation of $F(\vec{y},\vec{x})$ has been carried out, the corresponding solution is constructed based on the binarisation of the values $x_{ij}$ and $y_j$, with the aim of obtaining a feasible solution. However, note that there is no guarantee of finding feasible configurations. Subsequently, a classical post-processing step is performed to see if a better solution can be found, in which bins that contain only a single object, if there are in the initial solution, are traversed and reasigned to not full bins if possible.

% In order to adequately evaluate the quality of the results in the following section, a standard metric for this problem known as \textit{Normalised Packing Quality} (NPQ) will be used. This metric is used to assess the quality of a solution by measuring the overall utilisation of available capacity. Given the set of bins used $B_{sol}=\{b_1,\ldots,b_{s_{\text{min}}}\}$ and let $W_{b_k}$ be the total weight assigned to bin $b_k \in B_{sol}$, the total waste is computed as $\sum_{b_k \in B_{sol}} \max(0, C - W_{b_k})$ and the metric is defined as
% \begin{equation*}
% \mathrm{NPQ} = \left( 1 - \frac{\sum_{b_k \in B_{sol}} \max(0, C - W_b)}{\mathcal{N}\cdot C} \right) \in [0,1].
% \end{equation*}
% A value close to $1$ indicates efficient use of total capacity, whilst low values reflect greater waste in the assignment of objects to bins.

\subsubsection{Experiments and results}

\begin{figure}[h]
    \centering
     \subfloat[$k\in\{1,2,3,4\}$]{\includegraphics[width = 0.5\textwidth]{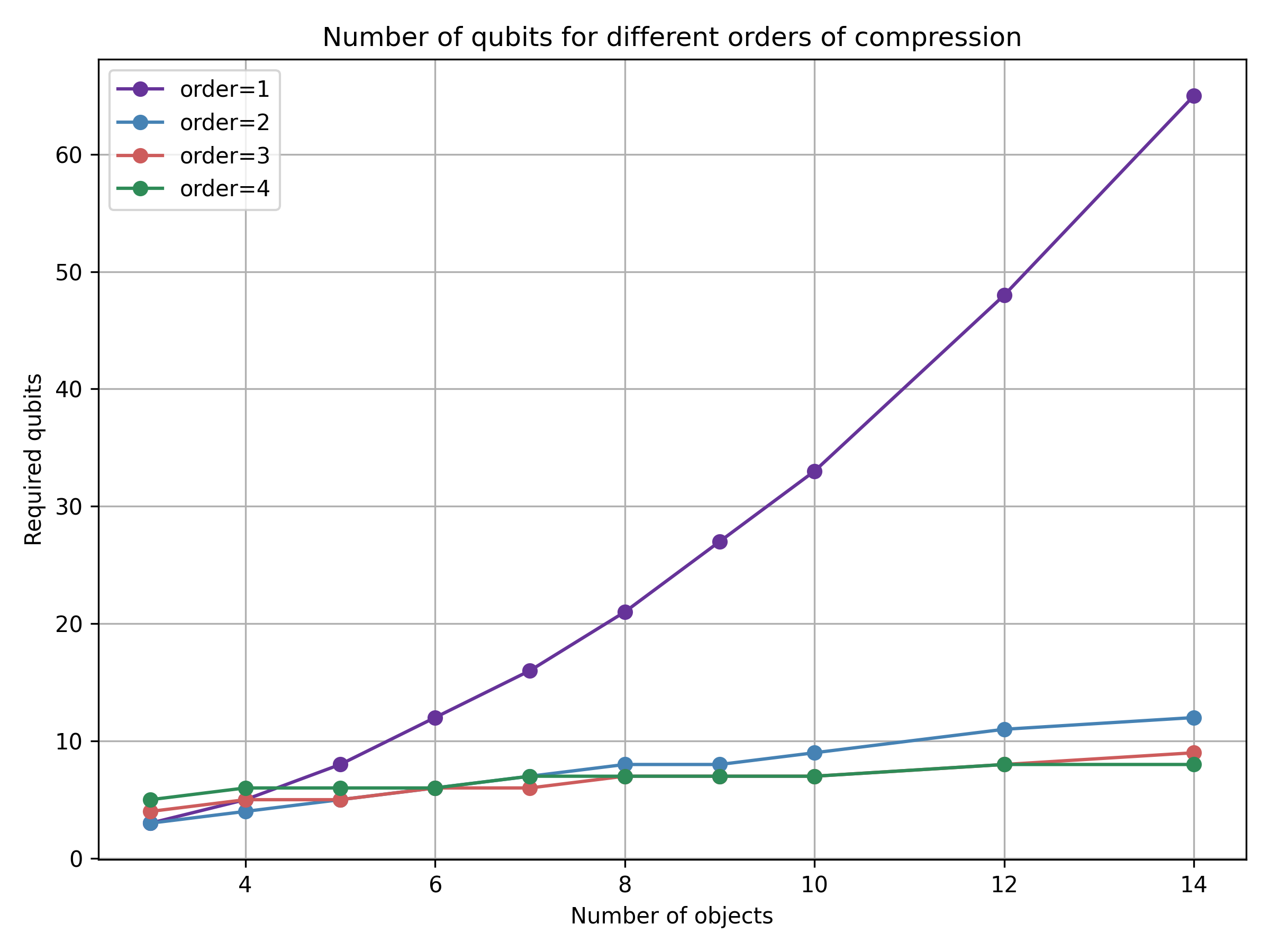}}
     \subfloat[$k\in\{2,3,4\}$]{\includegraphics[width = 0.5\textwidth]{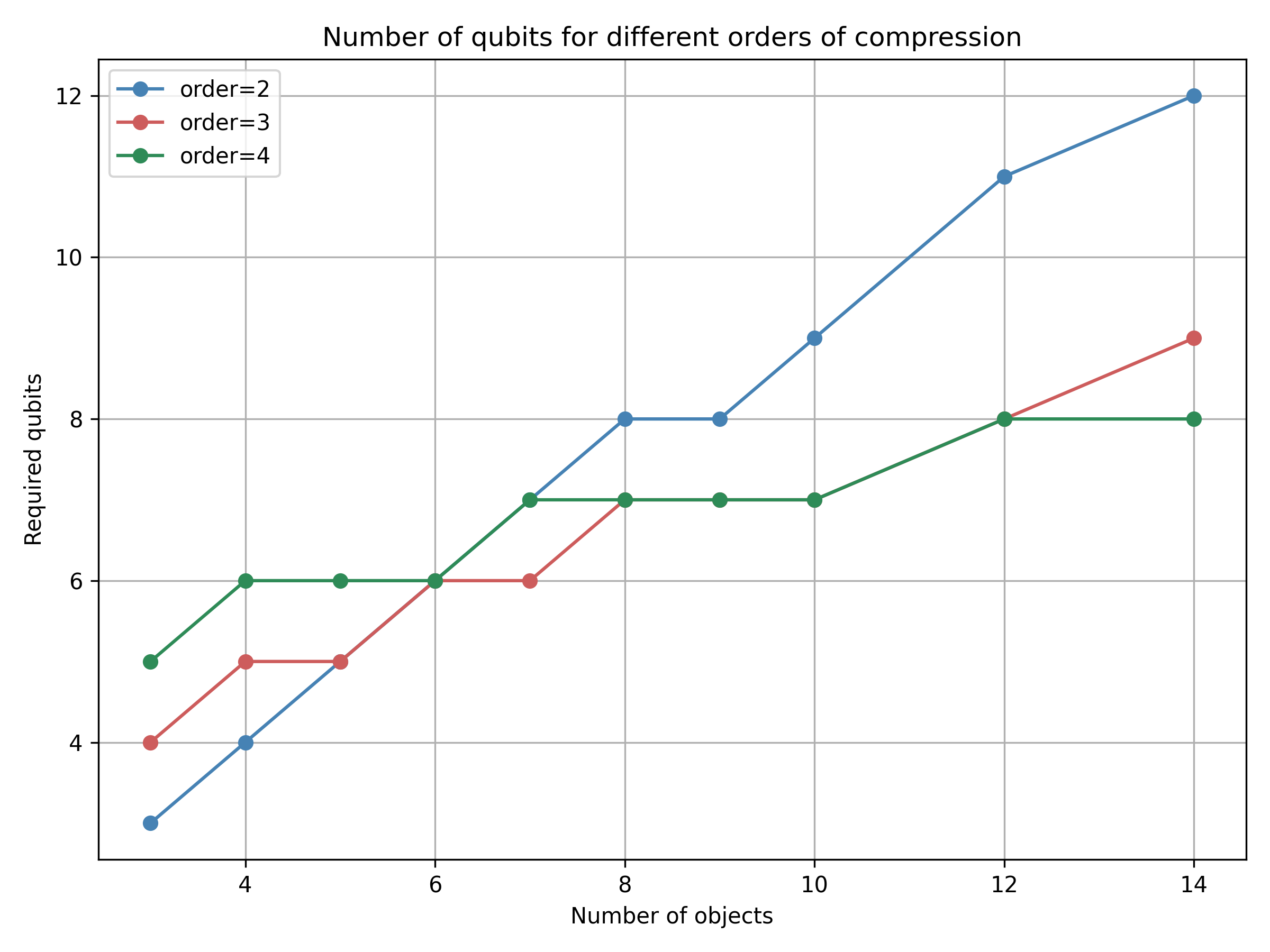}}
    \caption{Scaling of the number of qubits as a function of the number of variables for the BPP.}
    \label{fig: BPP_num_qubits}
\end{figure}

The \texttt{QOPTLib} dataset considers 10 instances of the BPP, characterised by a number of nodes ${\mathcal{N}}\in \{3,4,5,6,7,8,9,10,12,14\}$. Since the model requires encoding $m=({\mathcal{N}}^2+{\mathcal{N}})$ binary variables, the number of qubits required scales according to Figure~\ref{fig: BPP_num_qubits}.

Among the various studies carried out to characterise the problem, the first consisted of analysing the influence of the penalty parameters associated with the assignment and capacity constraints, namely $\lambda_2$ and $\lambda_3$, together with the binarisation hyperparameter $\alpha$. In practice, $\lambda_1$ was fixed to unity and several parameter sweeps were performed over $(\lambda_2,\lambda_3,\alpha)$ for different problem instances and the resulting solutions were compared according to their feasibility and the number of bins obtained after reconstruction and post-processing. These experiments showed that, although no single parameter combination consistently provides the best performance across all instances, there exists a relatively robust region in parameter space. In particular, the capacity constraint plays a more critical role in determining the feasibility of the solutions than the assignment constraint. Consequently, it is generally advisable to operate in a regime where
\begin{equation*}
    \lambda_3 \geq \lambda_2 > \lambda_1,
\end{equation*}
since the capacity constraint~\eqref{penal_bpp_3}, tuned by $\lambda_3$, is the main determinant of feasibility and should therefore dominate the optimisation process. Violations of this term lead to overloaded bins and hence infeasible packings. The assignment constraint~\eqref{penal_bpp_2}, tuned by $\lambda_2$, which ensures that each item is assigned to exactly one bin, should receive a lower but still significant penalty. Finally, the objective term~\eqref{penal_bpp_1}, tuned by $\lambda_1$, which minimizes the number of used bins, should have the smallest weight so that feasibility is prioritised before bin-count reduction, even if this temporarily favours solutions using more bins than necessary. Empirically, it was observed that a range of values produced similar behaviour, with good performance generally obtained for $\lambda_2 \in (50,100)$ and $\lambda_3 \in (100,150)$. Based on this observation, we adopted
\begin{equation*}
\{\lambda_1, \lambda_2, \lambda_3\} = \{1, 50, 100\},
\end{equation*}
as the configuration for our experiments, although it should be noted again that other choices are also possible.

From the aforementioned tests, it was observed that the hyperparameter $\alpha$ influences the solution obtained after optimisation. This was later confirmed in a second analysis, which consisted of incorporating a regularisation term
\begin{equation}
Q_{reg}(\vec{x}) = \beta ~\lambda_{reg}\Big( \frac{1}{m}\sum_{i=1}^\mathcal{N} \sum_{j=1}^\mathcal{N}(x_{ij})^2\Big)^2,
\label{BPP_reg_term}
\end{equation}
dependent on a new hyperparameter $\beta$, and performing several sweeps over both hyperparameters in order to determine which values yield the best solutions. The choice $\lambda_{reg} =\lambda_1$ was made so that the contribution of the regularisation term is sufficiently significant to smooth and facilitate the optimisation process, without dominating the cost function.

\begin{figure}[H]
    \centering
    % ---------------- Row 1 ----------------
    \subfloat[Mean initial cost for $k=2$.]{\includegraphics[width=0.51\textwidth]{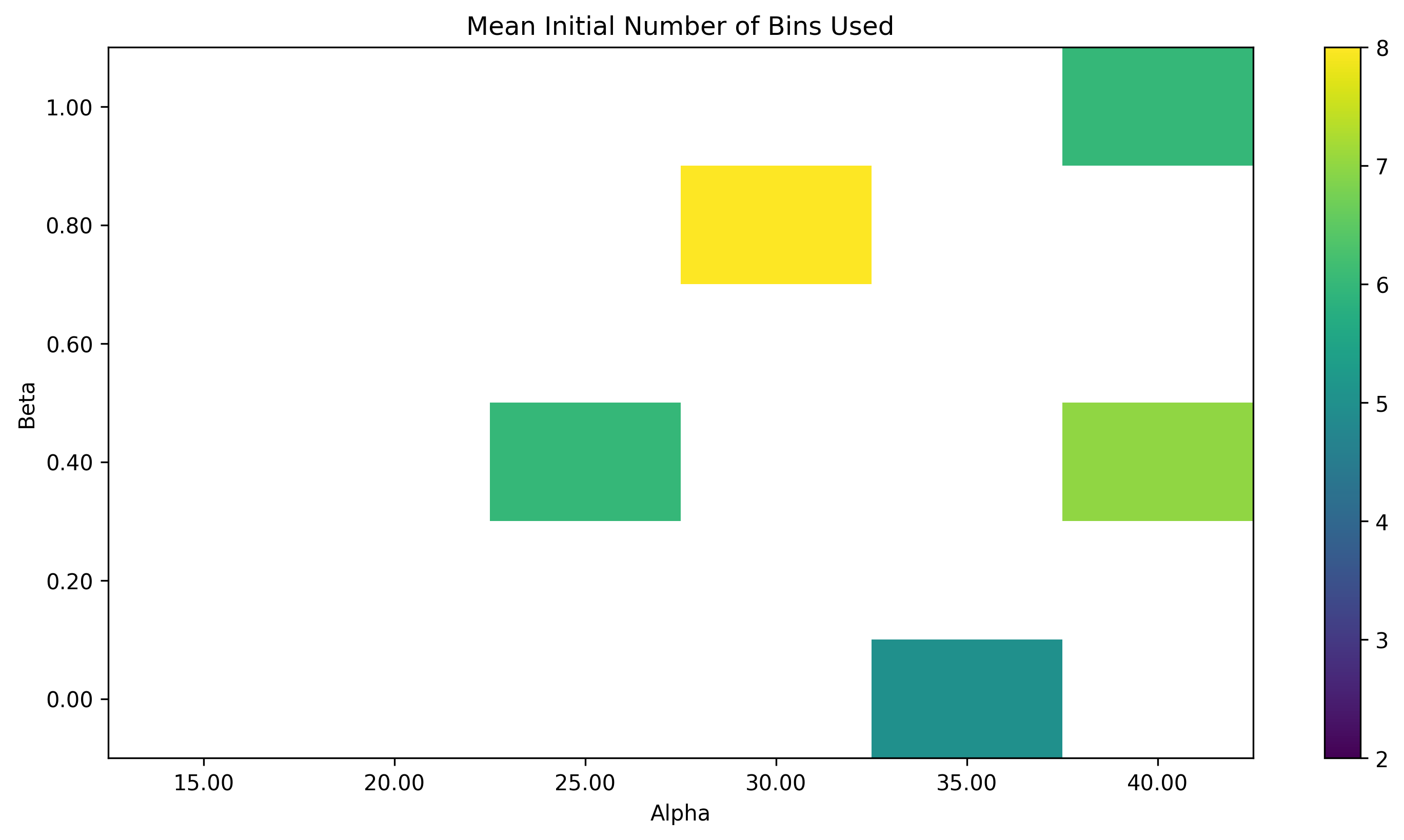}}
    \subfloat[Mean post processed cost for $k=2$.]{\includegraphics[width=0.51\textwidth]{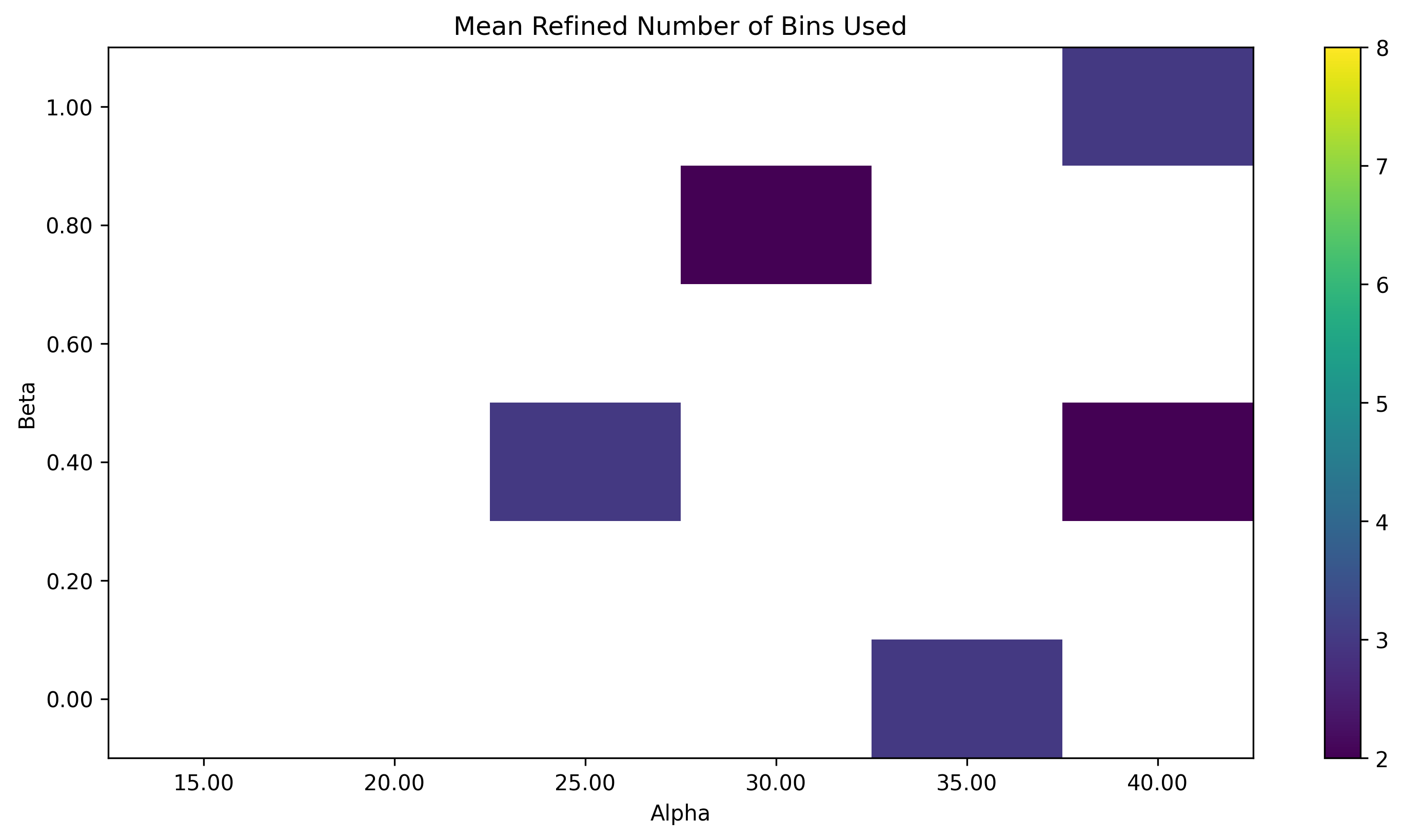}}\\[1ex]

    % ---------------- Row 2 ----------------
    \subfloat[Mean initial cost for $k=3$.]{\includegraphics[width=0.51\textwidth]{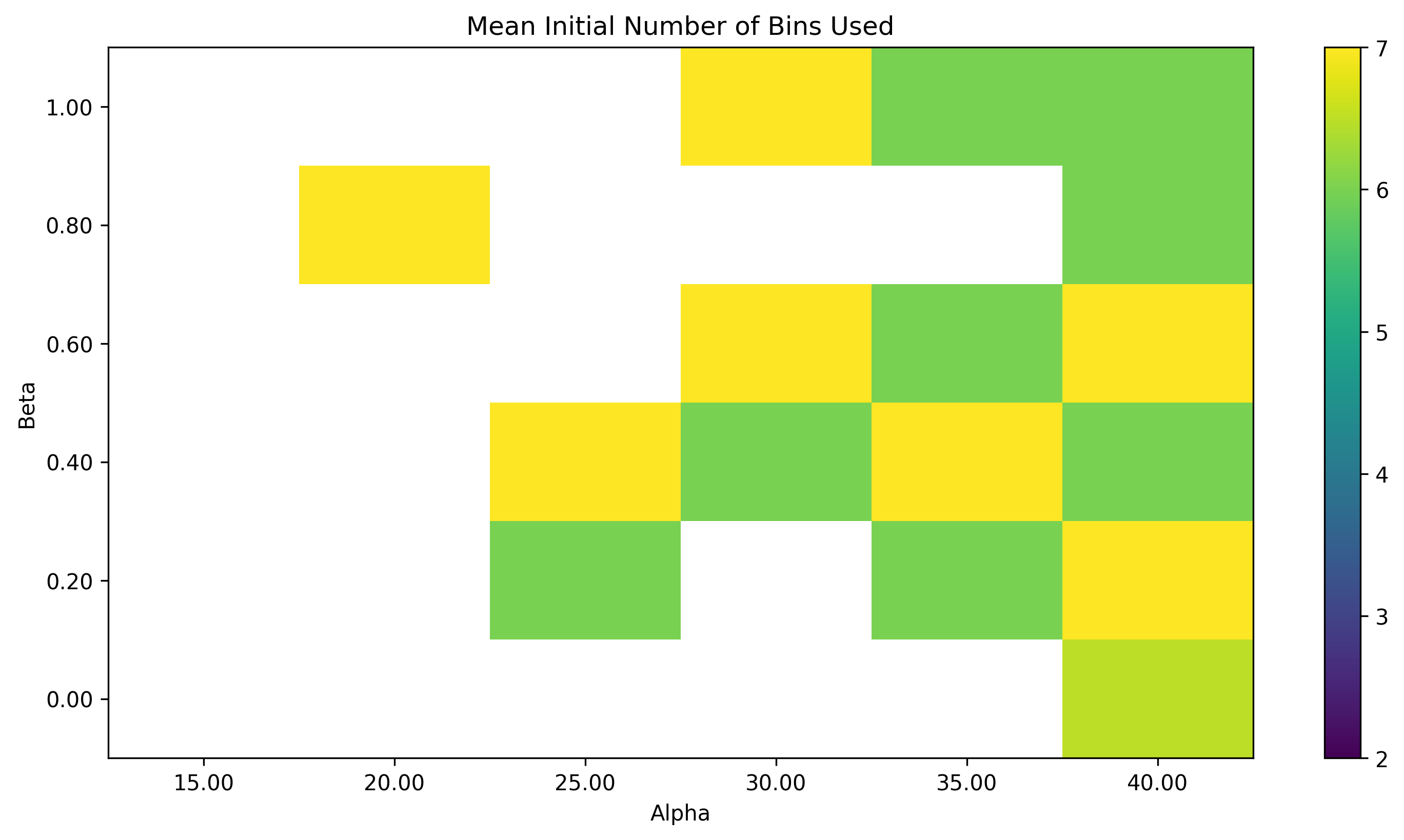}}
    \subfloat[Mean post processed cost for $k=3$.]{\includegraphics[width=0.51\textwidth]{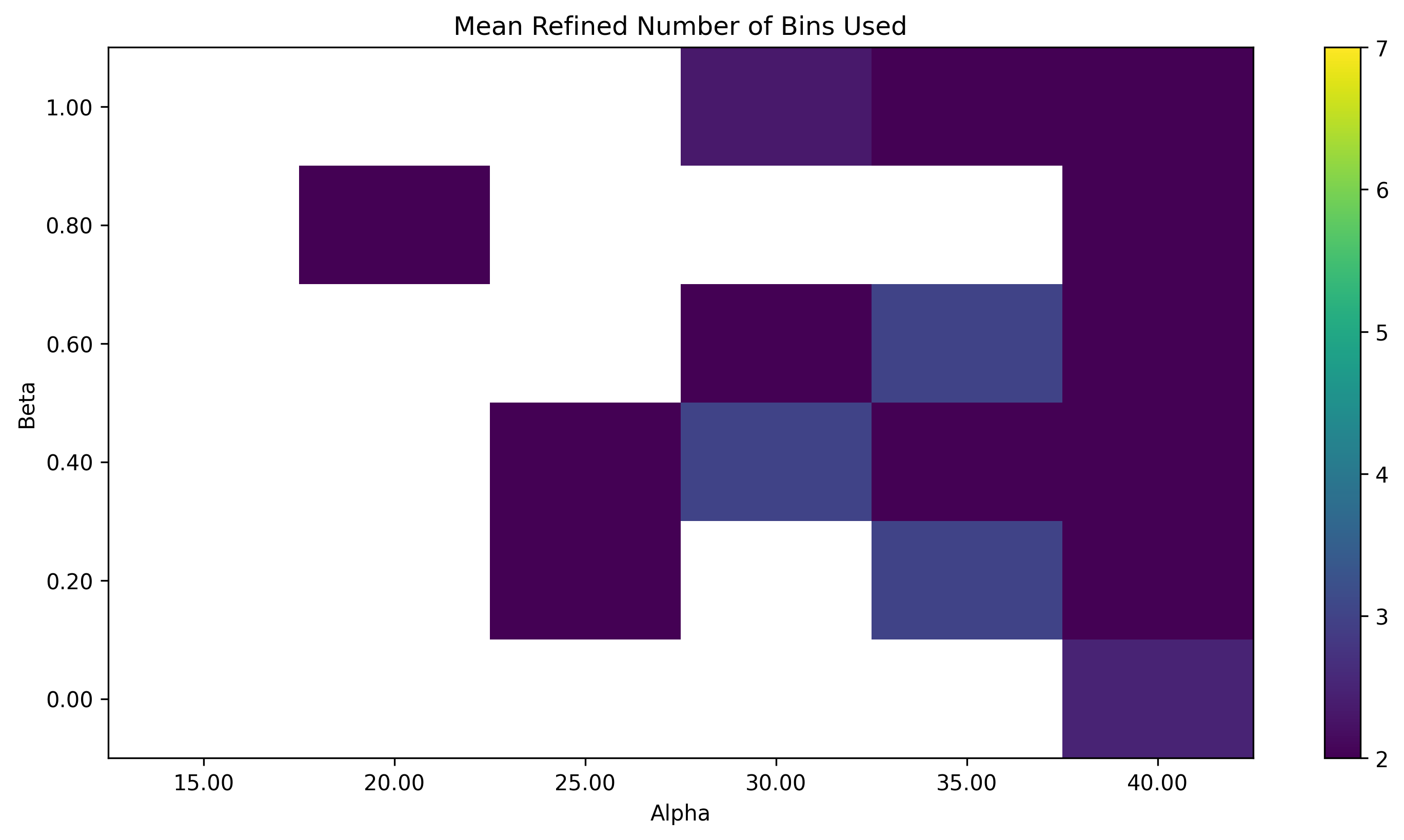}}\\[1ex]
    \caption{Heat map of solutions for instance ${\mathcal{N}}=8$ of the BPP, as a function of $\alpha$ and $\beta$, averaged over five random initialisations.}
    \label{fig: BPP_barrido}
\end{figure}

As an example, Figure~\ref{fig: BPP_barrido} presents a heatmap of the minimum number of bins obtained under feasibility constraints, where blank regions correspond to parameter combinations for which no feasible solution was found. Although it is not clear to extract a relationship between the number of variables and the optimal value of $\alpha$, as the best-performing value varies across instances. Nevertheless, $\alpha \in [30,40]$ tends to provide robust performance, while $\beta \in [0.4,0.8]$ improves the quality of the initial solutions in the vast majority of instances.

The post-processing stage further enhances the obtained packings, producing additional reductions in the number of bins. However, the quality of the refined solutions remains strongly dependent on the choice of $\alpha$ and $\beta$, which highlights the importance of properly characterising the hyperparameter landscape and suggests that a prior study remains highly valuable despite its associated computational cost.

% Recently, Padín-Martínez et al.~\cite{Padin_2026} introduced a variant of the PCE algorithm called Iterative-$\alpha$ PCE, which simultaneously optimises this algorithm parameter, showing that results are improved in the case of constrained problems. This possibility has not been implemented in this benchmark, remaining as an open option for future work.

\begin{figure}[H]
    \centering
    \subfloat[Best solution obtained before post-processing, for $k\in\{2,3,4\}$.]{%
        \includegraphics[width=0.87\textwidth]{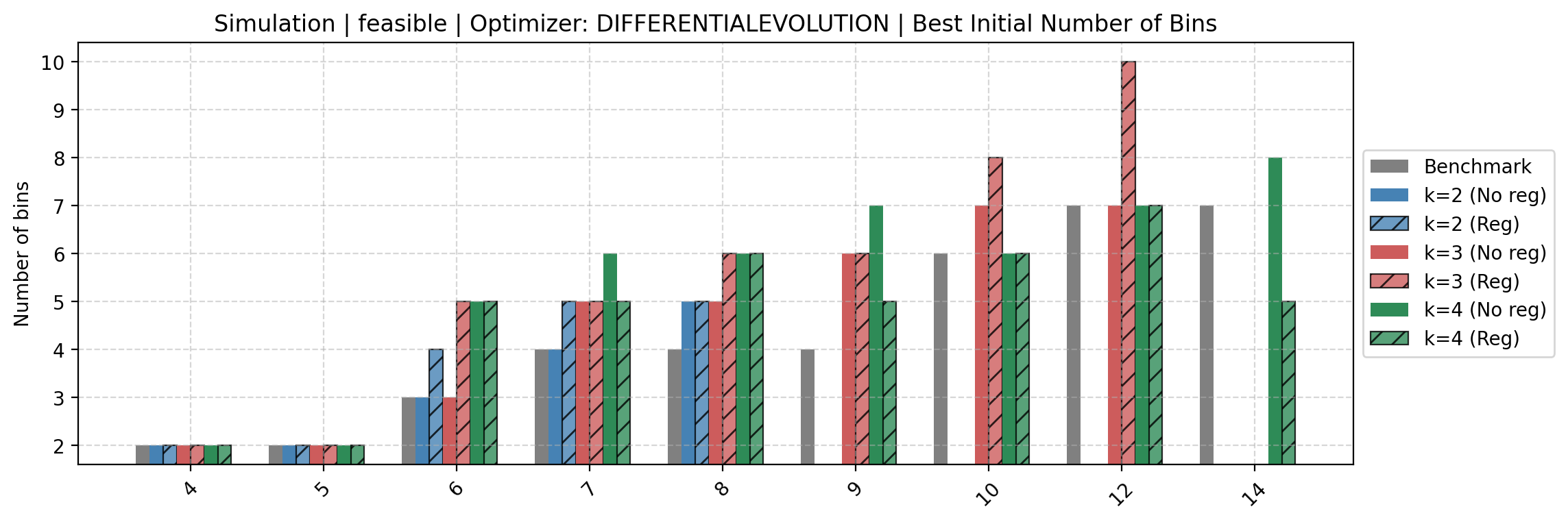}%
    }\\[0.5ex]
    \subfloat[Best solution obtained after post-processing, for $k\in\{2,3,4\}$.]{%
        \includegraphics[width=0.87\textwidth]{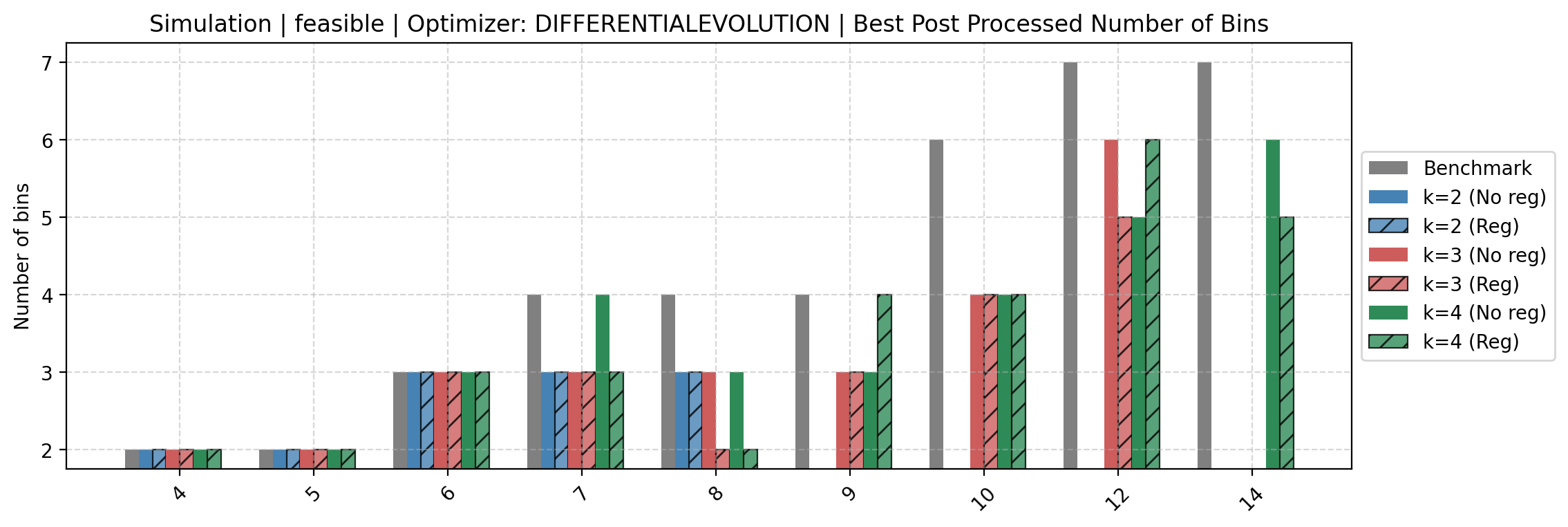}%
    }
    \caption{Comparison of the best solutions obtained with and without regularisation across the 10 BPP instances.}
    \label{fig:BPP_bins_results}
\end{figure}

Finally, 50 random initialisations of all different instances have been run for compression orders $k \in \{2,3,4\}$ using the optimal $\alpha$ and $\beta$ configurations \footnote{For further details about the hyperparameters configuration see \ref{Appendix_A}.}. The results are shown in Figures~\ref{fig:BPP_pct_results} and~\ref{fig:BPP_bins_results}, which report the percentage of feasible solutions and the frequency with which the minimum is reached without post-processing for each instance, together with the smallest number of bins obtained before and after post-processing.

% \begin{figure}[H]
%     \centering
%     \subfloat[Percentage of feasible solutions for $k\in\{2,3,4\}$.]{%
%         \includegraphics[width=0.87\textwidth]{images/BPP/BPP_DIFFERENTIALEVOLUTION_pct_feasible_all_k.png}%
%     }\\[0.5ex]
%     \subfloat[Percentage of feasible solutions that achieved the best objective value for $k\in\{2,3,4\}$.]{%
%         \includegraphics[width=0.87\textwidth]{images/BPP/BPP_DIFFERENTIALEVOLUTION_pct_best_feasible_initial_all_k.png}%
%     }
%     \caption{Comparison of the percentages of feasible and best feasible solutions obtained with and without regularisation across the 10 BPP instances.}
%     \label{fig:BPP_pct_results}
% \end{figure}

% \begin{figure}[H]
%     \centering
%     \subfloat[Best solution obtained before post-processing, for $k\in\{2,3,4\}$.]{%
%         \includegraphics[width=0.87\textwidth]{images/BPP/feasible_compare_best_initial_bins.png}%
%     }\\[0.5ex]
%     \subfloat[Best solution obtained after post-processing, for $k\in\{2,3,4\}$.]{%
%         \includegraphics[width=0.87\textwidth]{images/BPP/feasible_compare_post_best_bins.png}%
%     }
%     \caption{Comparison of the best solutions obtained with and without regularisation across the 10 BPP instances.}
%     \label{fig:BPP_bins_results}
% \end{figure}

The results reveal a clear dependence on the compression order $k$. Feasible solutions are not obtained for every configuration: for $k=2$, feasibility is lost beyond instance $\mathcal{N}=8$, whereas for $k=3$ no feasible solutions are found after instance $\mathcal{N}=12$. Only the $k=4$ configuration is capable of producing feasible solutions for all considered instances. This behaviour can be explained by the increased expressive capacity of the variational model as $k$ grows, since higher compression orders introduce a larger number of variational parameters and therefore provide a richer representation of the solution space.

A similar trend is observed as the size of the BPP instances increases. The percentage of feasible solutions decreases steadily, reflecting the rapid growth of the combinatorial search space and the increasing difficulty of identifying feasible packings. Consequently, the percentage of best solutions among feasible ones must be interpreted alongside the feasibility rate. For the largest instances, where only a small fraction of runs produce feasible solutions, this percentage often approaches $100\%$; however, this is largely a consequence of the limited number of feasible solutions obtained rather than an indication of superior performance. In contrast, instances with moderate feasibility rates typically exhibit percentages below $50\%$, suggesting that while feasible solutions can be found with reasonable frequency, attaining the best objective value remains substantially more challenging.

The analysis of solution quality further supports these observations. The initial solutions obtained require more bins than the benchmark solutions, with the exceptions of instances $m=4,5,10,12,$ and $14$. Nevertheless, as mentioned before, the post-processing procedure consistently improves the initial solutions, reducing the number of bins in every instance considered. These results suggest that the variational algorithm is effective at identifying promising regions of the search space, while the post-processing stage plays an important role in refining the solutions and recovering near-optimal packings.

The results obtained demonstrate that the proposed variational approach is capable of producing good-quality solutions for a broad range of BPP instances. Nevertheless, the experimental analysis also reveals a strong dependence on the specific characteristics of each instance. This behaviour is largely a consequence of the discrete and highly constrained nature of the BPP solution space, where small variations in the input data can lead to significant changes in the optimal packing configuration. As a result, the performance of the variational model is closely linked to its ability to capture the complex correlations and dependencies that govern feasible solutions.

A representative limitation arises from the continuous relaxation of the binary decision variables. Since the optimisation is performed over variables in the interval $[0,1]$, solutions in which an item is partially assigned to multiple bins may exhibit objective values that are nearly indistinguishable from those corresponding to a valid binary assignment. For example, assigning an item simultaneously to two bins with values close to $0.5$ may incur a loss comparable to that of a nearly discrete assignment with value $0.9999$ in a single bin. As a result, the optimisation process may fail to develop a clear preference for one configuration over the other, particularly in instances containing items with identical or very similar weights. However, these limitations do not diminish the overall potential of the approach.

\begin{figure}[H]
    \centering
    \subfloat[Percentage of feasible solutions for $k\in\{2,3,4\}$.]{%
        \includegraphics[width=0.87\textwidth]{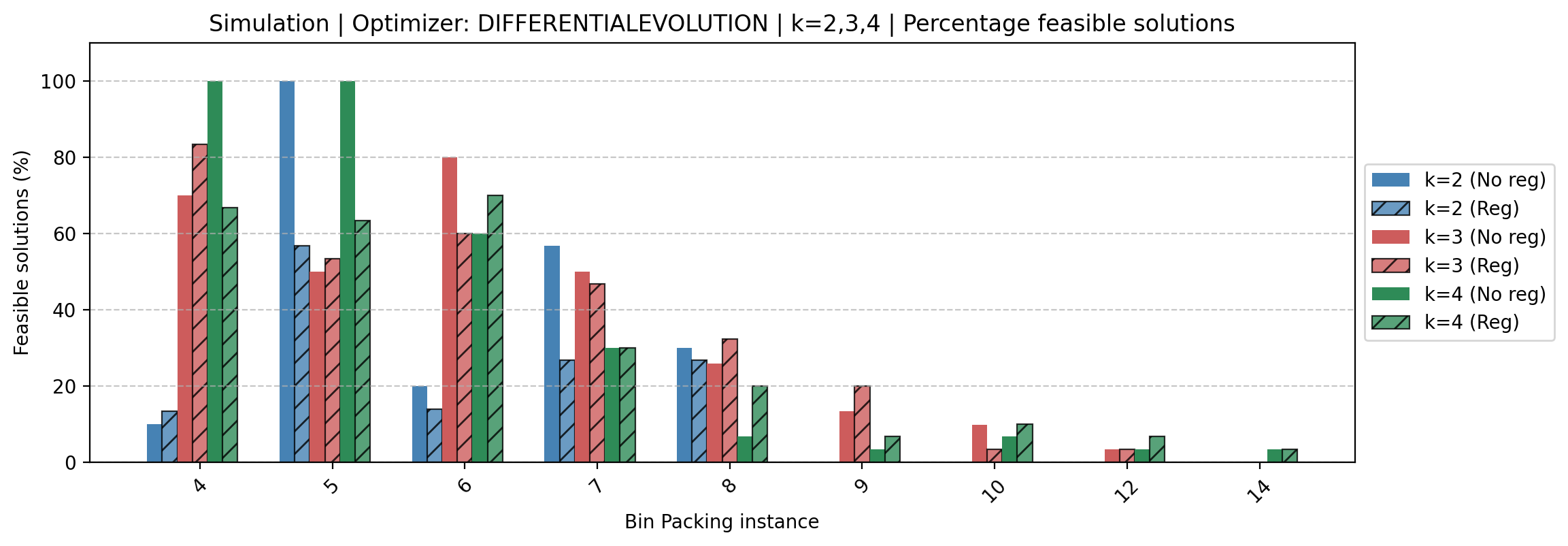}%
    }\\[0.5ex]
    \subfloat[Percentage of feasible solutions that achieved the best objective value for $k\in\{2,3,4\}$.]{%
        \includegraphics[width=0.87\textwidth]{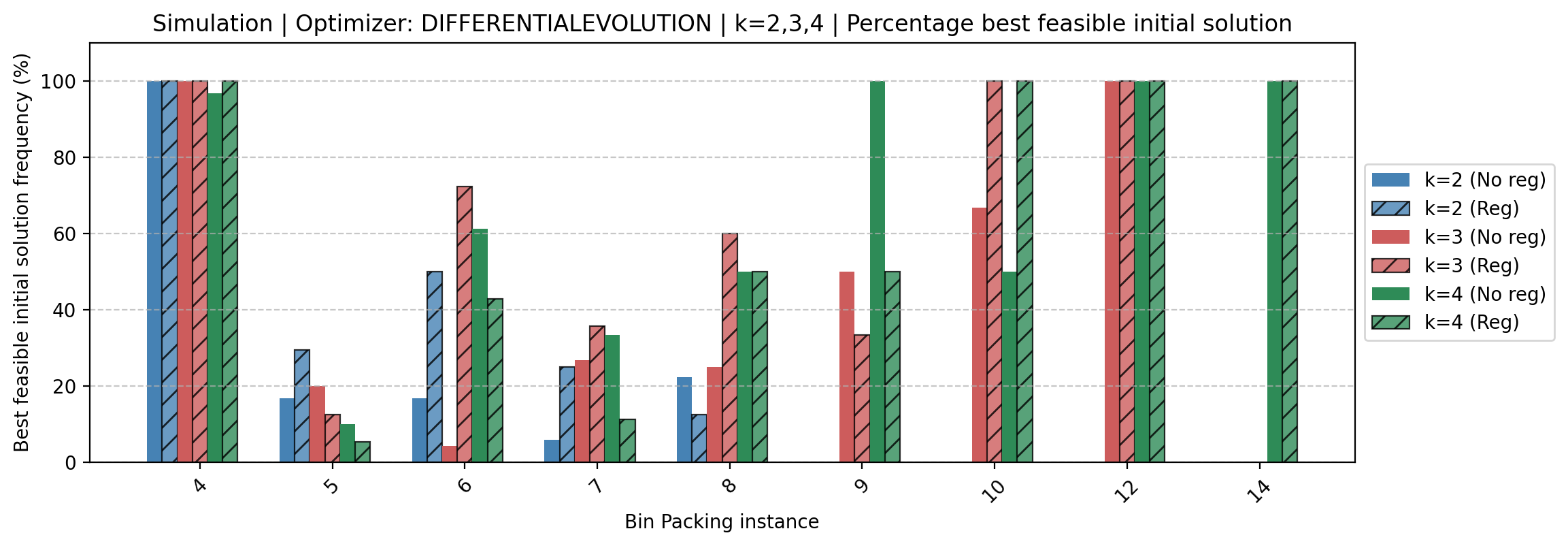}%
    }
    \caption{Comparison of the percentages of feasible and best feasible solutions obtained with and without regularisation across the 10 BPP instances.}
    \label{fig:BPP_pct_results}
\end{figure}

\subsection{Travelling Salesman Problem}\label{subsec_43}

The TSP is an optimisation problem whose main objective is to determine a route or tour of minimum length, starting and ending at the same location whilst visiting all others exactly once. Although the TSP is conceptually simple and can be solved for small graphs with well-defined edge connectivity, it involves strict and unbreakable constraints that make scaling in size computationally intractable. This is why it belongs to the class of NP-hard problems~\cite{Korte_2008, Feinstein_2012}.

Given the large number of real-world applications of this problem~\cite{Matai_2010}, significant efforts have been devoted to the development of efficient algorithms for solving the TSP, making it one of the first combinatorial optimisation problems to be tackled using quantum or hybrid approaches~\cite{SystematicLiteratureReview, Srinivasan_2019, Koken_2022}.

Mathematically, it can be formulated through a complete graph $G = (V, E)$, where $V = \{v_1, v_2, \ldots, v_{\mathcal{N}}\}$ is the set of nodes and $E = \{(v_i, v_j) ~|~ \forall v_i, v_j \in V,\; i \neq j\}$ is the set of edges between these nodes, where each node has an associated cost $c_{ij}$, which is independent of the direction of travel, that is, $c_{ij} = c_{ji} > 0$, with $c_{ij} \in \mathbb{R}$.

Thus, the problem consists of solving
\begin{equation}
\min_{x_{ij} \in \{0,1\}} \sum_{i=1}^{\mathcal{N}} \sum_{\substack{j=1 \\ j \neq i}}^{\mathcal{N}} c_{ij} x_{ij}
\label{obj_tsp}
\end{equation}
subject to
\begin{align}
&\sum_{\substack{i=1 \\ i \neq j}}^{n} x_{ij} = 1, 
\quad \forall j \in \{1, \dots,{\mathcal{N}}\}  \label{rest_tsp_1}\\
& \sum_{\substack{j=1 \\ j \neq i}}^{\mathcal{N}} x_{ij} = 1, 
\quad \forall i \in \{1, \dots,{\mathcal{N}}\} \label{rest_tsp_2}\\
& \sum_{i \in S} \sum_{\substack{j \in S \\ j \neq i}} x_{ij} \leq |S| - 1, 
\quad \forall S \subset V,\ S \neq \emptyset,\label{rest_tsp_3}
\end{align}
where $x_{ij} \in \{0,1\}$ takes the value 1 if edge $(v_i,v_j)$ is used in the solution. Apart from the cost function~\eqref{obj_tsp}, the problem comprises constraints~\eqref{rest_tsp_1} and~\eqref{rest_tsp_2}, which enforce that each node has exactly one incoming and one outgoing edge (equivalent to each node being visited exactly once, except the starting node), and constraint~\eqref{rest_tsp_3}, which guarantees the absence of sub-tours, that is, that any subset of nodes $S$ must be left at least once.

To define a cost function compatible with the PCE encoding and consistent with a standard QUBO formulation, it is necessary to reinterpret the meaning of $x_{ij}$ as in~\cite{Qian_2023}. In this setting, each binary variable $x_{ij} \in \{0,1\}$ indicates whether node $v_i$ is assigned to position $j$ in the resulting cycle and, if we consider $i \in \{1,\ldots,\mathcal{N}\}$ and $j \in \{1,\ldots,\mathcal{N}+1\}$ the total number of binary variables would be $m=(\mathcal{N}^2+\mathcal{N})$. In this case, the problem can be defined as
\begin{equation*}
   \min_{\vec{x}} F(\vec{x}) = \min_{\vec{x}} \lambda_1 Q_1(\vec{x}) + \lambda_{2}Q_{2}(\vec{x}) + \lambda_{3}Q_{3}(\vec{x}),
\label{obj_tsp_pce}
\end{equation*}
where
\begin{align}
& Q_1(\vec{x}) = \sum_{i=1}^{\mathcal{N}}\sum_{j=1}^{\mathcal{N}} \sum_{k=1}^{{\mathcal{N}}} c_{ij} \, x_{ik} \, x_{jk+1}\label{penal_tsp_1}\\
& Q_2(\vec{x})= \sum_{i=1}^{\mathcal{N}} \left(1- \sum_{k=1}^{\mathcal{N}} x_{ik} \right)^2 \label{penal_tsp_2}\\
& Q_3(\vec{ ºx})= \sum_{k=1}^{\mathcal{N}} \left(1- \sum_{i=1}^{\mathcal{N}} x_{ik} \right)^2, \label{penal_tsp_3}
\end{align}
where~\eqref{penal_tsp_1} is the equivalent transformation of~\eqref{obj_tsp},  and~\eqref{penal_tsp_2} and ~\eqref{penal_tsp_3} are the equivalent transformation of~\eqref{rest_tsp_1} and~\eqref{rest_tsp_2}, respectively.

Since the tour is cyclic, the successor of position $\mathcal{N}$ is identified with position $1$, so the transition from the last position back to the first closes the cycle implicitly. Moreover, optimal TSP solutions exhibit rotational symmetry, meaning that multiple configurations related by cyclic shifts represent the same tour.

Both observations allow a simplification of the formulation: the cyclic closure can be treated implicitly and the rotational redundancy can be removed by fixing the starting node, thereby reducing the size of the search space and the associated encoding cost. Without loss of generality, we set node $v_1$ as both the initial and terminal point of the tour and, under this constraint, 
we have that
\begin{equation*}
    x_{1,j}=
\begin{cases}
1 & \text{if }~ j=1\\[4pt]
0 & \text{other case}
\end{cases}
\end{equation*}
and thus the problem requires only $m=(\mathcal{N}-1)^2$ decision variables. Finally, the mathematical formulations is
\begin{equation*}
   \min_{\vec{x}} F(\vec{x}) =
   \min_{\vec{x}}
   \lambda_1 Q_1(\vec{x}) +
   \lambda_2 Q_2(\vec{x}) +
   \lambda_3 Q_3(\vec{x}),
\end{equation*}
where
\begin{align}
Q_1(\vec{x})
&=
\sum_{i=2}^{\mathcal{N}}
\sum_{j=2}^{\mathcal{N}}
\sum_{k=2}^{\mathcal{N}-1}
c_{ij}\, x_{ik}\, x_{j,k+1}
+
\sum_{i=2}^{\mathcal{N}}
c_{1i}\left(x_{i2}+x_{i,\mathcal{N}}\right),
\label{penal_tsp_fixed_1}
\\
Q_2(\vec{x})
&=
\sum_{i=2}^{\mathcal{N}}
\left(
1-\sum_{k=1}^{\mathcal{N}-1} x_{ik}
\right)^2,
\label{penal_tsp_fixed_2}
\\
Q_3(\vec{x})
&=
\sum_{k=1}^{\mathcal{N}-1}
\left(
1-\sum_{i=2}^{\mathcal{N}} x_{ik}
\right)^2 .
\label{penal_tsp_fixed_3}
\end{align}

As with the BPP, since the TSP variables take values $\{0,1\}$, the change of variable in \eqref{ising_to_binary} must be done on the PCE encoding values, to guarantee a correct problem formulation.

Once the optimisation of $F(\vec{x})$ has been carried out, a preliminary solution is reconstructed from the binarisation of the relaxed variables by only extracting the feasible cycles, generating a set of candidate cycles. However, note that there is no guarantee of finding feasible cycles. From this pool of candidate cycles the best is selected without requiring further modification of the underlying construction process. Subsequently, a classical post-processing step based on the 2-opt heuristic is applied to refine the reconstructed cycle. The method systematically examines pairs of non-adjacent edges and performs a 2-opt move by reversing the intermediate segment between them. In other words, given two edges $(v_i,v_{i+1})$ and $(v_j,v_{j+1})$, the reconnection
\begin{equation*}
    (v_i,v_{i+1}),(v_j,v_{j+1})
\rightarrow
(v_i,v_j),(v_{i+1},v_{j+1}),
\end{equation*}
generates a new candidate tour while preserving the Hamiltonian structure of the cycle. If the resulting tour has a shorter total distance, the modification is accepted.

\subsubsection{Experiments and results}\label{Results_TSP}

\begin{figure}[H]
    \centering
     \subfloat[$k\in\{1,2,3,4,5\}$]{\includegraphics[width = 0.5\textwidth]{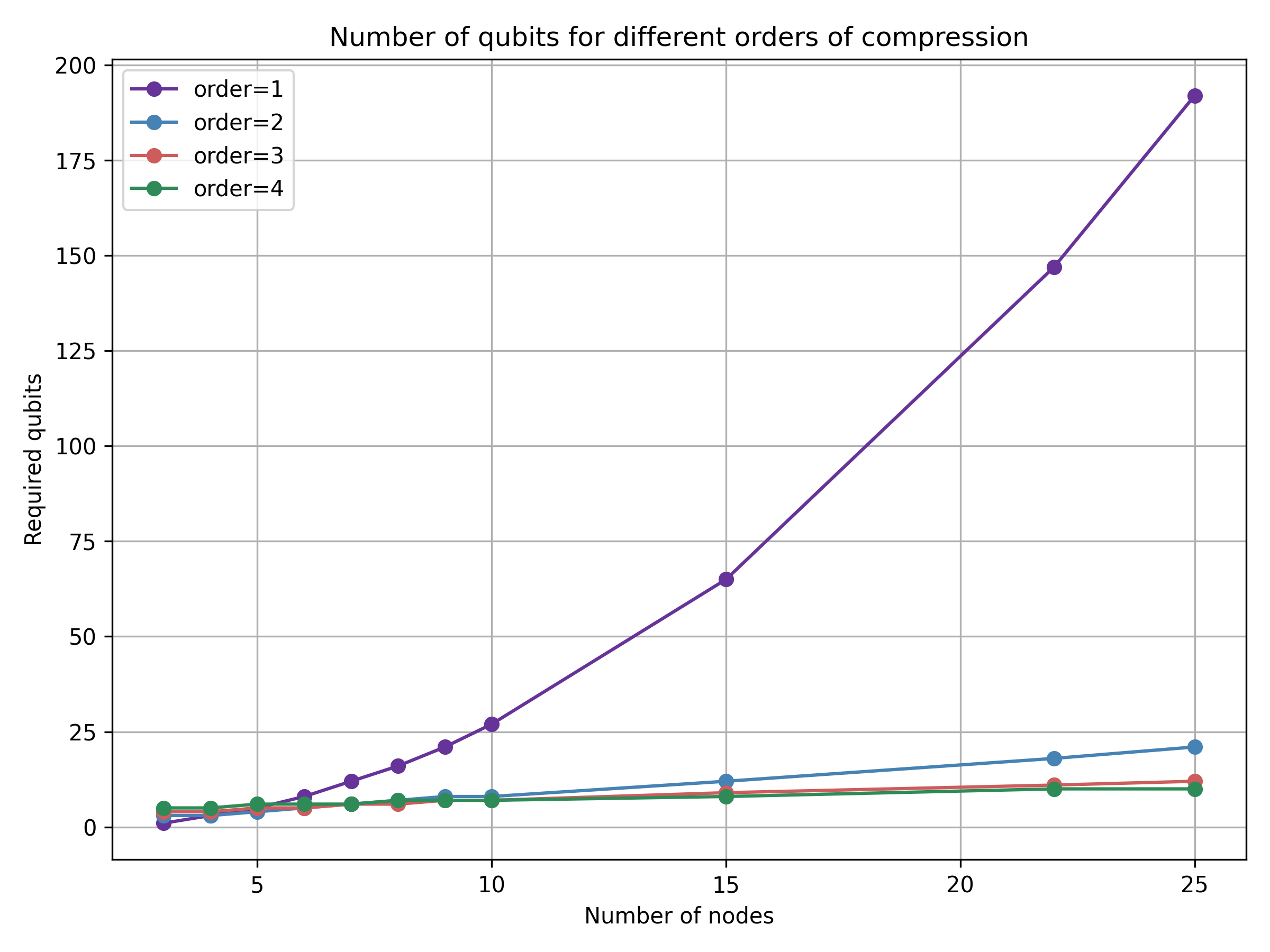}}
     \subfloat[$k\in\{2,3,4,5\}$]{\includegraphics[width = 0.5\textwidth]{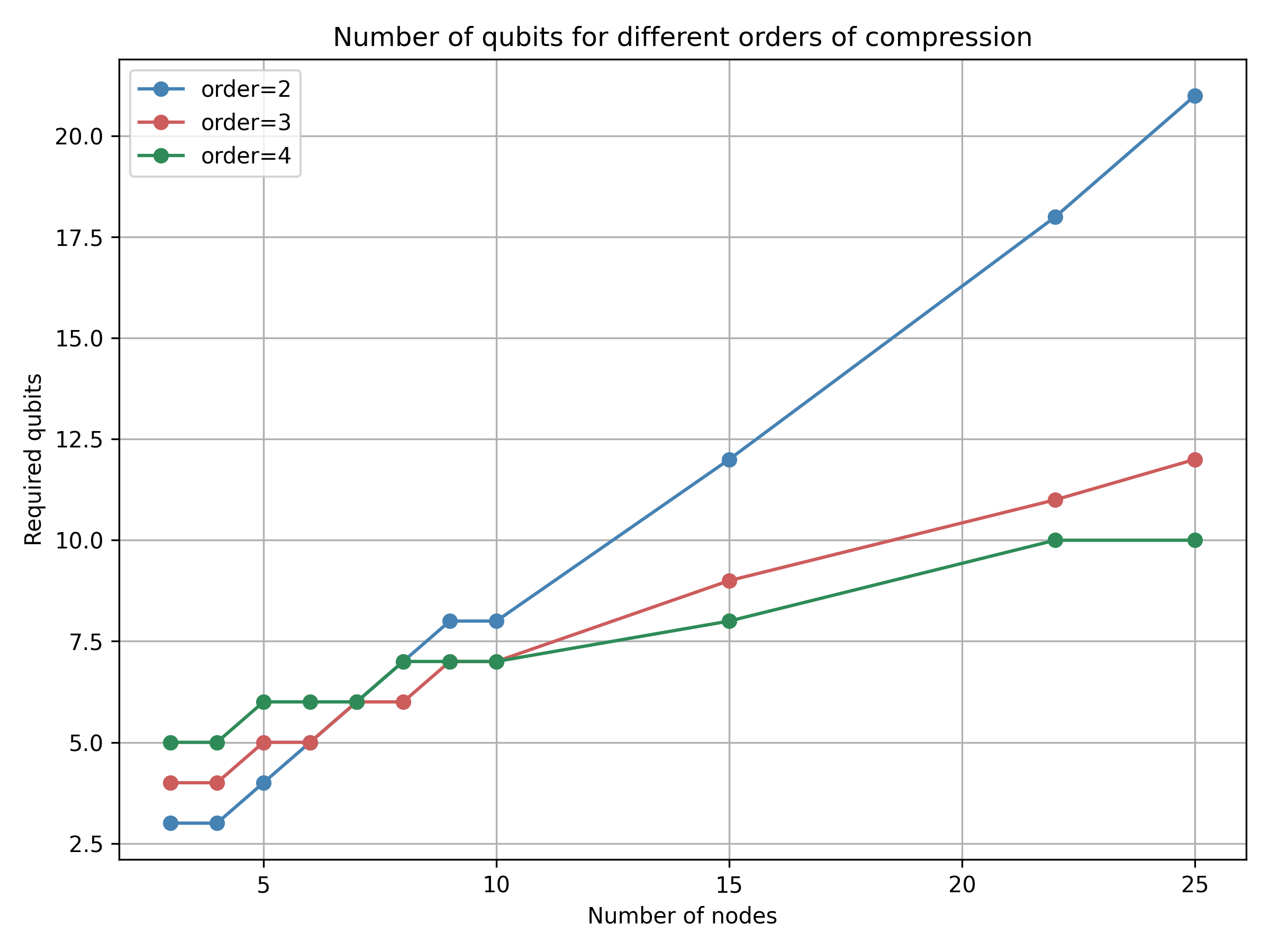}}
    \caption{Scaling of the number of qubits as a function of the number of variables for the TSP.}
    \label{fig: TSP_num_qubits}
\end{figure}

The \texttt{QOPTLib} dataset considers 10 instances of the TSP, characterised by a number of nodes ${\mathcal{N}}\in \{3,4,5,6,7,8,9,10,15,22,25\}$. Since the model requires encoding $m={\mathcal{N}}^2$ binary variables, the number of qubits required scales according to Figure~\ref{fig: TSP_num_qubits}.

\begin{figure}[H]
    \centering
    % ---------------- Row 1 ----------------
    \subfloat[Mean initial cost for $k=2$.]{\includegraphics[width=0.51\textwidth]{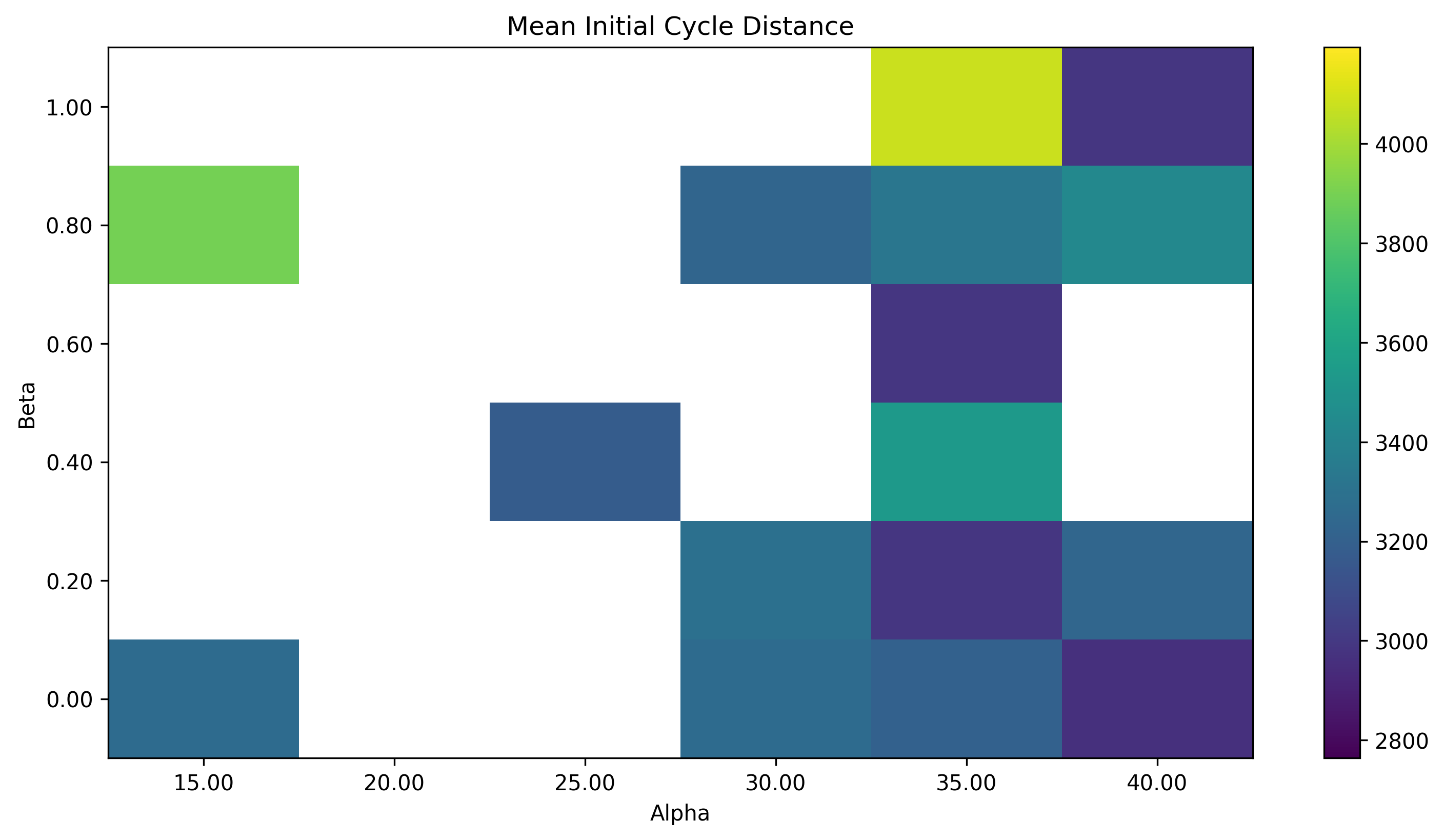}}
    \subfloat[Mean post processed cost for $k=2$.]{\includegraphics[width=0.51\textwidth]{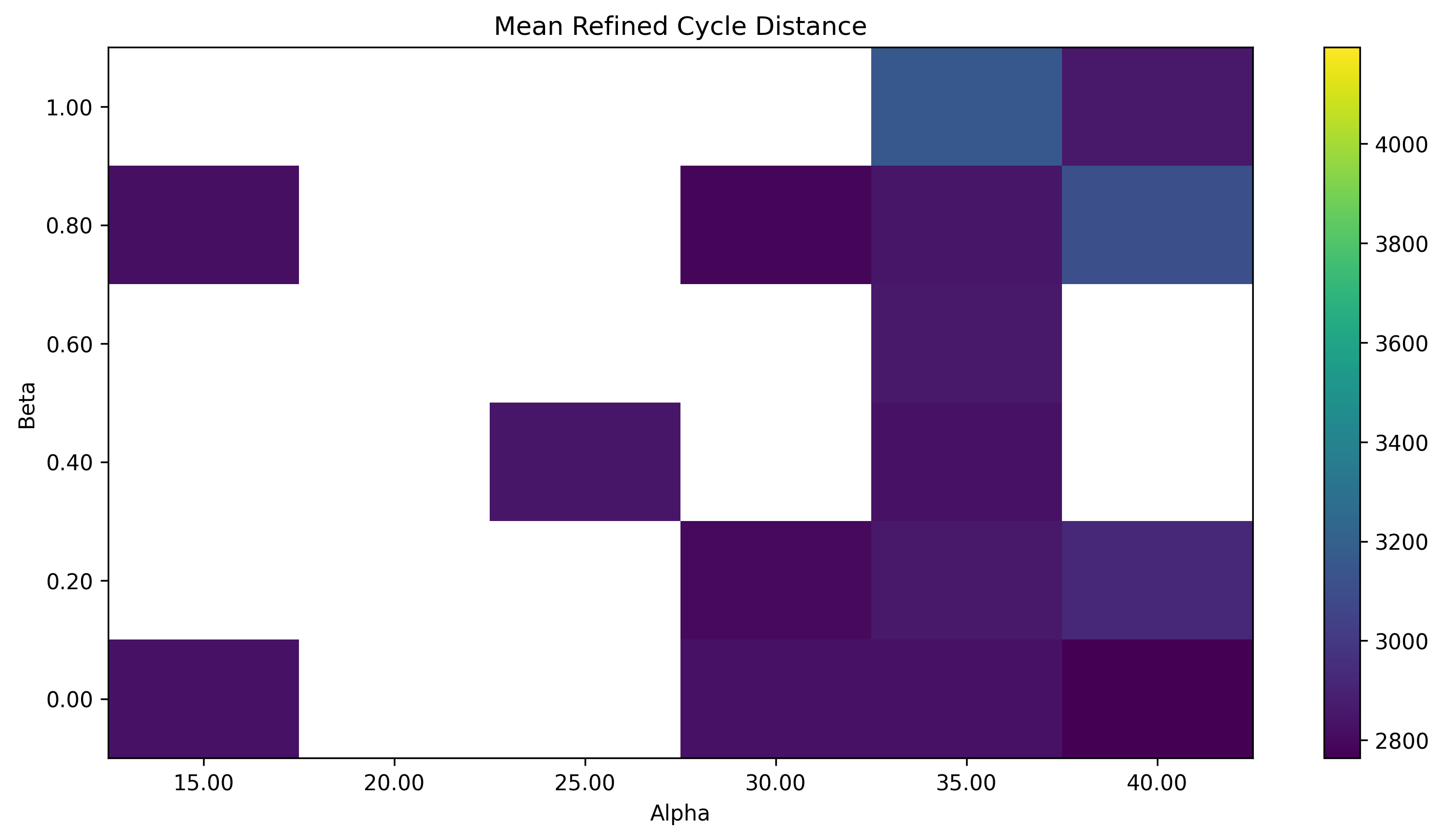}}\\[1ex]

    % ---------------- Row 2 ----------------
    \subfloat[Mean initial cost for $k=3$.]{\includegraphics[width=0.51\textwidth]{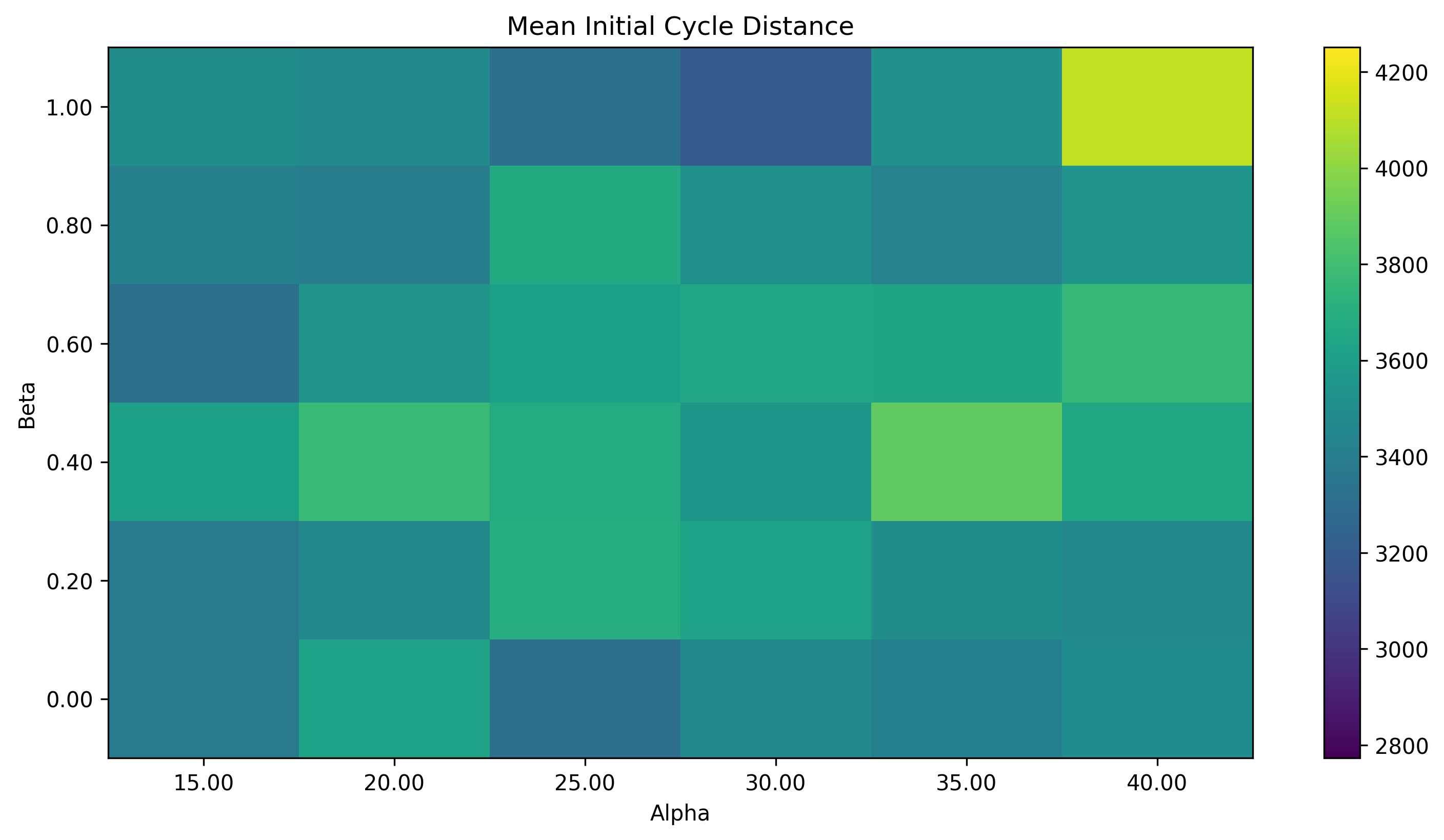}}
    \subfloat[Post processed cost variance for $k=3$.]{\includegraphics[width=0.51\textwidth]{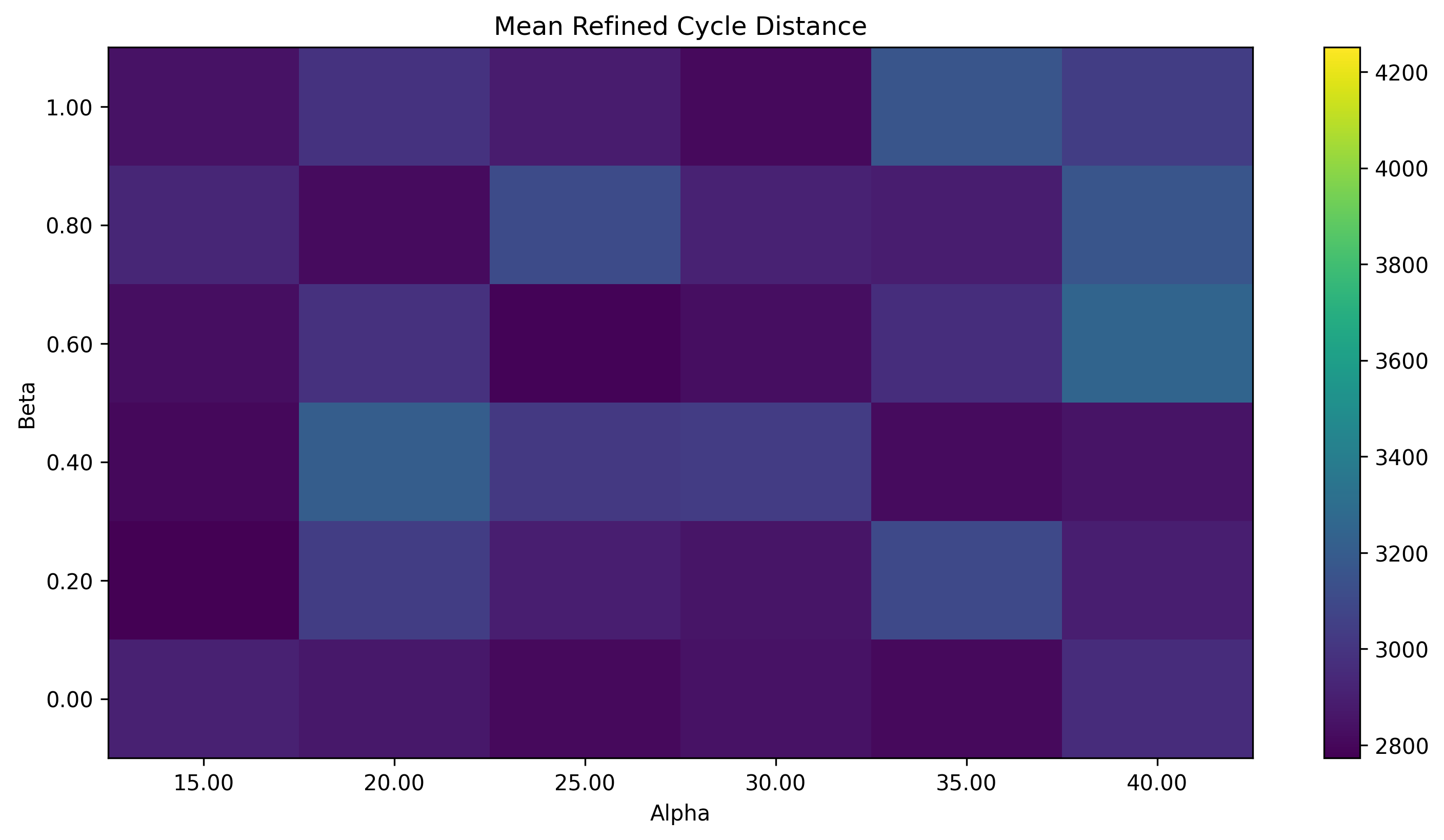}}\\[1ex]
    \caption{Heat map of solutions for instance ${\mathcal{N}}=8$ of the TSP, as a function of $\alpha$ and $\beta$, averaged over five random initialisations.}
    \label{fig: TSP_barrido}
\end{figure}

Among the various studies carried out to characterise the problem, the first followed the same methodology as that employed for the BPP, namely the analysis of the penalty parameters $\{\lambda_1,\lambda_2,\lambda_3\}$ as a function of the hyperparameter $\alpha$. In practice, $\lambda_1$ was fixed to unity, whilst several parameter sweeps were performed over $(\lambda_2,\lambda_3,\alpha)$ for the different TSP instances considered. The quality of the resulting solutions was evaluated according to their feasibility and tour length after reconstruction. From these experiments, it was observed that excessively large penalty values do not necessarily improve the quality of the solutions and may deteriorate them through soft or nearly inactive configurations. Based on the regions that consistently yielded the highest proportion of valid tours, the configuration
\begin{equation*}
    \{\lambda_1,\lambda_2,\lambda_3\}
    =
    \{1,~50\cdot c_{\max},~25\cdot c_{\max}\},
\end{equation*}
was adopted for the experiments reported in this work, where $\lambda_2$ and $\lambda_3$ are scaled by the maximum edge weight of the graph, $c_{\max}$. Nevertheless, it should be emphasised that this choice should not be regarded as universally optimal, since the most suitable penalties may still depend on the size and characteristics of the particular TSP instance under consideration.
As expected, it was also observed that the hyperparameter $\alpha$ strongly influences the solution obtained after optimisation. Following the same approach as before, this was later confirmed in a second analysis, which consisted of incorporating a regularisation term as in \ref{BPP_reg_term}, with $\lambda_{reg} =\lambda_1$, in order to assist the optimisation process rather than hinder it. Using the same analysis, Figure~\ref{fig: TSP_barrido} presents a heatmap of the total distance of the obtained tours under feasibility constraints, where blank regions correspond to parameter combinations for which no feasible solution was found. No clear relationship between the number of variables and the optimal value of $\alpha$ could be identified, as the best-performing value varies significantly across instances. Nevertheless, values of $\alpha \in [30,40]$ tend to provide consistently good performance, while values of $\beta \in [0.6,0.8]$ improve the quality of the initial solutions in the vast majority of cases.

The post-processing stage further improves the obtained tours, leading to additional reductions in the total travelled distance. However, the quality of the refined solutions remains strongly dependent on the choice of $\alpha$ and $\beta$, which highlights the importance of properly characterising the hyperparameter landscape and suggests that a prior hyperparameter study remains highly valuable despite its associated computational cost.

% \begin{figure}[H]
%     \centering
%     % ---------------- Row 1 ----------------
%     \subfloat[Mean initial cost for $k=2$.]{\includegraphics[width=0.51\textwidth]{images/TSP/TSP_m8_barrido_init_k2.png}}
%     \subfloat[Initial cost variance for $k=2$.]{\includegraphics[width=0.51\textwidth]{images/TSP/TSP_std_m8_barrido_init_k2.png}}\\[1ex]

%     % ---------------- Row 2 ----------------
%     \subfloat[Mean post processed cost for $k=2$.]{\includegraphics[width=0.51\textwidth]{images/TSP/TSP_m8_barrido_post_k2.png}}
%     \subfloat[Post processed cost variance for $k=2$.]{\includegraphics[width=0.51\textwidth]{images/TSP/TSP_std_m8_barrido_post_k2.png}}\\[1ex]
%     \caption{Heat map of solutions for instance ${\mathcal{N}}=8$ of the TSP, as a function of $\alpha$ and $\beta$, averaged over five random initialisations.}
%     \label{fig: TSP_barrido}
% \end{figure}

\begin{figure}[H]
    \centering
    \subfloat[Percentage of feasible solutions for $k\in\{2,3,4\}$.]{%
        \includegraphics[width=0.87\textwidth]{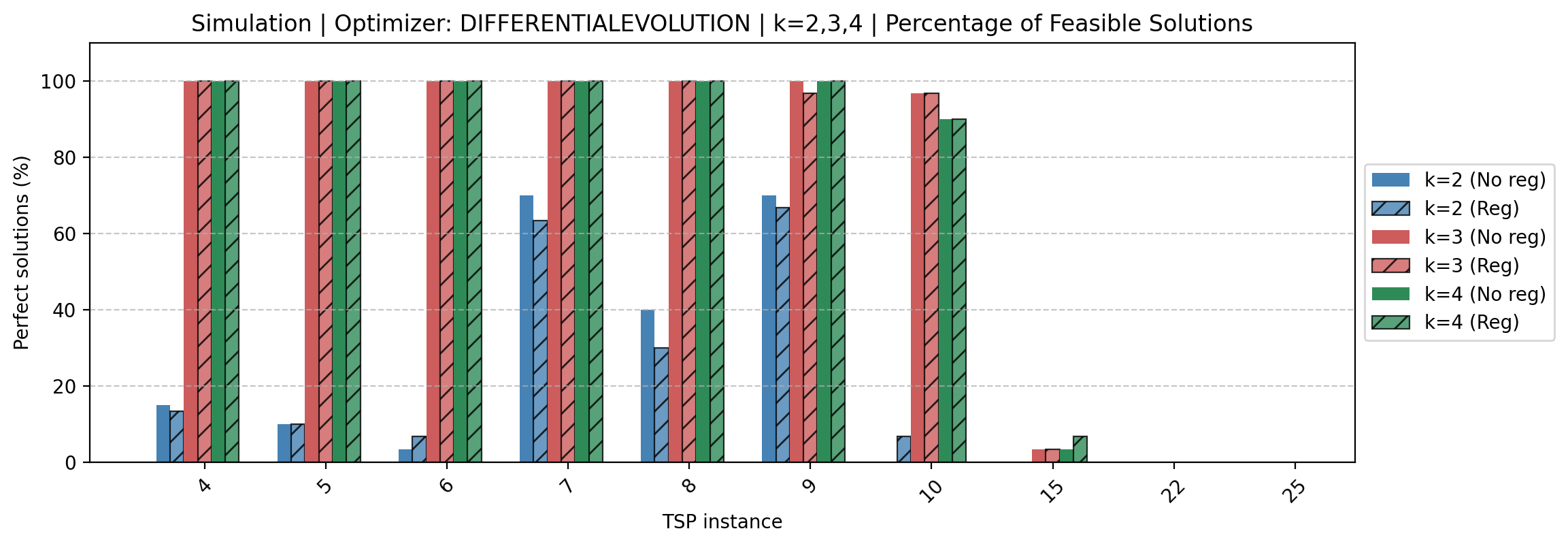}%
    }\\[0.5ex]
    \subfloat[Percentage of feasible solutions that achieved the best objective value for $k\in\{2,3,4\}$.]{%
        \includegraphics[width=0.87\textwidth]{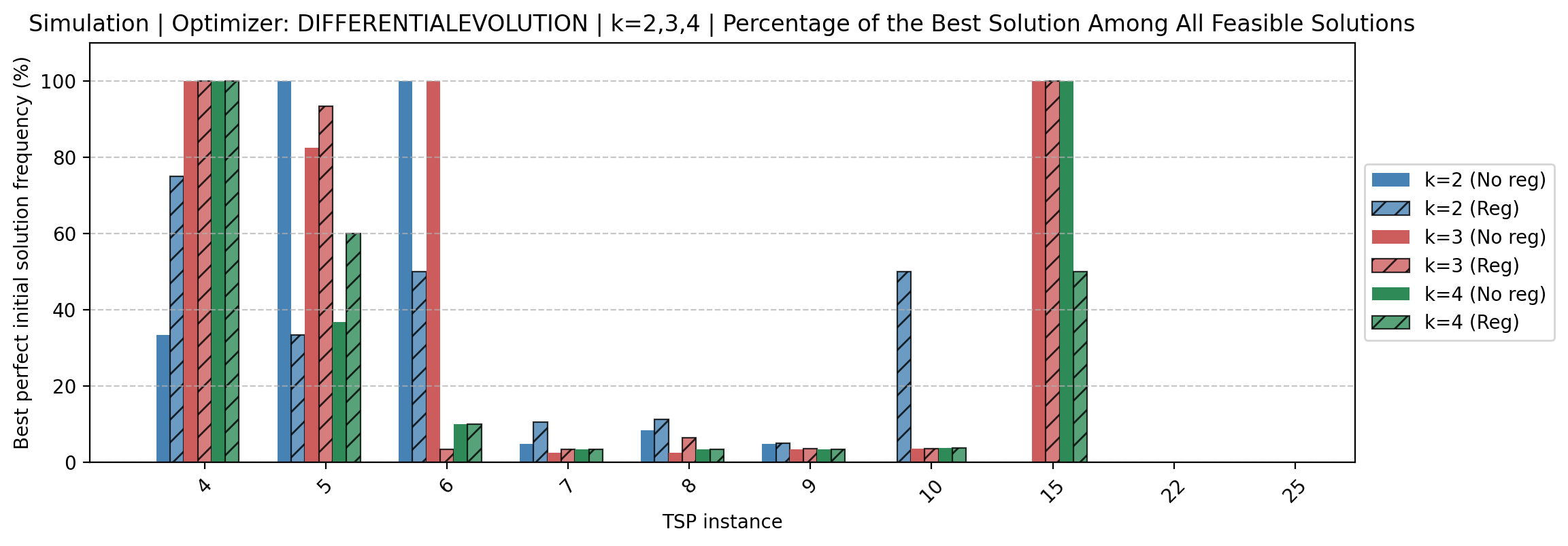}%
    }
    \caption{Comparison of the percentages of feasible and best feasible solutions obtained with and without regularisation across the 10 TSP instances.}
    \label{fig:TSP_pct_results}
\end{figure}

\begin{figure}[H]
    \centering
    \subfloat[Best results before post-processing for $k\in\{2,3,4\}$.]{%
        \includegraphics[width=0.87\textwidth]{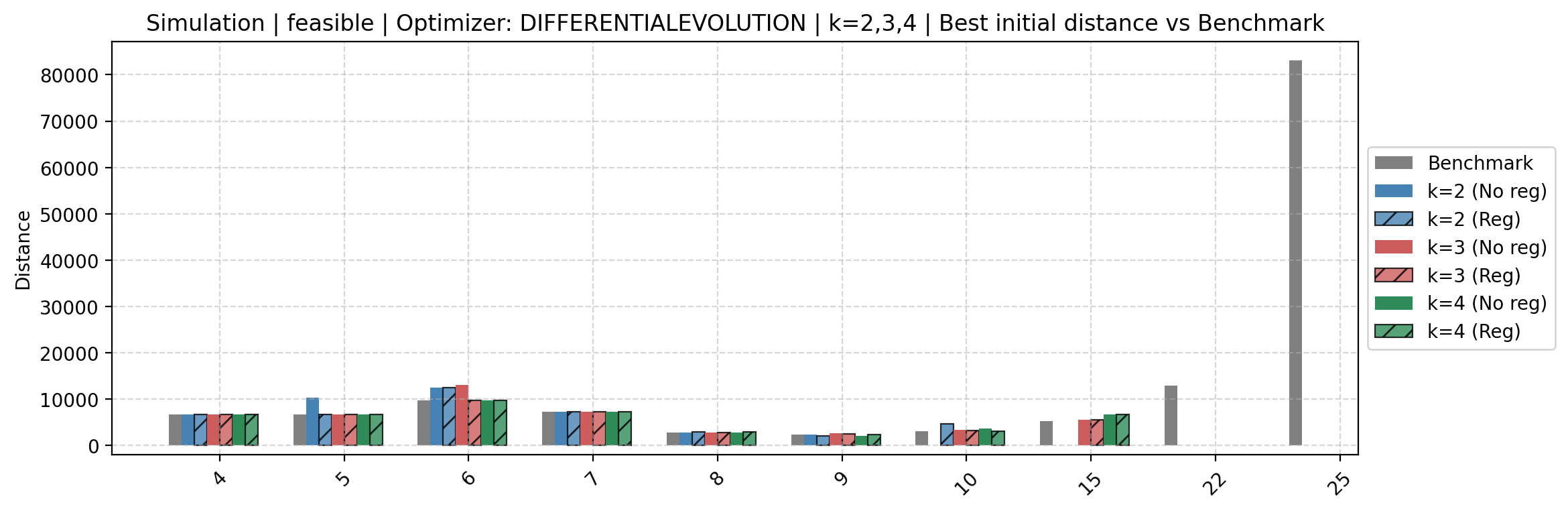}%
    }\\[0.5ex]
    \subfloat[Best results after post-processing for $k\in\{2,3,4\}$.]{%
        \includegraphics[width=0.87\textwidth]{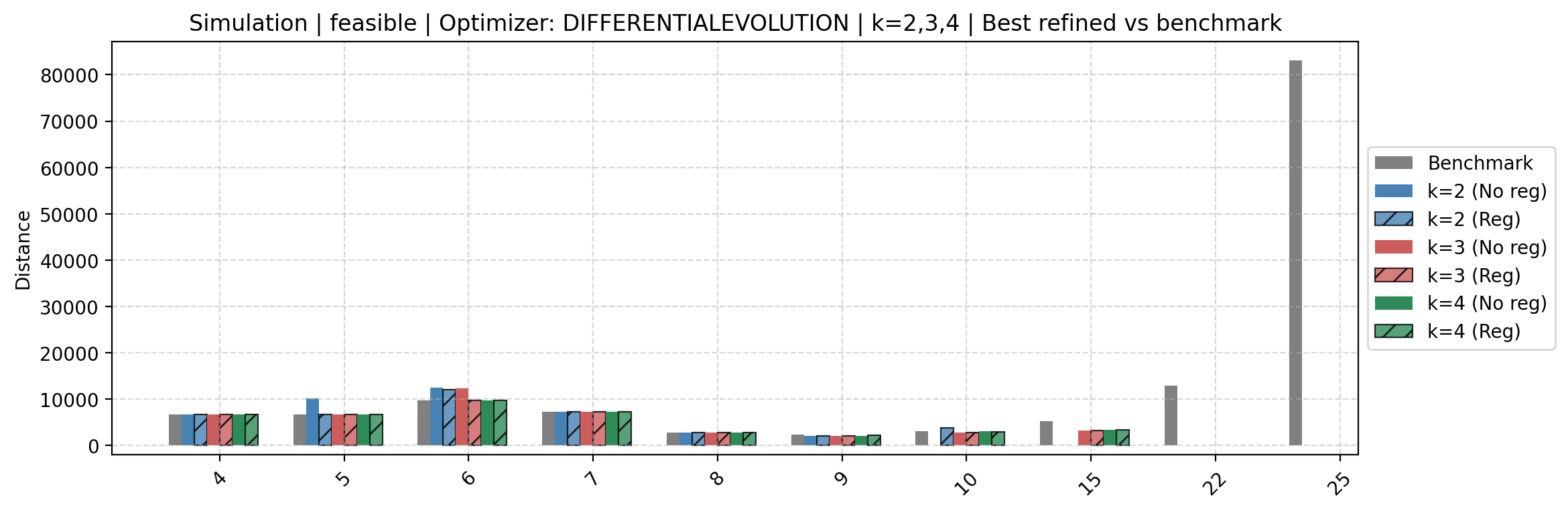}%
    }
    \caption{Comparison of the best solutions obtained with and without regularisation across the 10 TSP instances.}
    \label{fig:TSP_results_1}
\end{figure}

Finally, 50 independent random initialisations were performed for each problem instance and for compression orders $k \in {2,3,4}$, using the optimal hyperparameter configuration for $\alpha$ and $\beta$\footnote{For further details regarding the hyperparameter configuration, see Appendix~\ref{Appendix_B}.}. The corresponding results are presented in Figures~\ref{fig:TSP_pct_results} and~\ref{fig:TSP_results_1}. These figures report, for each instance, the percentage of feasible solutions obtained, the frequency with which the optimal solution is reached without applying post-processing, and the shortest tour distance obtained both before and after the post-processing stage.

\begin{figure}[H]
    \centering
    \subfloat[Best percentage improvement before post-processing for $k\in\{2,3,4\}$.]{%
        \includegraphics[width=0.87\textwidth]{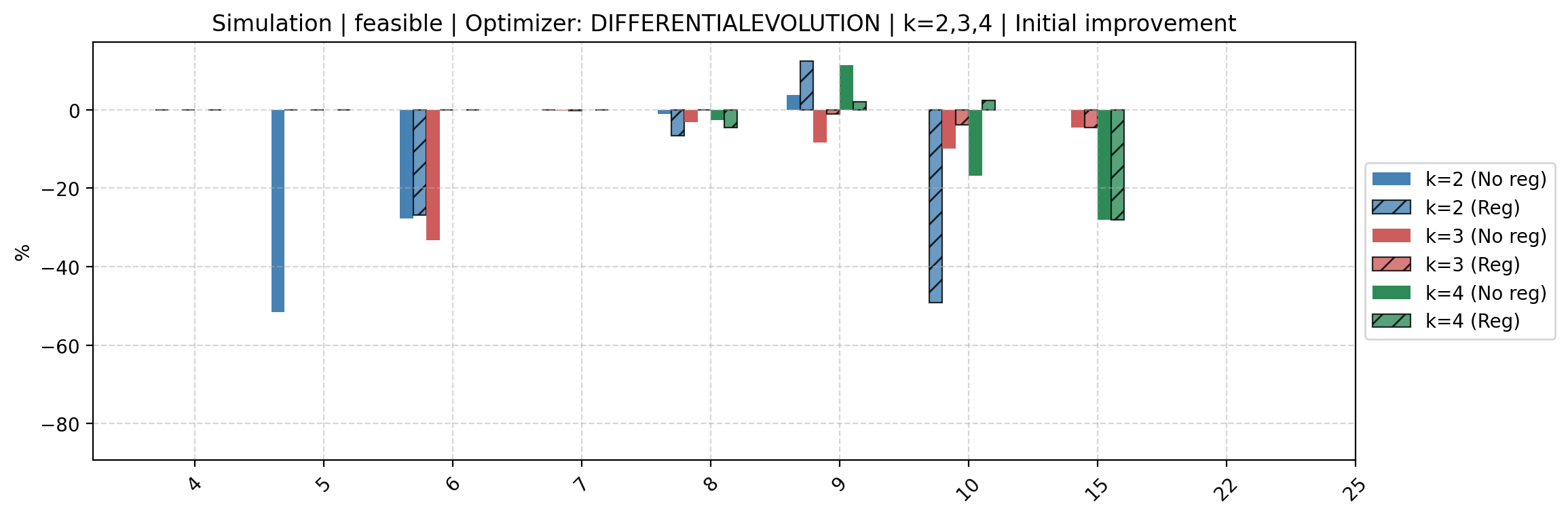}%
    }\\[1ex]
    \subfloat[Best percentage improvement after post-processing for $k\in\{2,3,4\}$.]{%
        \includegraphics[width=0.87\textwidth]{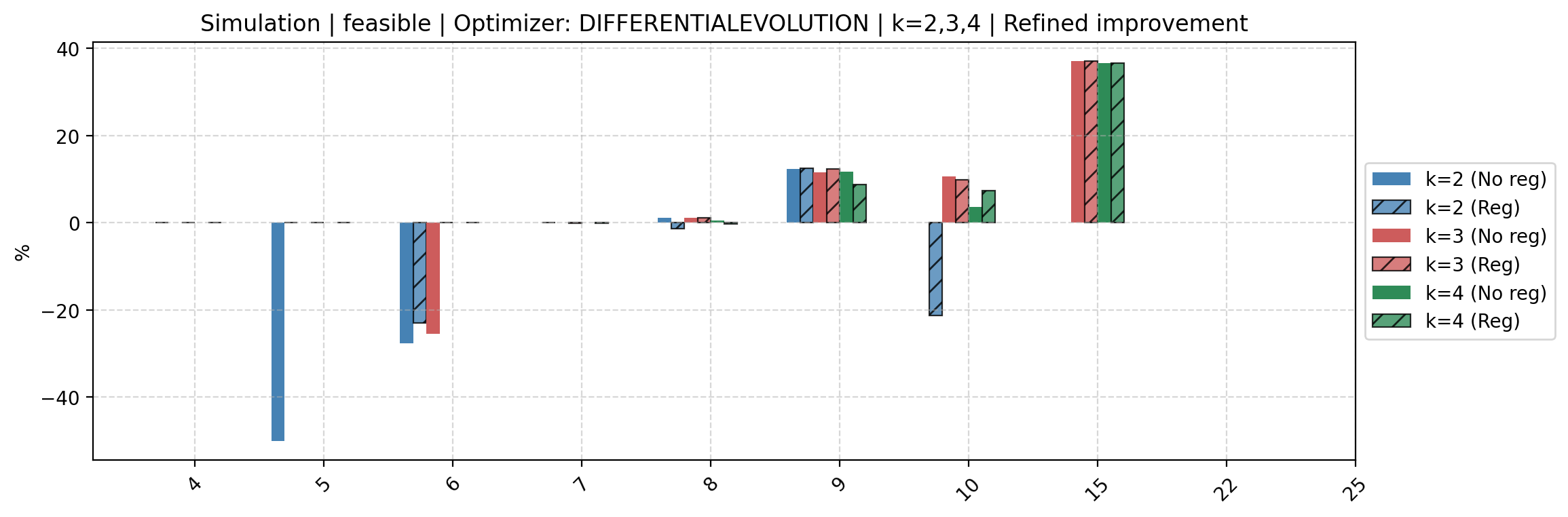}%
    }
    \caption{Comparison of the percentage improvemente of the solutions obtained with and without regularisation across the 10 TSP instances.}
    \label{fig:TSP_results_2}
\end{figure}

To provide a more intuitive measure for evaluating solution quality, the \textit{percentage improvement} metric is introduced, defined as
\begin{equation*}
\Delta_{\%} = \Bigg( \frac{C_{\text{bench}} - C_{\text{sol}}}{C_{\text{bench}}} \Bigg)\cdot 100,
\end{equation*}
where positive values indicate solutions that outperform the benchmark, while negative values correspond to solutions whose quality is inferior to the benchmark solution. The results according to this metric are presented in Figure~\ref{fig:TSP_results_2}.

Consistent with the behaviour observed in the BPP, the results reveal a strong dependence on the compression order $k$. Not all configurations produce feasible solutions across the full range of problem instances. In particular, for $k=2$, feasible solutions are no longer obtained beyond instance $\mathcal{N}=10$, while for $k\in\{3,4\}$ no feasible solutions are found for instances larger than $\mathcal{N}=15$. This behaviour reflects the increased expressive capacity of the variational model as $k$ increases, improving the optimisation process and its ability to satisfy feasibility constraints.

A more detailed analysis of feasibility rates shows a similar trend. For $k=2$, the percentage of feasible solutions is lower than for $k \in \{3,4\}$, where feasibility remains close to $100\%$ for small and medium-sized instances up to $\mathcal{N}=15$, before decreasing as problem complexity grows. Consequently, the percentage of runs reaching the best solution must be interpreted together with feasibility: for larger instances, fewer feasible solutions naturally increase this percentage, sometimes approaching $100\%$, without necessarily indicating better performance.

Beyond feasibility, the quality of the solutions obtained provides further insight into the performance of the method. In general, the initial solutions generated match the quality of those provided by the benchmark and, in several cases, even outperform it, particularly for compression orders $k \in \{3,4\}$. Some isolated exceptions can nevertheless be observed, including cases where lower compression orders yield better solutions, such as $\mathcal{N}=10$ for $k=2$, as well as instances where higher compression orders produce slightly inferior results, notably for $\mathcal{N} \in \{8,10,15\}$. Despite these variations, the post-processing procedure consistently leads to a substantial improvement in solution quality across all considered instances.

Overall, the results demonstrate that the proposed variational approach is capable of producing competitive solutions across a broad range of TSP instances. The algorithm is able to generate good-quality initial solutions, while the post-processing stage further refines them and significantly enhances the final performance. Nevertheless, the experimental analysis also highlights an increasing dependence on instance size and problem complexity. As the dimension of the TSP instances grows, the number of optimisation variables increases considerably, making the search process progressively more challenging and limiting the scalability of the approach.

\section{Noisy model executions}\label{sec_5}

The previous results have been obtained with classical simulations that allow the expected values to be computed with great precision through the obtained state vector. However, in real quantum processors, it is not possible to obtain that state vector straightforwardly, being restricted to performing multiple measurements or \emph{shots}. Therefore, for each of the problems in the \textit{benchmark}, the question naturally arises as to what the practical limitations of the algorithm are when executed on real quantum hardware. In particular, it is fundamental to determine the number of \textit{shots} required to estimate the expected values with sufficient precision, and to analyse how this impacts an iterative optimisation procedure such as the one considered thus far. For this reason, this section presents two complementary strategies: the first consists of studying the scaling of the number of \textit{shots}, both in the absence and presence of noise (gate inefficiencies, physical qubit behaviour such as decay, etc.), with the aim of evaluating the quality of expected value reconstruction; the second focuses on analysing how this error influences the optimisation process, that is, to what extent the estimation affects the performance and convergence of the algorithm.

For these executions, CUNQA has been used, a Distributed Quantum Computing (DQC)~\cite{Barral_2025} emulator developed at the \textit{Galicia Supercomputing Center} (CESGA), which allows distributed quantum computing architectures to be studied and evaluated in high-performance computing environments. CUNQA defines the concept of a virtual QPU (vQPU) --- classical processes that emulate the behaviour of quantum processing units within an HPC system --- and offers an API for managing these vQPUs, designing and submitting quantum tasks, and managing communication and synchronisation between them. In this case, CUNQA has been used to parallelise the execution of multiple circuits. This is because, in order to compute the expected values of the estimator in PCE, it is necessary to measure the resulting circuit in the $\{X, Y, Z\}$ bases. Each of the three circuits was therefore assigned to a vQPU with an identical default configuration, namely, each comprising two cores with the corresponding backend. %This setup was intended to enable parallel execution and thereby significantly reduce computation time compared to sequential execution.

\subsection{Expected values estimation}\label{subsec_51}

Firstly, the reconstruction of expectation values has been analysed under ideal execution conditions,i.e., assuming all-to-all connectivity and no noise, thereby establishing a theoretical reference for the statistical behaviour of the algorithm. The same procedure was then followed considering the conditions of a real quantum device, employing a vQPU configuration that emulates the \textit{IBM Sherbrooke} backend of \textit{IBM Quantum}, a QPU with more than 127 superconducting qubits and a fixed connectivity topology based on couplings between neighbouring qubits that supports the gate set 
$\{\texttt{Id}, \texttt{Rz}, \texttt{SX}, \texttt{X}, \texttt{ECR}\}$ . This topology determines the transpilation of the circuits, reflecting more realistically the experimental limitations associated with execution on a real QPU. For these measurements, exclusively the MCP problem has been used, employing the same experimental configurations described in \ref{subsec_41}.

The study considers 10 different random initialisation seeds of the variational parameters, with the aim of characterising the variability of the executions, as well as different compression orders $k$. Regarding metrics, both the \textit{Mean Absolute Error} (MAE) and the \textit{Maximum Error} (MaxErr) have been employed to identify outliers that might be diluted in the mean and mask significant deviations in specific components.

\begin{figure}[H]
    \centering
    \subfloat[Ideal - $k=2$]{\includegraphics[width=0.5\textwidth]{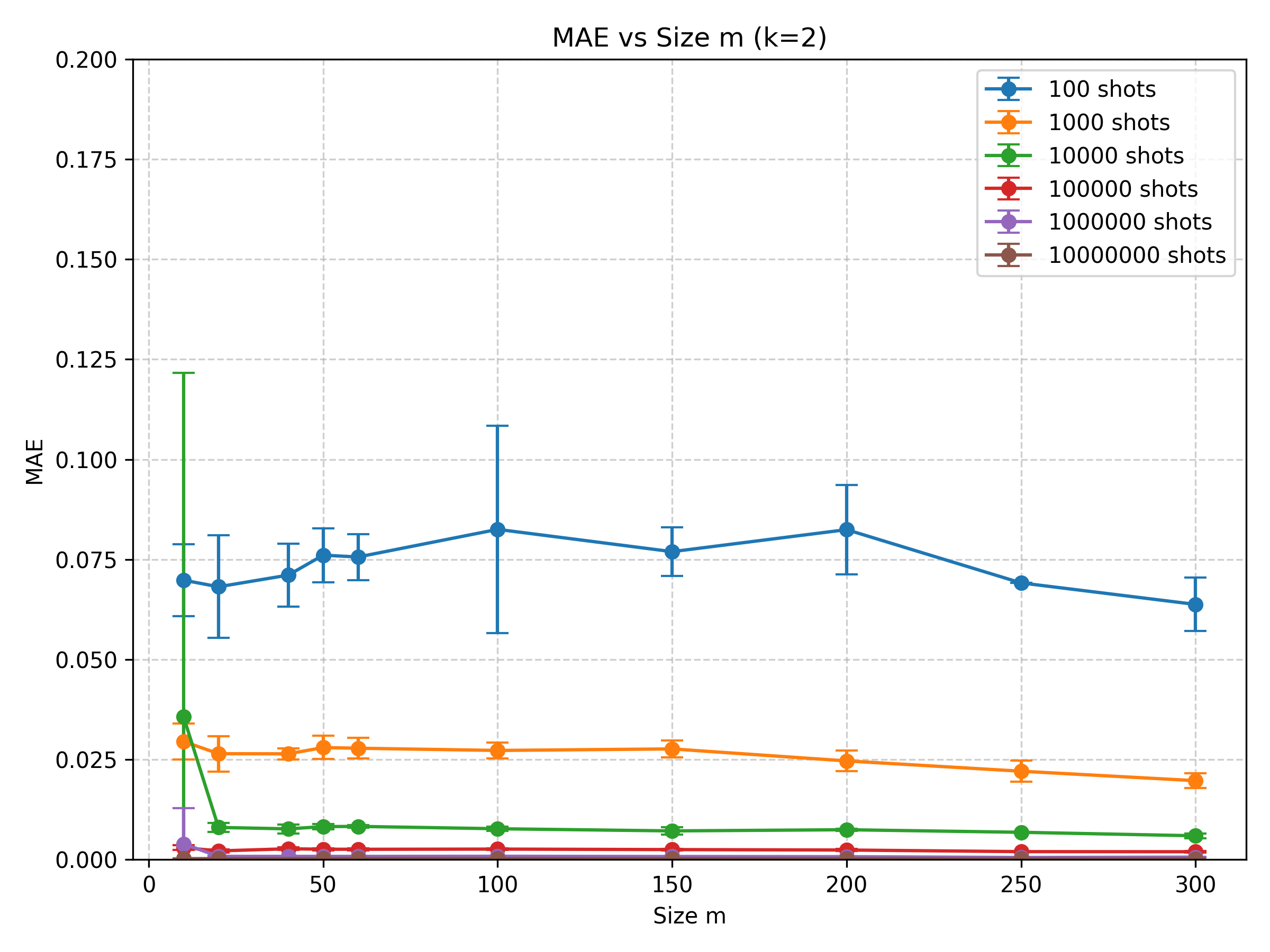}}
    \subfloat[Noisy - $k=2$]{\includegraphics[width=0.5\textwidth]{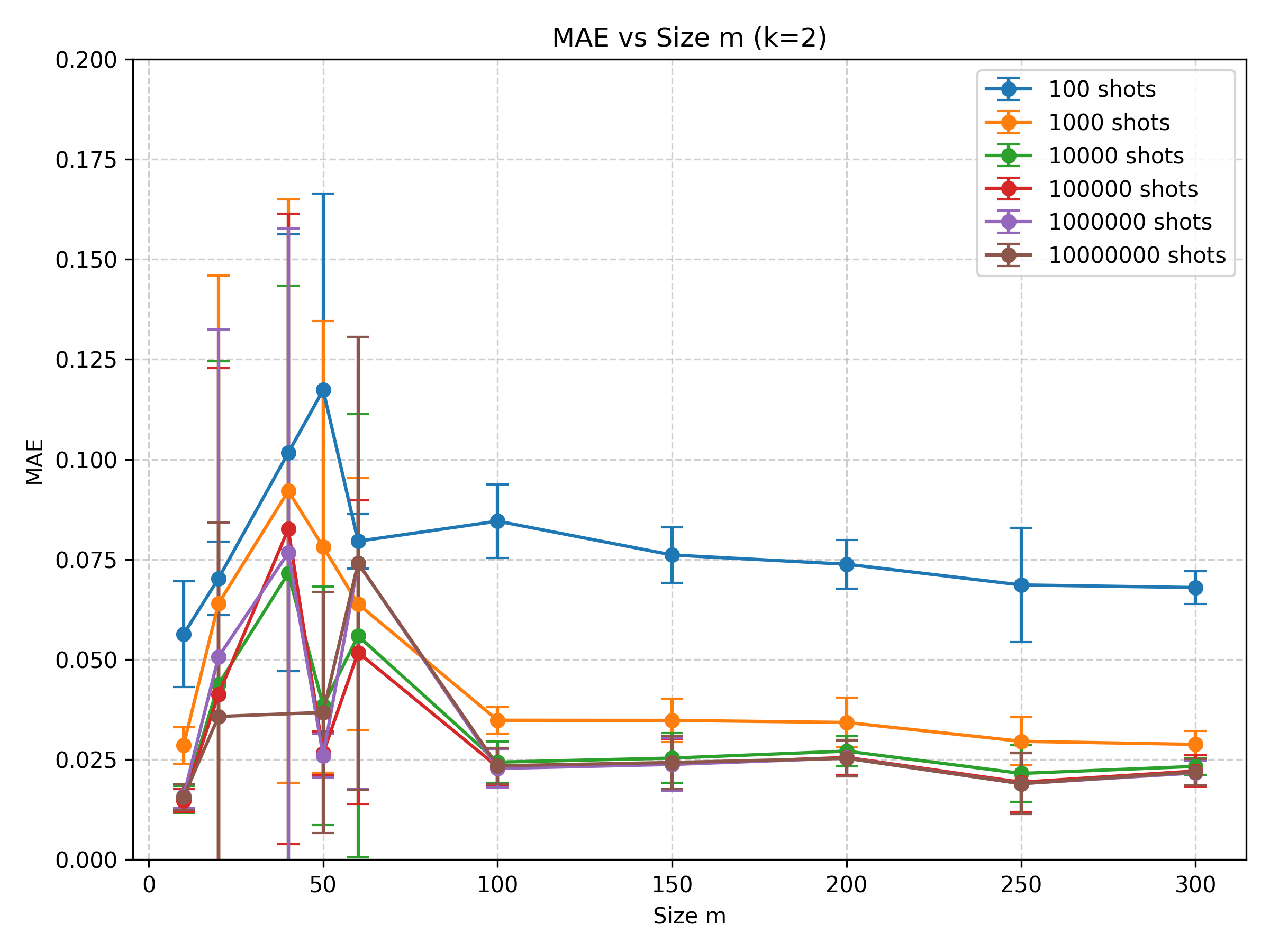}}

    \caption{MAE as a function of the number of \textit{shots} for 10 MCP instances with $k=2$.}
    \label{fig:mae_shots_k2}
\end{figure}

\begin{figure}[H]
    \centering
    \subfloat[Ideal - $k=3$]{\includegraphics[width=0.5\textwidth]{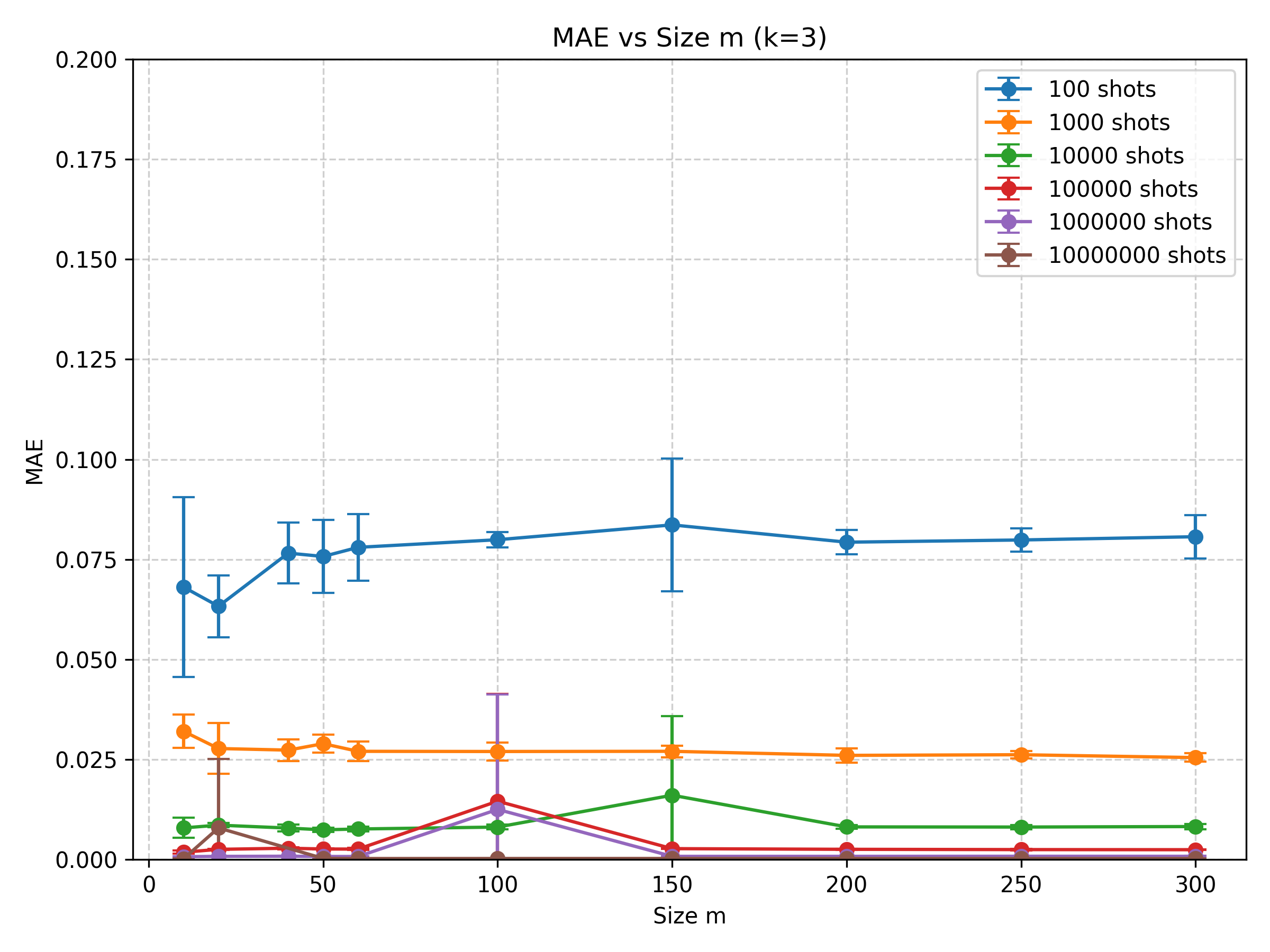}}
    \subfloat[Noisy - $k=3$]{\includegraphics[width=0.5\textwidth]{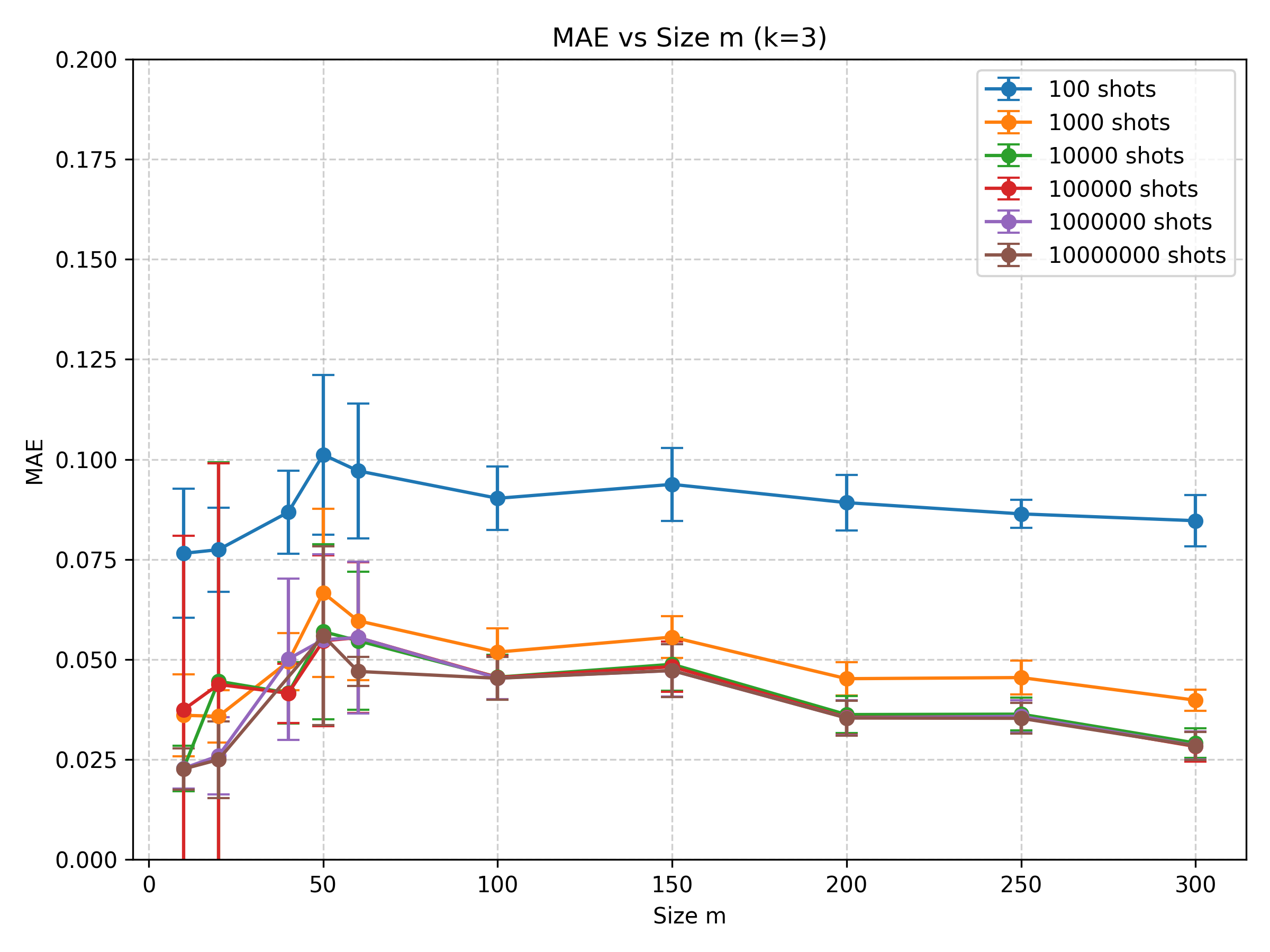}}

    \caption{MAE as a function of the number of \textit{shots} for 10 MCP instances with $k=3$.}
    \label{fig:mae_shots_k3}
\end{figure}

\begin{figure}[H]
    \centering
    \subfloat[Ideal - $k=4$]{\includegraphics[width=0.5\textwidth]{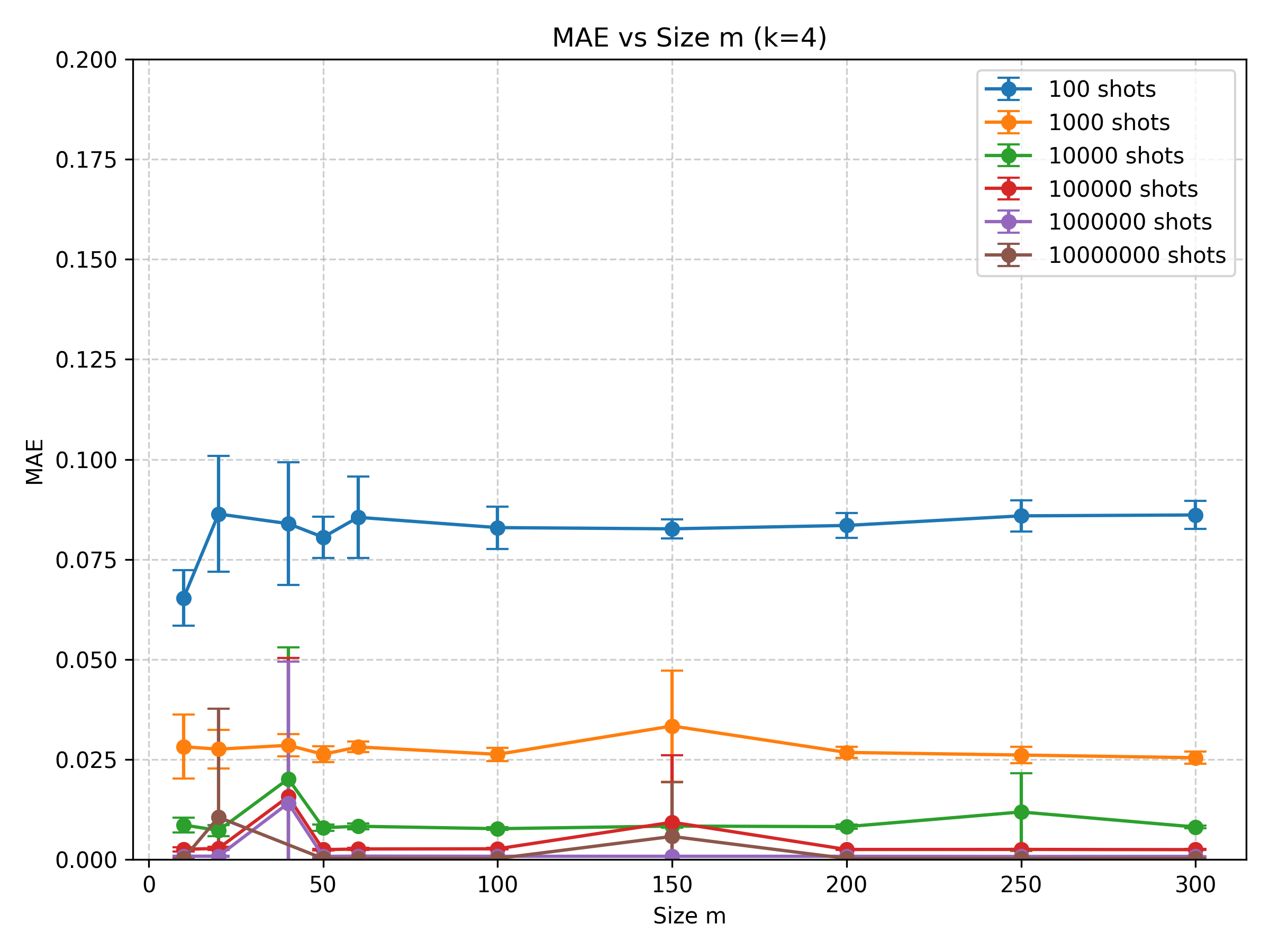}}
    \subfloat[Noisy - $k=4$]{\includegraphics[width=0.5\textwidth]{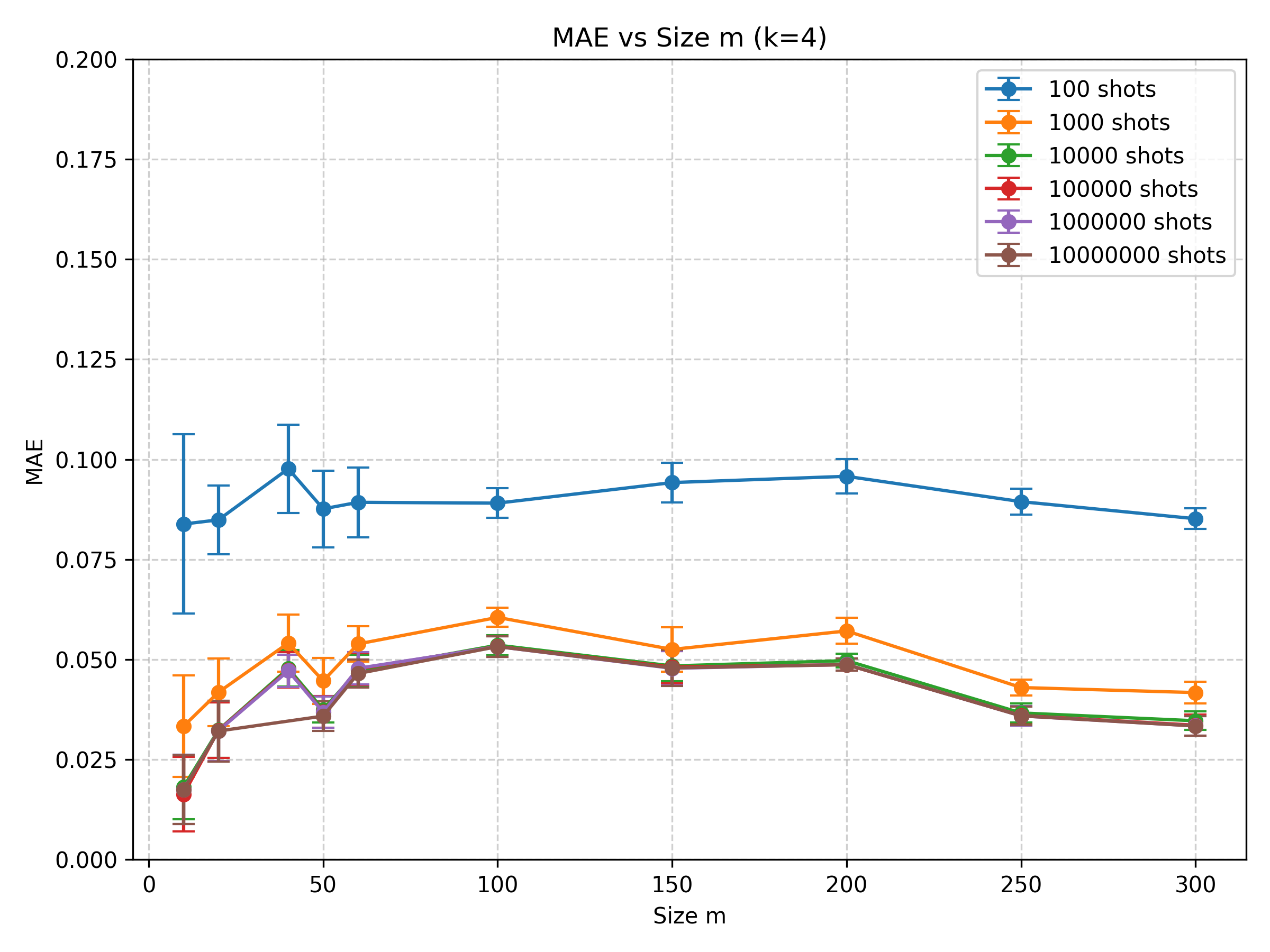}}

    \caption{MAE as a function of the number of \textit{shots} for 10 MCP instances with $k=4$.}
    \label{fig:mae_shots_k4}
\end{figure}

\begin{figure}[H]
    \centering
    \subfloat[Ideal - $k=2$]{\includegraphics[width=0.5\textwidth]{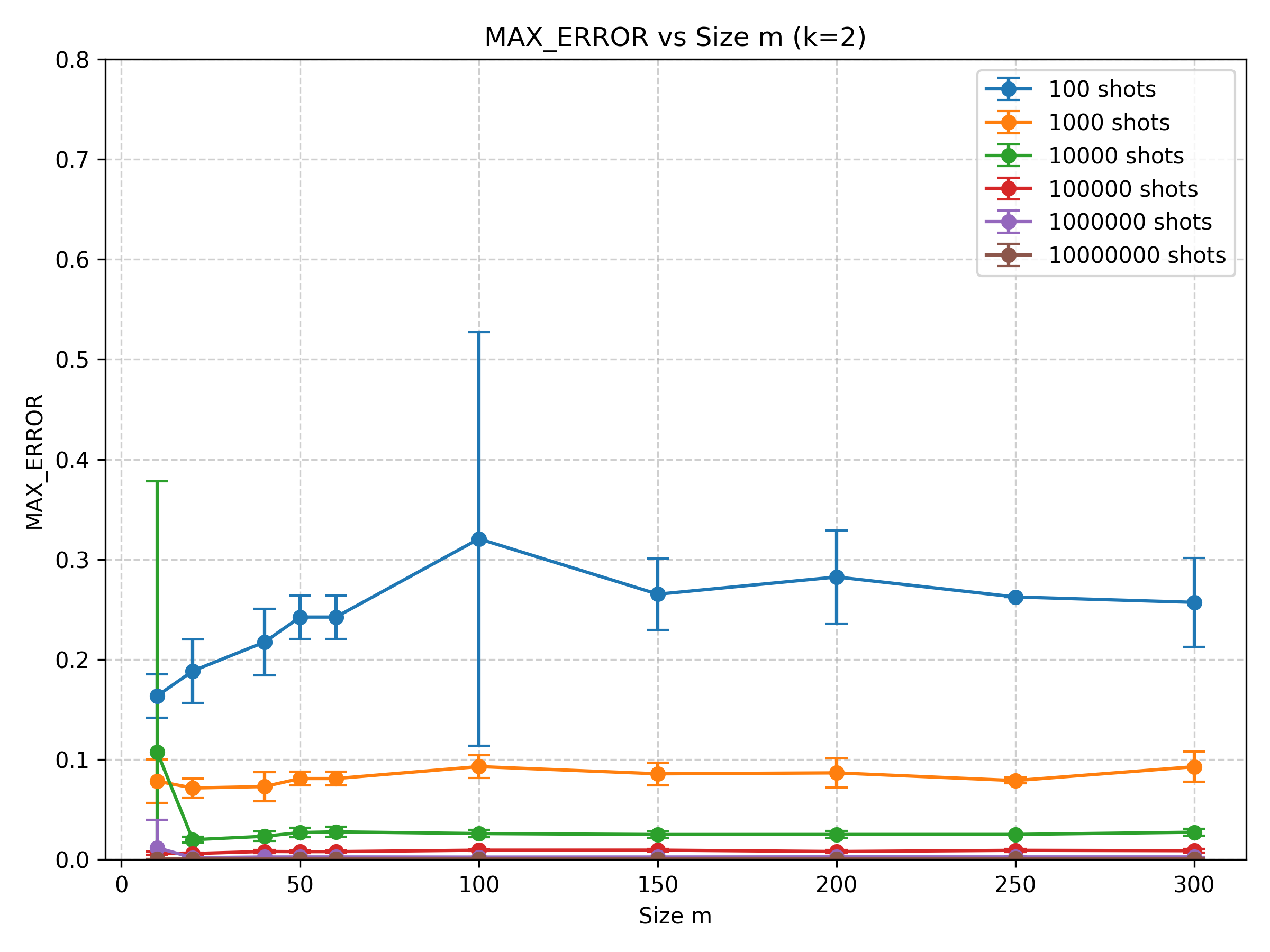}}
    \subfloat[Noisy - $k=2$]{\includegraphics[width=0.5\textwidth]{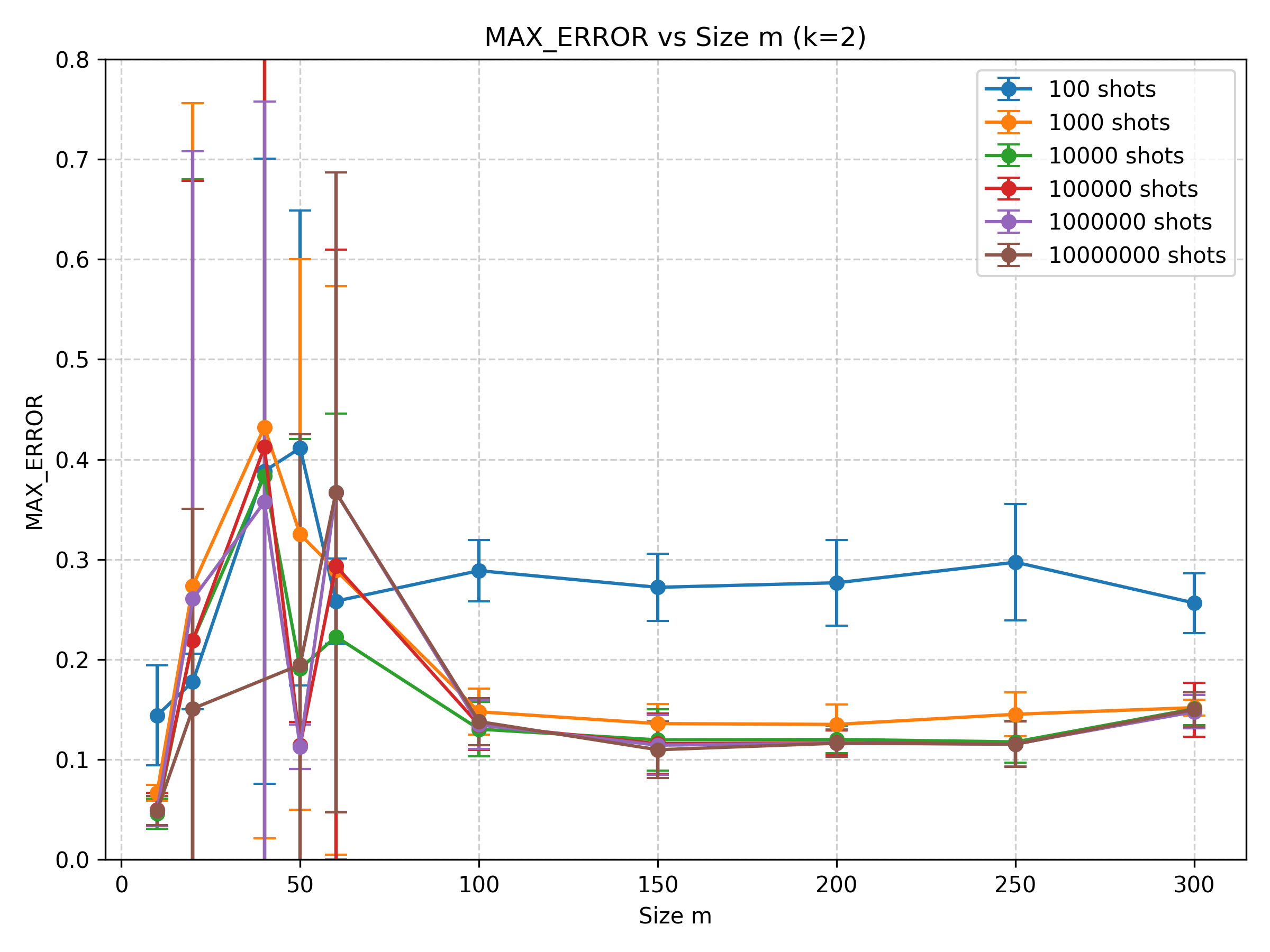}}

    \caption{MaxErr as a function of the number of \textit{shots} for 10 MCP instances with $k=2$.}
    \label{fig:max_error_shots_k2}
\end{figure}

\begin{figure}[H]
    \centering
    \subfloat[Ideal - $k=3$]{\includegraphics[width=0.5\textwidth]{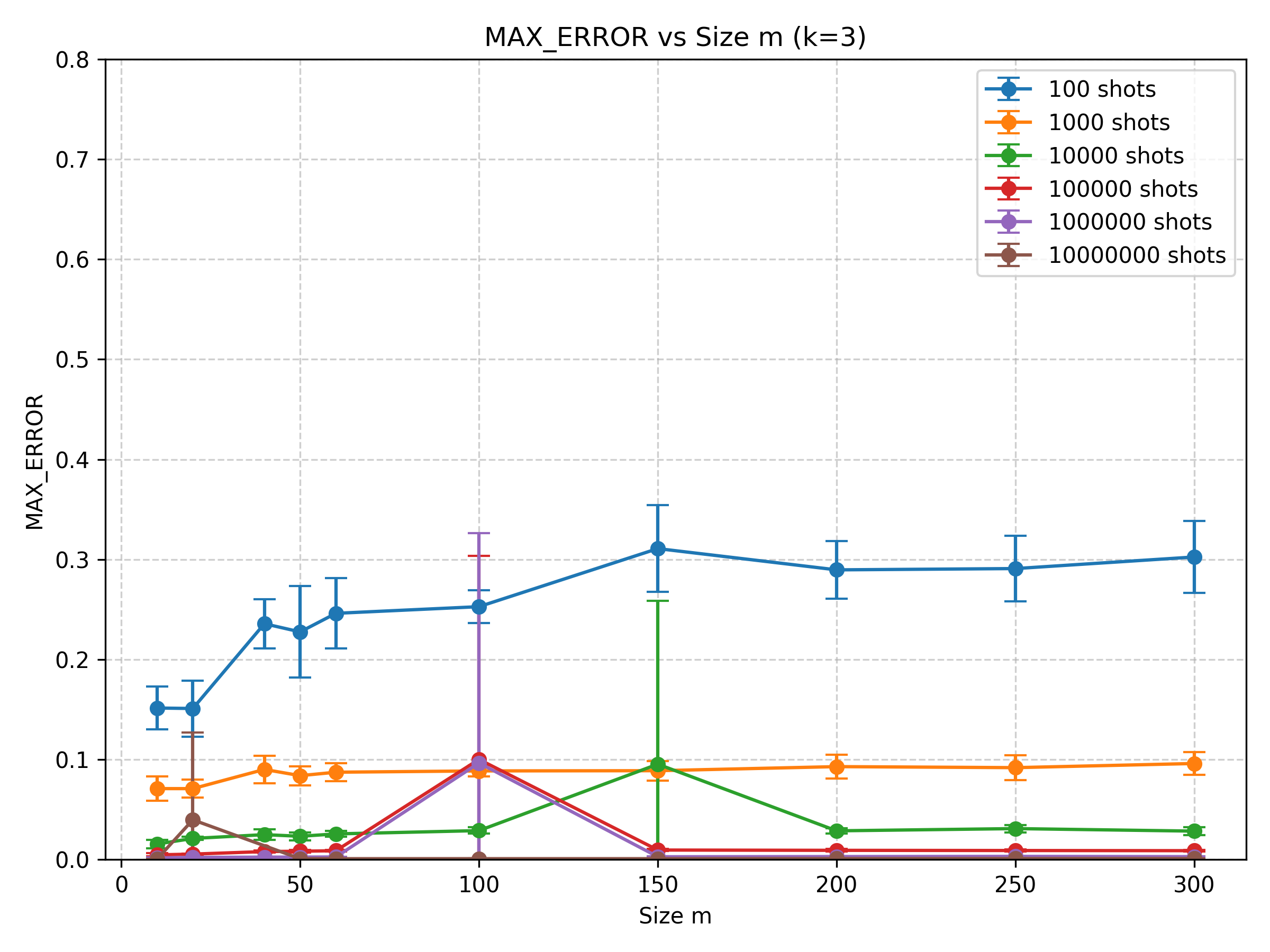}}
    \subfloat[Noisy - $k=3$]{\includegraphics[width=0.5\textwidth]{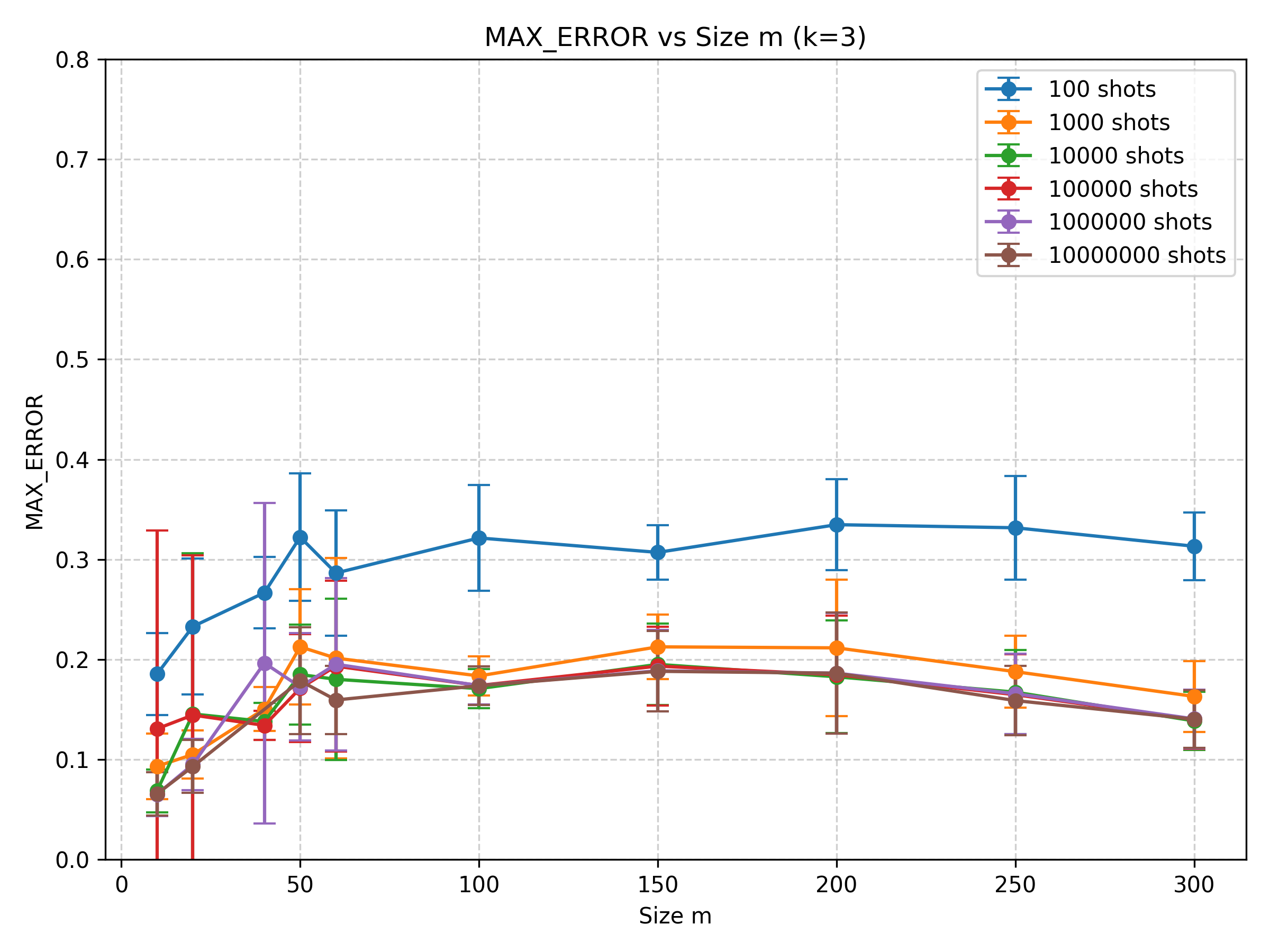}}

    \caption{MaxErr as a function of the number of \textit{shots} for 10 MCP instances with $k=3$.}
    \label{fig:max_error_shots_k3}
\end{figure}

\begin{figure}[H]
    \centering
    \subfloat[Ideal - $k=4$]{\includegraphics[width=0.5\textwidth]{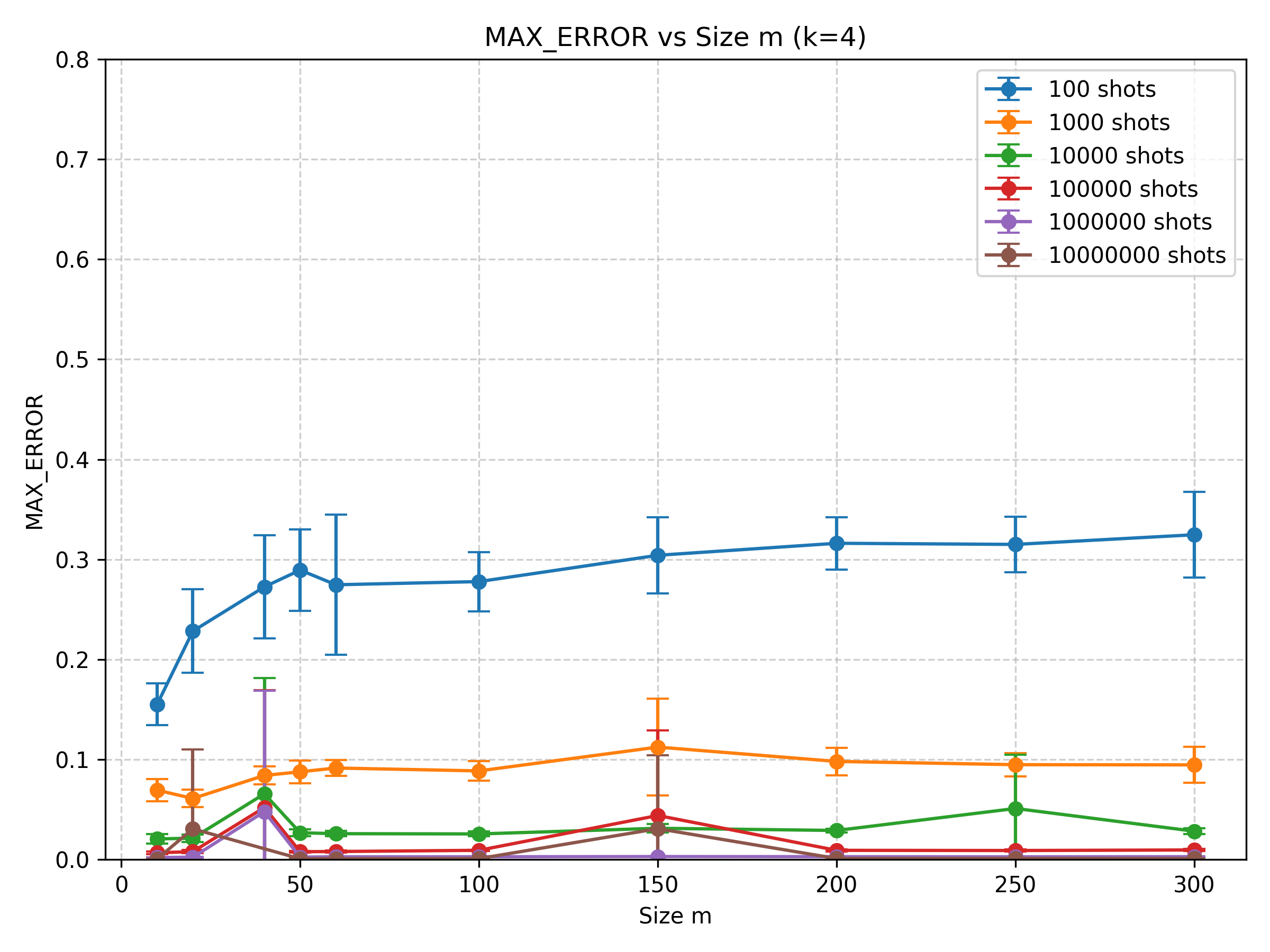}}
    \subfloat[Noisy - $k=4$]{\includegraphics[width=0.5\textwidth]{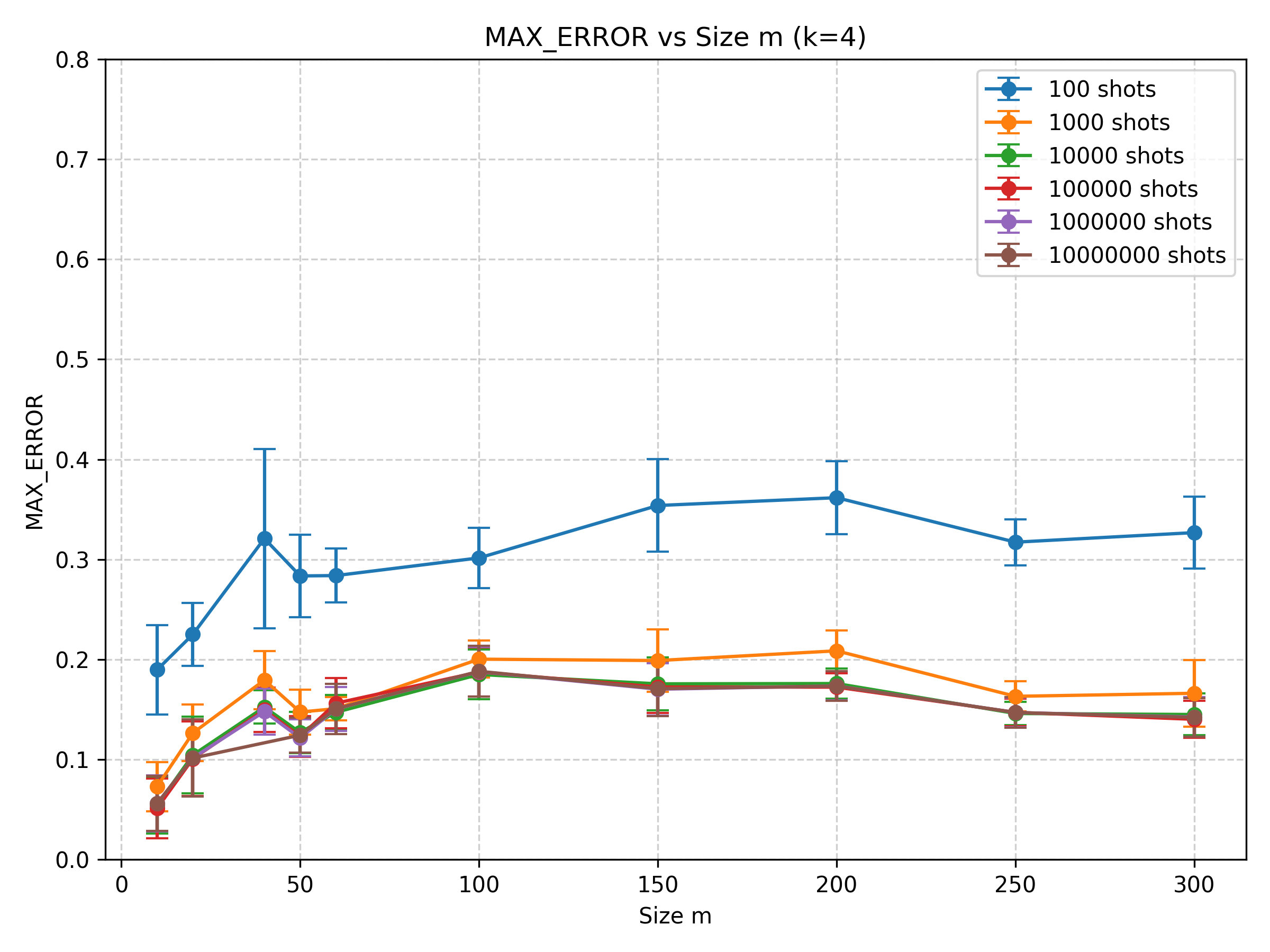}}

    \caption{MaxErr as a function of the number of \textit{shots} for 10 MCP instances with $k=4$.}
    \label{fig:max_error_shots_k4}
\end{figure}

From the results shown from Figures~\ref{fig:mae_shots_k2} to~\ref{fig:max_error_shots_k4}, it is observed that, in general, increasing the number of \textit{shots} reduces both the mean error and its variability. This is consistent with the statistical nature of sampling: the greater the number of repetitions, the more precise the estimation of the probabilities associated with the quantum state, which is equivalent to a more faithful reconstruction of the expected values.

% ================= k = 2 =================

% ================= k = 3 =================

% ================= k = 4 =================

% ================= k = 2 =================

% ================= k = 3 =================

% ================= k = 4 =================

In the ideal emulation scenario, from $10^5$ \textit{shots} onwards, the error reaches convergence on the order of $10^{-3}$ and even $10^{-4}$ in some cases. However, increasing the compression order $k$ introduces a certain level of instability across different seeds for shot counts of $10^6$ and $10^7$, although the error remains consistently within the range of $10^{-2}$ to $10^{-3}$. This behaviour may be explained by the presence of very small amplitudes in the final statevector for some initialisations, which cannot be accurately estimated through finite sampling. As a result, the corresponding expectation values may be biased, ultimately affecting the computed error metrics.

\begin{figure}[H]
    \centering
     \subfloat[Ideal backend]{\includegraphics[width = 0.5\textwidth]{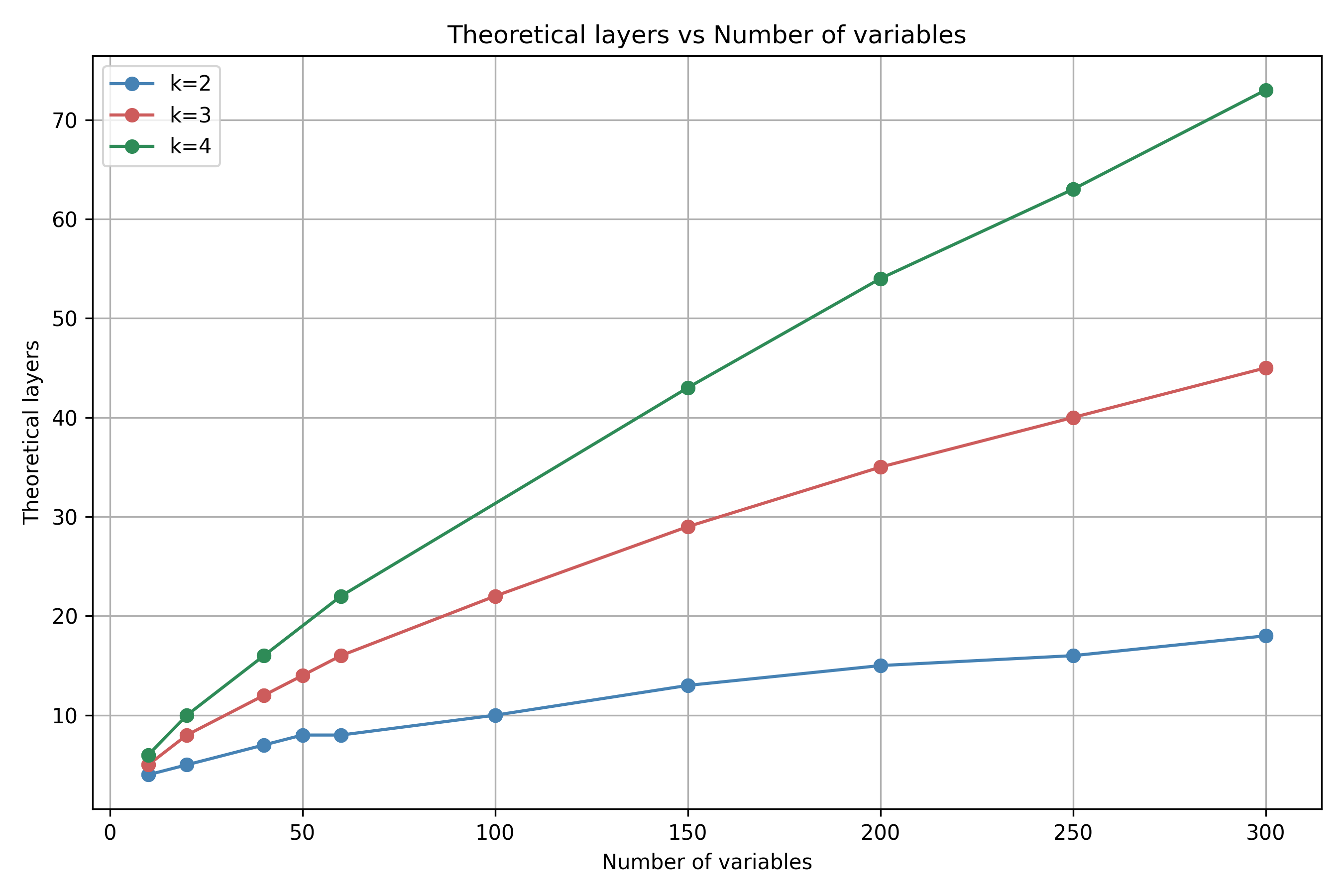}}
     \subfloat[IBM Sherbrooke backend]{\includegraphics[width = 0.5\textwidth]{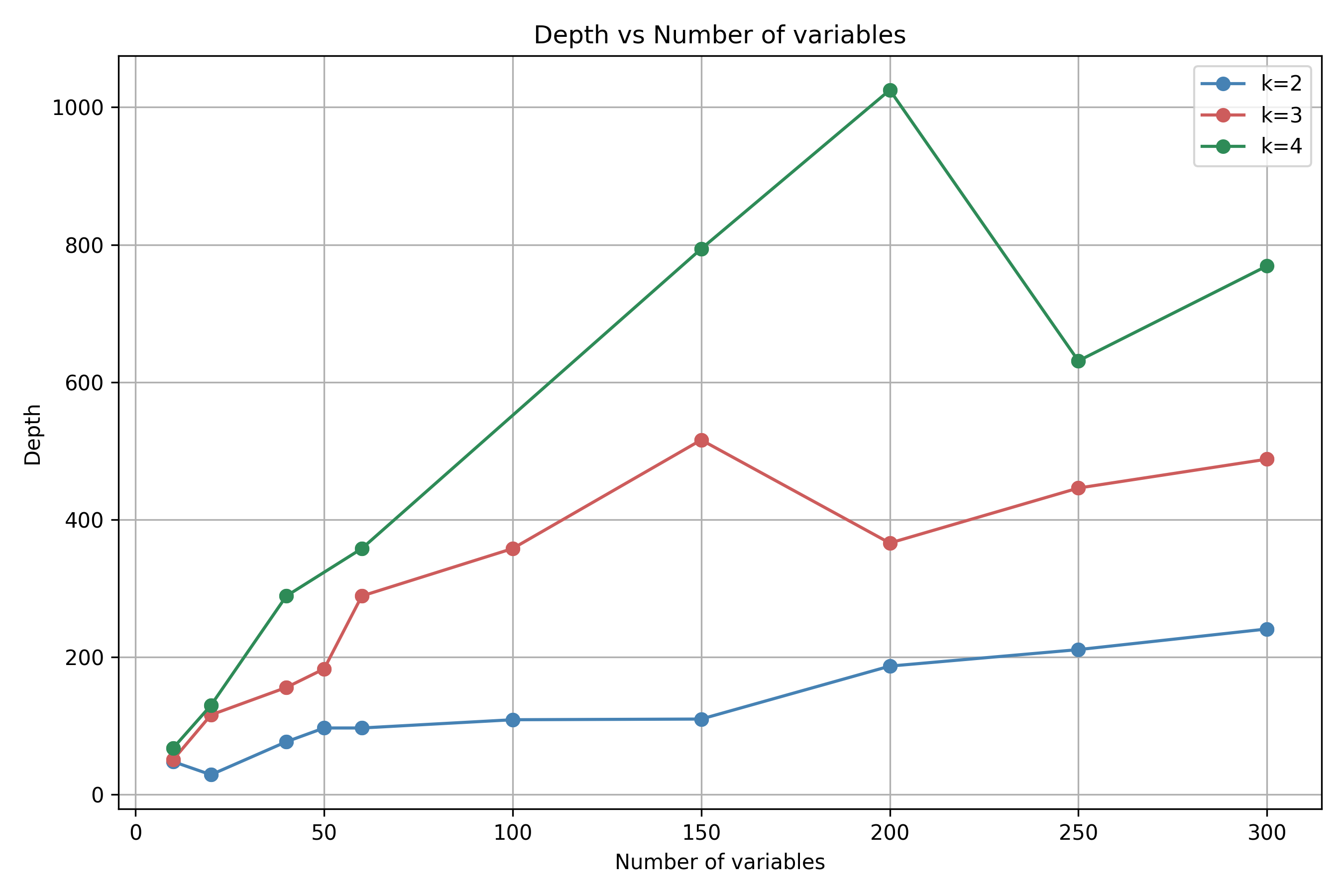}}
    \caption{Scaling of the transpiled circuit depth as a function of the number of variables for $k\in\{2,3,4\}$ under differente backends.}
    \label{fig: Noisy_depth}
\end{figure}

This situation changes markedly when considering executions on a real \textit{backend}. As shown in Figure~\ref{fig: Noisy_depth}, the transpilation to native gates increases the effective circuit depth, while physical noise further accumulates as $k$ increases. Here, noise is so dominant that it effectively enforces a nearly constant error across different shot counts that masks the aforementioned problem of the small statevector amplitudes. While this may appear as a stabilisation, it actually reflects a limitation: the estimation cannot be further improved by increasing the number of shots, as noise prevents any meaningful gain in precision. %Even with up to $10^7$ \textit{shots}, the achieved precision remains on the order of $10^{-1}$, highlighting the current limitations of quantum hardware.

\subsection{Runtime optimisation estimation}\label{subsec_52}

\begin{figure}[H]
    \centering
    \includegraphics[width=0.75\textwidth]
    {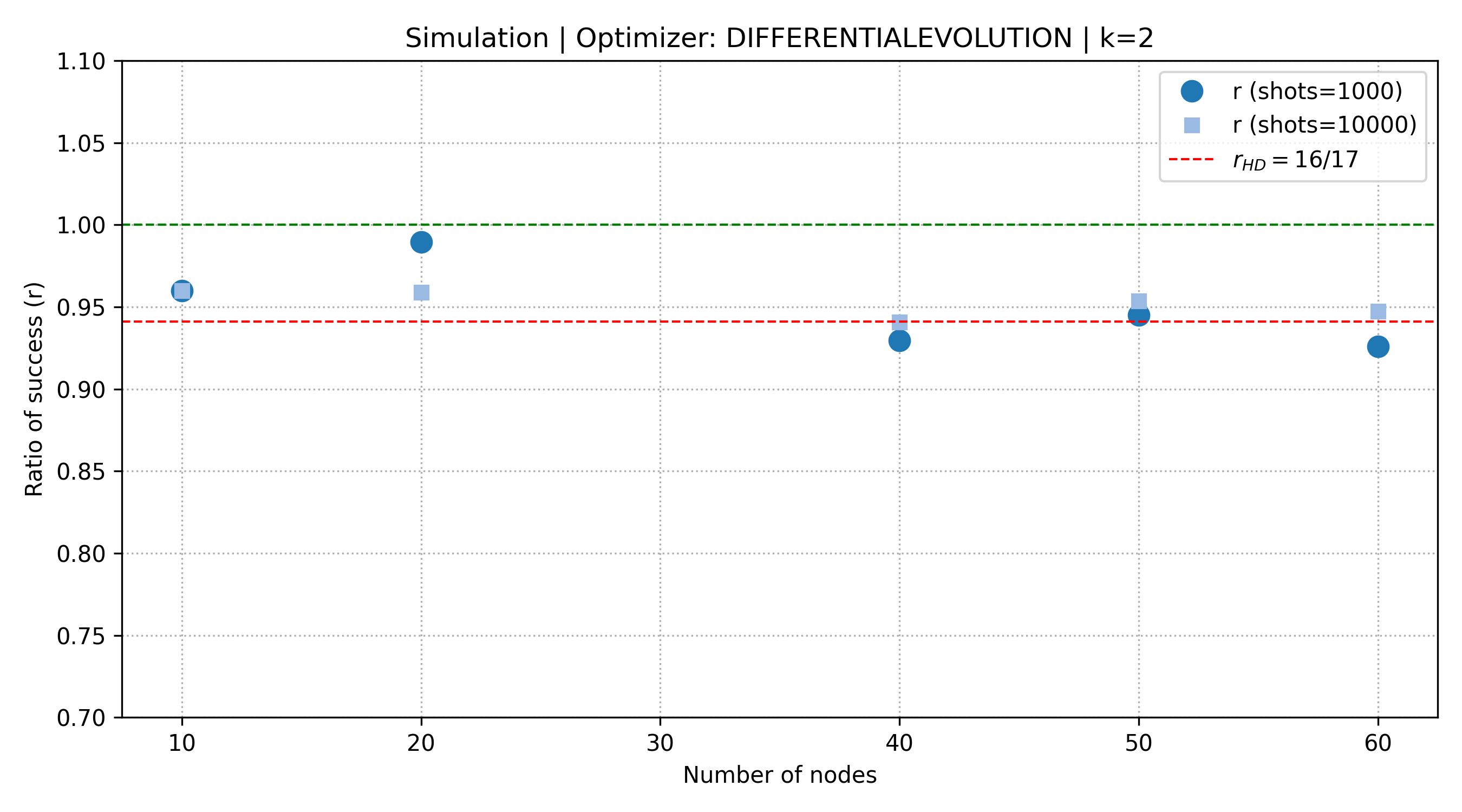}

    \caption{Solution ratio for $k=2$ on ideal backend, using shot-based executions for MCP.}
    \label{fig:mc_noise_ideal_k2}
\end{figure}

\begin{figure}[H]
    \centering
    \includegraphics[width=0.75\textwidth]   {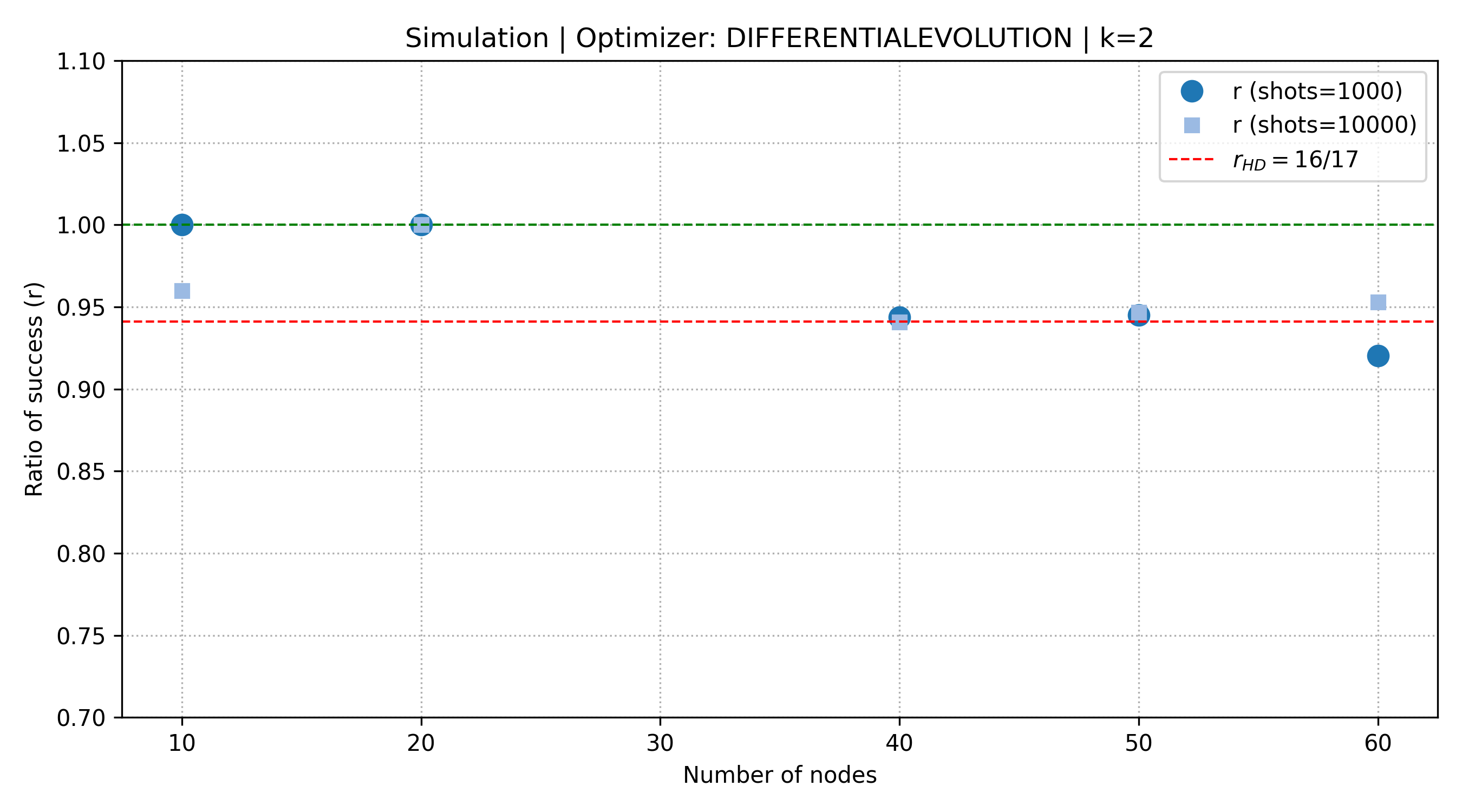}

    \caption{Solution ratio for $k=2$ on noisy backend, using shot-based executions for MCP.}
    \label{fig:mc_noise_noisy_k2}
\end{figure}

Although a thorough study of error propagation throughout the iterative optimisation process is beyond the scope of this work, a proof of concept is presented to illustrate how a certain level of noise may, counterintuitively, benefit the optimisation procedure. Based on the error metrics analysed in the previous sections, a value of 1000 \textit{shots} was identified as a sufficient sampling to ensure reasonably accurate expectation value estimation while maintaining a manageable computational cost. Using this configuration, several simulations were carried out for 5 MCP instances under both ideal and noisy backend conditions, keeping the same random seeds across all runs to ensure a fair comparison. Figures~\ref{fig:mc_noise_ideal_k2} and~\ref{fig:mc_noise_noisy_k2} show the results for ideal and noisy backend configurations, respectively, suggesting that hardware noise may help the optimiser escape suboptimal local minima, leading in some cases to improved optimisation performance.

% ================= Ideal backend =================

% ================= Noisy backend =================

% \begin{figure}[H]
%     \centering
%     \subfloat[Solution ratio for $k=2$ on ideal backend.]{%
%         \includegraphics[width=0.65
%         \textwidth]{images/Noisy/MaxCut_grafico_r_DIFFERENTIALEVOLUTION_k2_not_noisy_shots.png}%
%     }\\[0.5ex]
%     \subfloat[Solution ratio for $k=2$ on noisy backend.]{%
%         \includegraphics[width=0.65\textwidth]{images/Noisy/MaxCut_grafico_r_DIFFERENTIALEVOLUTION_k2_noisy_shots.png}%
%     }
%     \caption{Results obtained using shot-based executions for 5 MCP instances, over the same seeds.}
%     \label{fig: MC_noise_example}
% \end{figure}

\begin{figure}[H]
    \centering
    % Upper row: k=2 and k=3 side by side
    \subfloat[$k=2$]{%
        \includegraphics[width=0.5\textwidth]{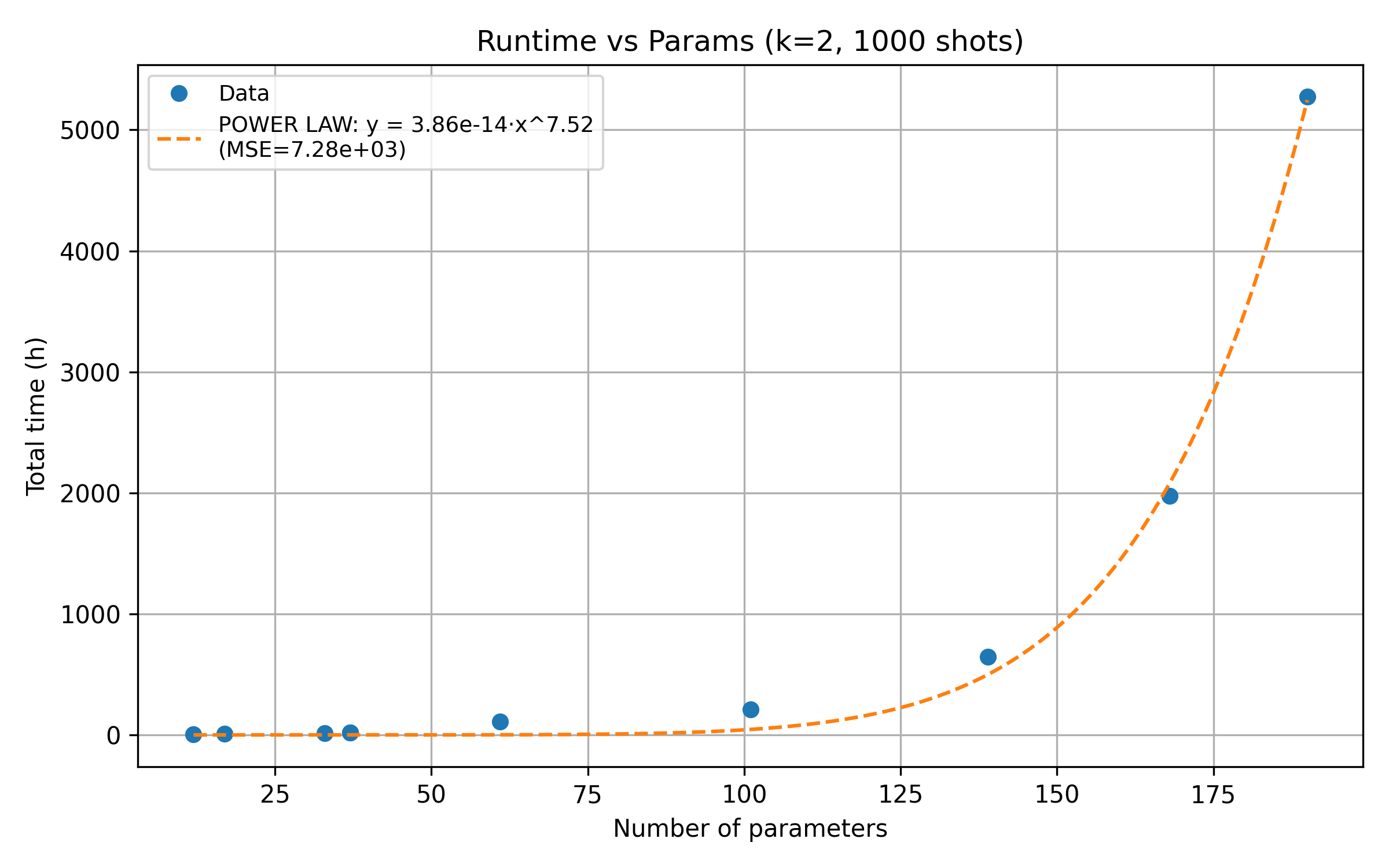}%
    }\hfill
    \subfloat[$k=3$]{%
        \includegraphics[width=0.5\textwidth]{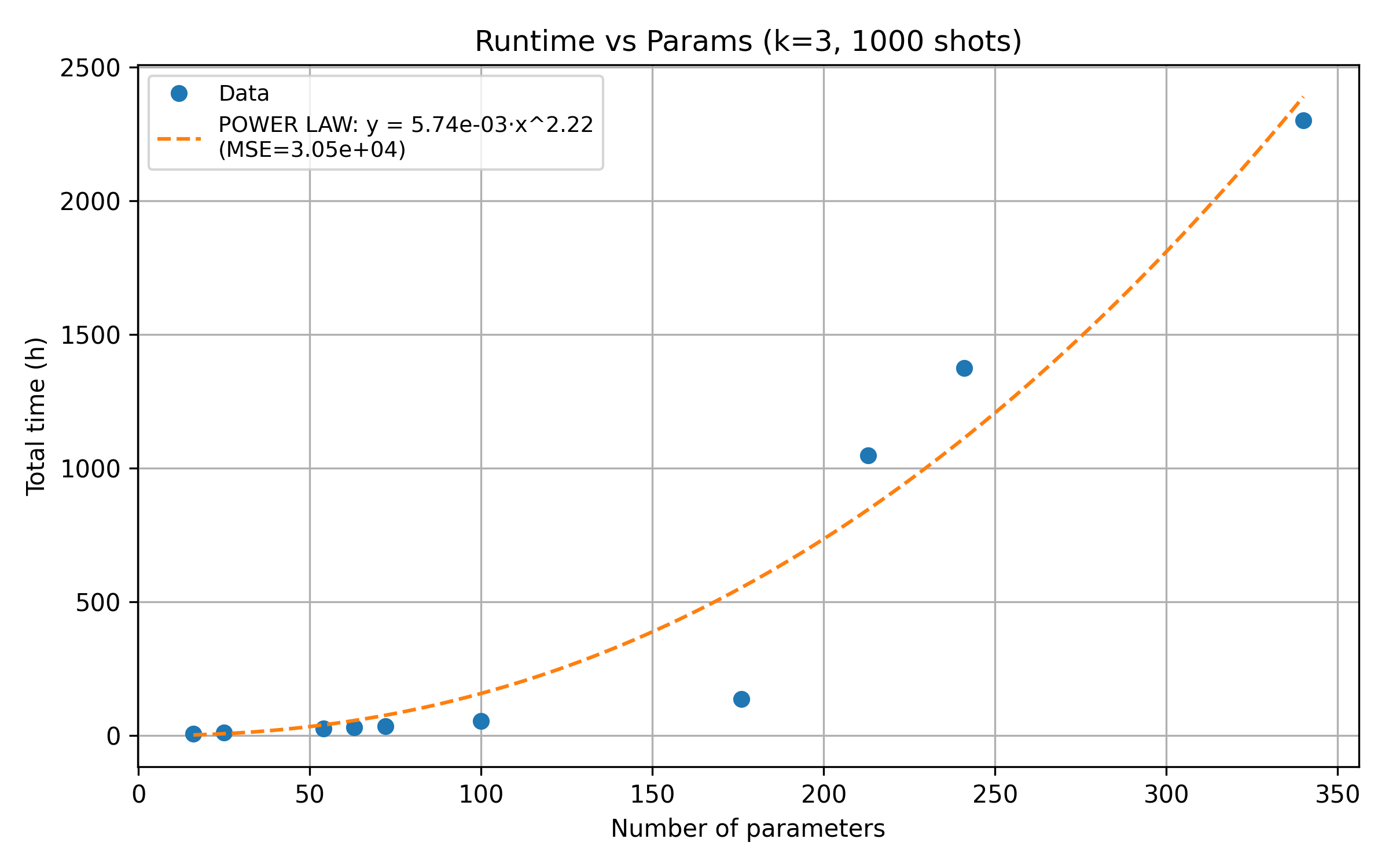}%
    }\\[0.5ex]
    % Lower row: k=4 centred
    \subfloat[$k=4$]{%
        \includegraphics[width=0.5\textwidth]{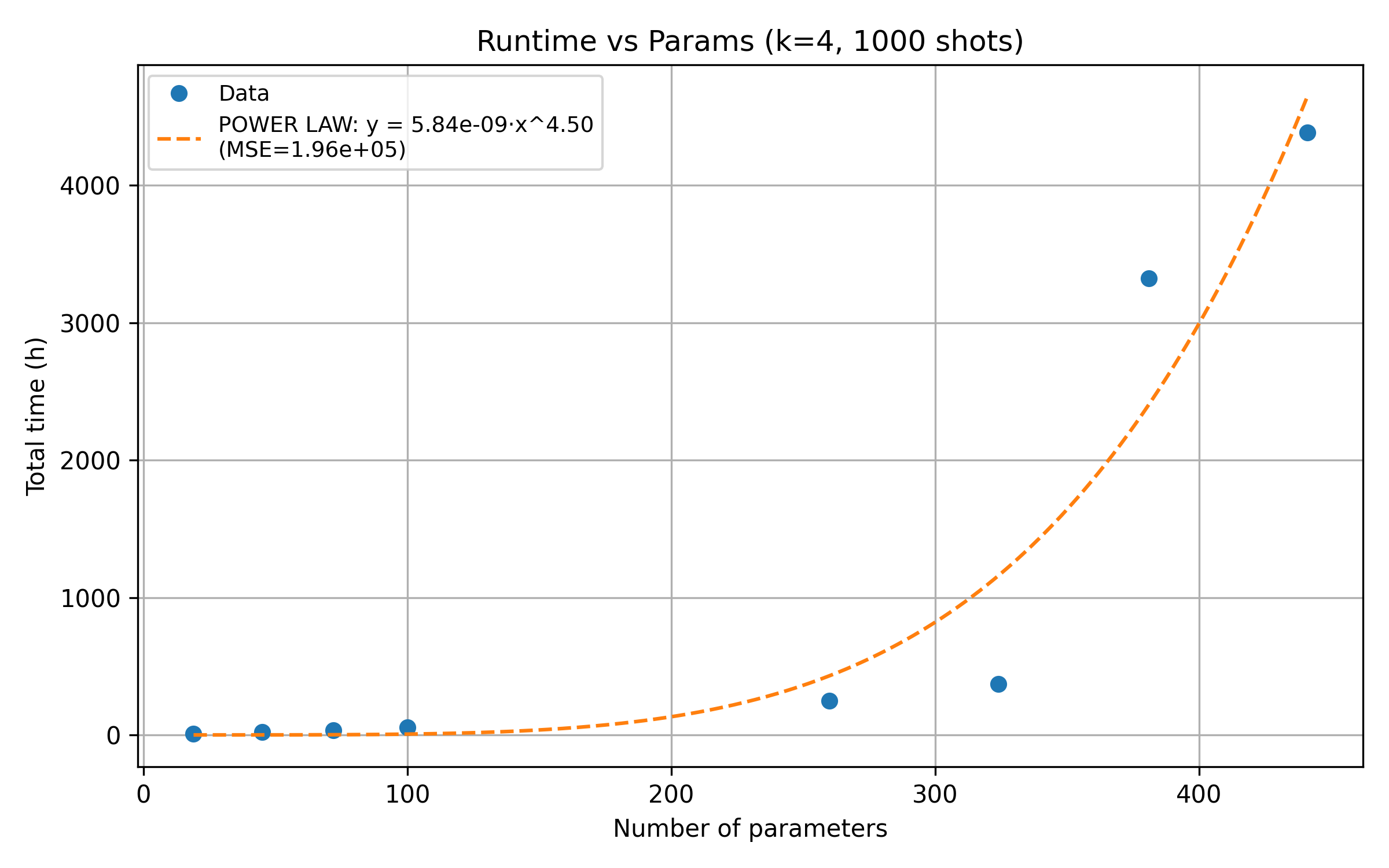}%
    }
    \caption{Estimated execution time as a function of the number of parameters for different $k$.}
    \label{fig: Runtime_estimation}
\end{figure}

However, reproducing the full optimisation with shot-based sampling is extremely costly: when using \textit{Differential Evolution}, the number of evaluations is given by the product of the \textit{popsize} and the number of circuit parameters, resulting in tens of millions of executions in the most complex cases. For this reason, an estimate of the optimisation runtime was performed assuming $10^3$ \textit{shots} per evaluation, as shown in Figure~\ref{fig: Runtime_estimation}. This value was selected based on the results presented in Figures~\ref{fig:mae_shots_k2} to~\ref{fig:max_error_shots_k4}, from which it was identified as a suitable trade-off between estimation precision and execution cost.

It is also worth noting that, as mentioned at the beginning of this section, these calculations were carried out under the assumption that three vQPUs with two cores each can be used to parallelise the execution. If additional computational resources were available, such as vQPUs with a larger number of cores, the execution times would be expected to decrease accordingly.

Whilst this approximation does not allow for strictly precise comparisons between optimisers under realistic sampling conditions, it provides a useful estimate of the execution constraints that would arise in practical implementations. The results suggest that, by employing less computationally demanding optimisers or more aggressive parallelisation strategies, it would be possible to perform complete optimisations without exceeding the available computational resources, thereby establishing a practical reference framework for future experiments.

Although the improvement in sampling efficiency must be weighed against the hardware limitations, such as noise and gate imperfections, which may adversely affect the accuracy and stability of the obtained results, in this context, it is also important to consider that real QPUs will offer a potential advantage, as they will be capable of performing sampling directly at the hardware level, which will be significantly faster than classical simulation.

% \begin{figure}[H]
%     \centering
%     % Upper row: k=2 and k=3 side by side
%     \subfloat[$k=2$]{%
%         \includegraphics[width=0.5\textwidth]{images/Noisy/runtime_vs_params_k2_1000_shots.png}%
%     }
%     \hfill
%     \subfloat[$k=3$]{%
%         \includegraphics[width=0.5\textwidth]{images/Noisy/runtime_vs_params_k3_1000_shots.png}%
%     }\\[2mm]
%     % Lower row: k=4 centred
%     \subfloat[$k=4$]{%
%         \includegraphics[width=0.5\textwidth]{images/Noisy/runtime_vs_params_k4_1000_shots.png}%
%     }
%     \caption{Estimated execution time as a function of the number of parameters for different compression orders.}
%     \label{fig: Runtime_estimation}
% \end{figure}

\section{Discussion}\label{sec_6}

Throughout the development of this work, aspects that affect the optimisation landscape and process have become particularly relevant with respect to the quality, feasibility, and stability of the resulting solutions. These include, on the one hand, the relaxation of the binary variables ---and, within this relaxation, the role of the hyperparameters $\alpha$ and $\beta$---, the (relative) weights of the penalty terms, and the combination of these aspects; and, on the other hand, the configuration of the optimiser itself.

Firstly, the relaxation of the binary variables allows the introduction of an additional term ---the regularisation term--- weighted by the hyperparameter $\beta$, which, as noted in the original proposal, forces the expected values to tend towards zero and has been observed to improve solver's performance. However, in our case, experiments with $\beta = 0$ yielded results similar to those obtained with larger $\beta$ values, likely because the small values of $\lambda_{\text{reg}}$ effectively attenuate the influence of $\beta$.

In turn, the election of the hyperbolic tangent over the sign function requires the introduction of the hyperparameter $\alpha$ in order to avoid the linear regime and to control the binarisation degree of the relaxed variables. In this case, the hyperparameter $\alpha$ has been shown to have a decisive influence on the quality of the solutions, as larger values of $\alpha$ generally lead to better results. A plausible explanation is that, when $\alpha$ is not sufficiently large, the relaxed variables remain close to the linear region of the hyperbolic tangent and therefore tend to take values near zero. If the variable transformation is applied, this behaviour translates into values concentrated around $0.5$. In such a regime, the variables are poorly binarised, which leads to a discrepancy between the value of the cost function during optimisation, particularly at the final optimisation step, and its value after reconstructing the binary solution. As a consequence, the ambiguity of the reconstruction process increases and the quality of the recovered combinatorial solution deteriorates.

Secondly, it is well known that determining appropriate values for the weights of the penalty terms is a difficult problem, classically known as the Big-M problem~\cite{Alessandroni_2025}. In the QUBO formulations considered here, this issue becomes particularly relevant once the binary variables are relaxed. For instance, when large weights are assigned to penalty terms of the form
\begin{equation*}
    \left(1-\sum_i x_i\right)^2,
\end{equation*}
the optimisation process may favour configurations that satisfy the constraint only in an averaged or distributed sense, that is, configurations in which several variables take fractional values whose sum remains close to one. Consequently, increasing the corresponding penalty weight does not necessarily favour binarisation; rather, it reinforces satisfaction of the aggregate constraint while still allowing ambiguous configurations. In \ref{Appendix_B} we study this issue in more detail, noting that when working with relaxed variables, special care must be taken in order to select the appropriate weights of the penalty terms.

On the other hand, regarding the optimiser configuration, it is worth noting that, in general, increasing the compression order $k$ leads to a larger number of circuit layers and, consequently, a higher number of parameters $p$. In turn, in the SciPy implementation of DE, an increase in the number of parameters results in a larger population size and, therefore, a greater number of function evaluations. Although this may, in principle, facilitate the discovery of better solutions, if the penalty weights are not properly chosen, it may also lead to a deterioration in solution quality and feasibility, yielding more ambiguous solutions even when the value of the cost function is effectively decreasing.

In summary, our results indicate that adequately addressing the discrete nature of the problems considered requires a careful selection of the hyperparameters $\alpha$ and $\beta$, as well as the penalty weights. In particular, excessively large penalty weights can be especially detrimental. Consequently, a detailed instance-specific analysis is required to identify suitable parameter configurations. While such an analysis is computationally expensive, it provides valuable insight into the parameter regimes that lead to satisfactory performance.

\section{Conclusions}\label{sec_7}

In this work, the use of the PCE encoding scheme as a tool for solving optimisation problems has been analysed, and it has been observed that it demonstrates strong computational performance for some optimization problems, provided that the cost function is suitably reformulated beforehand. To evaluate its effectiveness, three classical optimisation problems have been considered using the \emph{ansatz} proposed by the authors in~\cite{Sciorilli_2025} and the \emph{Differential Evolution} algorithm as the classical optimiser. The results obtained have been compared with the \texttt{QOPTLib} \textit{benchmark}, yielding nearly equivalent results for MC and for certain TSP instances, while improved solutions were obtained for BPP and other TSP instances.

The final performance remains strongly dependent on the choice of $\alpha$ and $\beta$, highlighting the importance of a prior characterisation of these hyperparameters despite the associated computational cost. In particular, $\alpha$ has been shown to play a decisive role in determining solution quality, with larger values generally leading to superior results by promoting a stronger binarisation of the relaxed variables. By contrast, although $\beta$ was introduced to encourage the expected values of the relaxed variables towards zero, setting $\beta = 0$ often yields results comparable to those obtained with larger values. Furthermore, classical post-processing can further refine the solutions obtained during the construction phase, yielding additional improvements in many instances.

Another important observation concerns the selection of appropriate penalty coefficients in the problem formulations, a challenge classically known as the Big-M problem. Our results suggest that increasing these weights does not necessarily favour binarisation; instead, it may reinforce constraint satisfaction while still allowing fractional and ambiguous configurations, thereby hindering the reconstruction of high-quality combinatorial solutions.

Furthermore, the compression order has been shown to play a significant role in the overall optimisation process. Under the considered configuration, increasing the compression order $k$ results in deeper variational circuits and a larger parameter space, which also increases the optimisation effort required by the classical solver. While this additional expressive capacity may facilitate the discovery of improved solutions, the obtained performance still remains strongly dependent on the underlying problem formulation and the calibration of its associated parameters.

Finally, a study based on shots has been carried out, analysing both the statistical noise and the hardware noise introduced in the computation of expected values. For the problem analysed and the emulated hardware configuration, it is found that the solution in the presence of noise stagnates at a MAE of approximately 0.025, failing to match the ideal solution. This analysis has allowed reference bounds to be established for the estimation of execution time within the CUNQA framework. In this context, it has furthermore been observed that the presence of noise, far from being solely detrimental, can in certain cases contribute to improving the exploration of the solution space and, therefore, favour the optimisation process.

\section{Future work}\label{sec_8}

This work has analysed the capabilities of the proposed PCE algorithm for solving optimisation problems using three benchmark problems in the field of quantum computing, and the results obtained suggest that the algorithm has significant potential for tackling large-scale combinatorial problems.

The work has focused solely on the use of the \emph{ansatz} proposed by the authors, together with Differential Evolution as implemented in the \textit{SciPy} software package, a classical optimisation algorithm that has previously demonstrated robustness against local minima in variational settings. Although this choice provides a reliable baseline, recent studies such as~\cite{Novak_2025} have shown that more advanced variants, including iL-SHADE, can achieve superior performance, further highlighting the potential benefits of adaptive optimisation strategies in this class of variational algorithms. Consequently, it would be of considerable interest to investigate whether alternative \emph{ansatzes} or more sophisticated classical optimisers could further improve the results obtained in this work. Such an analysis is particularly relevant given that the proposed compression technique enables the treatment of large-scale problems with a relatively small number of qubits, in contrast to approaches such as QAOA, which typically require a number of qubits that scales linearly with the number of binary decision variables. It is therefore necessary to investigate the combination of hyperparameters of this algorithm such as \emph{ansatz}, $\alpha$, $\beta$, classical optimisation --- with its own parameters ---, number of shots, etc. Indeed, one possible avenue that has shown improvements for constrained problems is the one proposed by~\cite{Padin_2026}, where the simultaneous optimisation of the \emph{ansatz} parameters and $\alpha$ is included. It would also be of great interest for accelerating the search for solutions to develop algorithms for a good initial parameter selection using \emph{warm-starting} techniques such as that proposed by~\cite{doCarmo2025WarmStartingPCE} for the TSP.

Another interesting direction for future research would be to determine what level of hardware noise is acceptable to obtain solutions identical or sufficiently close to those obtained under ideal conditions. Such an analysis would provide a clearer indication of the effective capabilities of the algorithm on near-term quantum devices, while also offering insight into its potential performance as fault-tolerant quantum computers begin to emerge.

However, it must also be considered that this algorithm offers possibilities beyond execution on real quantum hardware. In particular, it can be employed through classical simulation, forming part of a broader family of quantum-inspired algorithms. Given its capacity to reduce the number of qubits through different compression strategies, it becomes amenable to solving large-scale combinatorial problems using only classical computational resources. It is therefore of interest to better understand the behaviour of the algorithm across different problem classes as the level of compression increases.

In this context, future work could also focus on improvements within the training and reconstruction process itself. For instance, intermediate candidate solutions could be stored at selected stages of the optimisation procedure in order to identify feasible or higher-quality solutions before full convergence, particularly in cases where strong compression significantly reduces the number of binary variables. Moreover, when the relaxed solution contains conflicting active variables, a dedicated post-processing strategy could be developed to evaluate the combinatorial alternatives induced by these activations, in a way analogous to exploring a superposition of candidate solutions.

In summary, this work has presented an evaluation of the PCE algorithm across three benchmark optimisation problems, demonstrating both its capabilities and some of its limitations on current quantum hardware. Although a better understanding of the role of PCE hyperparameters as well as its potential for parallelisation within DQC environments requires future research, the algorithm shows strong potential compared to existing quantum approaches due to its remarkable capacity to compress combinatorial problems.

\section*{Acknowledgements}
This work has been mainly financed by  Spanish Ministry for Digital Transformation and of Civil Service of the Spanish Government through the QUANTUM ENIA project call - QuantumSpain, EU through the Recovery, Transformation and Resilience Plan – NextGenerationEU within the framework of the Digital Spain 2026. Also, Andrés Gómez acknowledges the grant PID2024-159713OB-I00 funded by MICIU/AEI/10.13039/501100011033 and by ERDF/EU. This work is also supported by the European Union’s Horizon Research
and Innovation Programme under Grant Agreement No. 10119256 (GRAVITEQA). 

Additionally, this research project was made possible through the access granted by the Galicia Supercomputing Center (CESGA) to two key parts of its infrastructure. Firstly, its Qmio quantum computing infrastructure with funding from the European Union, through the Operational Programme Galicia 2014-2020 of ERDF\_REACT EU, as part of the European Union's response to the COVID-19 pandemic. Secondly, the supercomputer FinisTerrae III and its permanent data storage system, which have been funded by the NextGeneration EU 2021 Recovery, Transformation and Resilience Plan, ICT2021-006904, and also from the Pluriregional Operational Programme of Spain 2014-2020 of the European Regional Development Fund (ERDF), ICTS-2019-02-CESGA-3, and from the State Programme for the Promotion of Scientific and Technical Research of Excellence of the State Plan for Scientific and Technical Research and Innovation 2013-2016 State subprogramme for scientific and technical infrastructures and equipment of ERDF, CESG15-DE-3114.

\section*{AI Usage}
The original text was written in Spanish and translated by Claude AI to British English. The authors reviewed the generated text in English.
\section*{Conflict of interest}
The authors declare no conflict of interest.

\section*{Availability of data and code}
The data and code can be downloaded from 
\href{https://github.com/csampron/CESGA-Quantum-Spain-PCE-Benchmark}{CESGA-Quantum-Spain-PCE-Benchmark}.

\printbibliography

\newpage

\appendix

\section{Experiment's optimal hyperparameter settings}\label{Appendix_A}

The following tables report the optimal hyperparameter settings employed for each experimental configuration considered in this study. In particular, the values of $\alpha$ and $\beta$ correspond to the parameter combinations yielding the best performance for each problem size and instance.

%---------------- MAXCUT ----------------%
\begin{table}[H]
\centering
\begin{tabular}{|c|cc|cc|cc|}
\hline
\multirow{2}{*}{\textbf{Nodes}}
& \multicolumn{2}{|c|}{\textbf{$k=2$}}
& \multicolumn{2}{|c|}{\textbf{$k=3$}}
& \multicolumn{2}{|c|}{\textbf{$k=4$}} \\
\cline{2-7}
& $\alpha$ & $\beta$
& $\alpha$ & $\beta$
& $\alpha$ & $\beta$ \\
\hline
10  & 6.0  & 0.5 & 6.0  & 0.5 & 7.5  & 0.5 \\
20  & 7.5  & 0.5 & 7.5  & 0.5 & 9.0  & 0.5 \\
40  & 9.0  & 0.5 & 9.0 & 0.5 & 9.0 & 0.5 \\
50  & 9.0  & 0.5 & 9.0 & 0.5 & 10.5 & 0.5 \\
60  & 9.0  & 0.5 & 9.0 & 0.5 & 10.5 & 0.5 \\
100 & 10.5 & 0.5 & 10.5 & 0.5 & 10.5 & 0.5 \\
150 & 12.0 & 0.5 & 12.0 & 0.5 & 12.0 & 0.5 \\
200 & 13.5 & 0.5 & 13.5 & 0.5 & 12.0 & 0.5 \\
250 & 13.5 & 0.5 & 13.5 & 0.5 & 13.5 & 0.5 \\
300 & 15.0 & 0.5 & 15.0 & 0.5 & 13.5 & 0.5 \\
\hline
\end{tabular}
\caption{MCP hyperparameter configuration for different graph sizes and values of $k$.}
\label{tab:maxcut_parameters}
\end{table}

%---------------- BPP ----------------%
\begin{table}[H]
\centering
\begin{tabular}{|c|cc|cc|cc|}
\hline
\multirow{2}{*}{\textbf{Objects}}
& \multicolumn{2}{c|}{$k=2$}
& \multicolumn{2}{c|}{$k=3$}
& \multicolumn{2}{c|}{$k=4$} \\
\cline{2-7}
& $\alpha$ & $\beta$
& $\alpha$ & $\beta$
& $\alpha$ & $\beta$ \\
\hline
4  & 35.0  & 0.6 & 35.0  & 0.6  & 35.0  & 0.6  \\
5  & 40.0  & 0.6 & 40.0  & 0.4  & 40.0  & 1.0  \\
6  & 40.0  & 0.2 & 40.0  & 0.8  & 30.0  & 0.4   \\
7  & 40.0  & 0.6 & 35.0  & 0.6  & 30.0  & 0.6  \\
8  & 40.0  & 0.4 & 35.0  & 0.6  & 30.0  & 0.6 \\
9  &  -    &  -  & 40.0  & 0.8  & 40.0  & 0.4  \\
10 &  -    &  -  & 35.0  & 0.4  & 50.0  & 0.8  \\
12 &  -    &  -  & 40.0  & 0.4  & 30.0  & 0.6 \\
14 &  -    &  -  &  -    &  -   & 40.0  & 0.4  \\
\hline
\end{tabular}
\caption{BPP hyperparameter configuration for different problem sizes and values of $k$.}
\label{tab:bpp_parameters}
\end{table}

%---------------- TSP ----------------%
\begin{table}[H]
\centering
\begin{tabular}{|c|cc|cc|cc|}
\hline
\multirow{2}{*}{\textbf{Nodes}}
& \multicolumn{2}{c|}{$k=2$}
& \multicolumn{2}{c|}{$k=3$}
& \multicolumn{2}{c|}{$k=4$} \\
\cline{2-7}
& $\alpha$ & $\beta$
& $\alpha$ & $\beta$
& $\alpha$ & $\beta$ \\
\hline
4  & 35.0 & 0.8 & 35.0 & 0.8  & 35.0 & 0.8  \\
5  & 35.0 & 0.8 & 35.0 & 0.8  & 35.0 & 0.8  \\
6  & 35.0 & 0.8 & 30.0 & 0.4  & 35.0 & 0.4  \\
7  & 30.0 & 0.8 & 30.0 & 1.0  & 25.0 & 1.0  \\
8  & 35.0 & 0.8 & 30.0 & 1.0  & 20.0 & 0.2  \\
9  & 40.0 & 0.6 & 30.0 & 0.8  & 25.0 & 0.2  \\
10 & 35.0 & 0.8 & 30.0 & 0.2  & 30.0 & 0.6   \\
15 &  -   &  -  & 30.0 & 1.0  & 30.0 & 1.0 \\
22 &  -   &  -  &  -   &  -   & -   &  -    \\
25 &  -   &  -  &  -   &  -   &  -   &  -    \\
\hline
\end{tabular}
\caption{TSP hyperparameter configuration for different graph sizes and values of $k$.}
\label{tab:tsp_parameters}
\end{table}

\section{Analysis of relaxed variables in TSP}\label{Appendix_B}

In this section, several metrics are introduced for the TSP in order to illustrate and quantify some of the observations discussed previously. The objective is to characterise general properties of the relaxed variables, such as their activation level and their degree of separability. Analogous metrics can naturally be defined for other combinatorial problems by adapting these ideas to the particular structure of their decision variables. Nevertheless, since the TSP constitutes one of the main case studies of this work and provides a particularly illustrative framework for analysing the phenomena described above, the discussion will henceforth focus exclusively on this problem.

First, given an index $i\in\{1,\ldots,\mathcal{N}\}$, we define the set
\begin{equation*}
\mathcal{X}_i := \Big\{x_{ij}~\Big|~j \in \{1,\ldots,\mathcal{N}+1\} \Big\}.
\end{equation*}
Furthermore, let
\begin{equation*}
{\max}_{(k)}(\mathcal{X}_i)
\end{equation*}
denote the $k$-th largest element of $\mathcal{X}_i$. In particular,
\begin{equation*}
{\max}_{(1)}(\mathcal{X}_i)
\quad \text{and} \quad
{\max}_{(2)}(\mathcal{X}_i),
\end{equation*}
represent the largest and second-largest values of $\mathcal{X}_i$, respectively.

Based on this definition, the activation level of the relaxed variables in each row can be evaluated through their maximum value. To this end, the metrics
\begin{equation*}
\texttt{mean}_\texttt{rowmax} =
\frac{1}{n}\sum_{i=1}^{n} {\max}_{(1)}(\mathcal{X}_i)
\quad \text{and} \quad
\texttt{min}_\texttt{rowmax} =
\min_i\Big({\max}_{(1)}(\mathcal{X}_i)\Big)
\end{equation*}
are introduced. While $\texttt{mean}_\texttt{rowmax}$ provides a global measure of the binarisation level of the relaxed variables, $\texttt{min}_\texttt{rowmax}$ identifies the weakest row and is particularly useful for detecting partially inactive variables after the optimisation process.

On the other hand, the degree of separability between the relaxed variables within a given row can be quantified through the difference between its two largest values,
\begin{equation*}
\texttt{gap}_i :=
{\max}_{(1)}(\mathcal{X}_i) - {\max}_{(2)}(\mathcal{X}_i).
\end{equation*}
From this quantity, the metrics
\begin{equation*}
\texttt{mean}_\texttt{gap} =
\frac{1}{n}\sum_{i=1}^{n} \texttt{gap}_i
\quad \text{and} \quad
\texttt{min}_\texttt{gap} =
\min_i(\texttt{gap}_i)
\end{equation*}
are defined. While $\texttt{mean}_\texttt{gap}$ provides a global measure of the separability of the relaxed variables, $\texttt{min}_\texttt{gap}$ identifies the most ambiguous row and is particularly useful for detecting ties or indecision between candidate variables after the optimisation process.

The previous metrics allow the different behavioural regimes observed during the experiments to be interpreted quantitatively. Two representative configurations are presented below, corresponding respectively to an unfavourable and a favourable parameter regime. The example considered corresponds to the 10-node TSP with $k=3$, since, according to the results presented in Section~\ref{Results_TSP}, a region of the parameter space has been identified in which the combination of penalties and hyperparameters consistently leads to feasible solutions. Taking this configuration as a reference case, a comparison is established with a scenario in which the penalty terms are increased by one order of magnitude, specifically by a factor of ten. The aim is to analyse how this modification affects the previously defined activation and separability metrics, thereby characterising the differences between the regimes and providing a deeper understanding of the effect of the penalty parameters on the solution-selection dynamics.

\begin{figure}[H]
    \centering
    % ---------------- Row 1 ----------------
    \subfloat[$\texttt{mean}_\texttt{gap}$ ]{\includegraphics[width=0.5\textwidth]{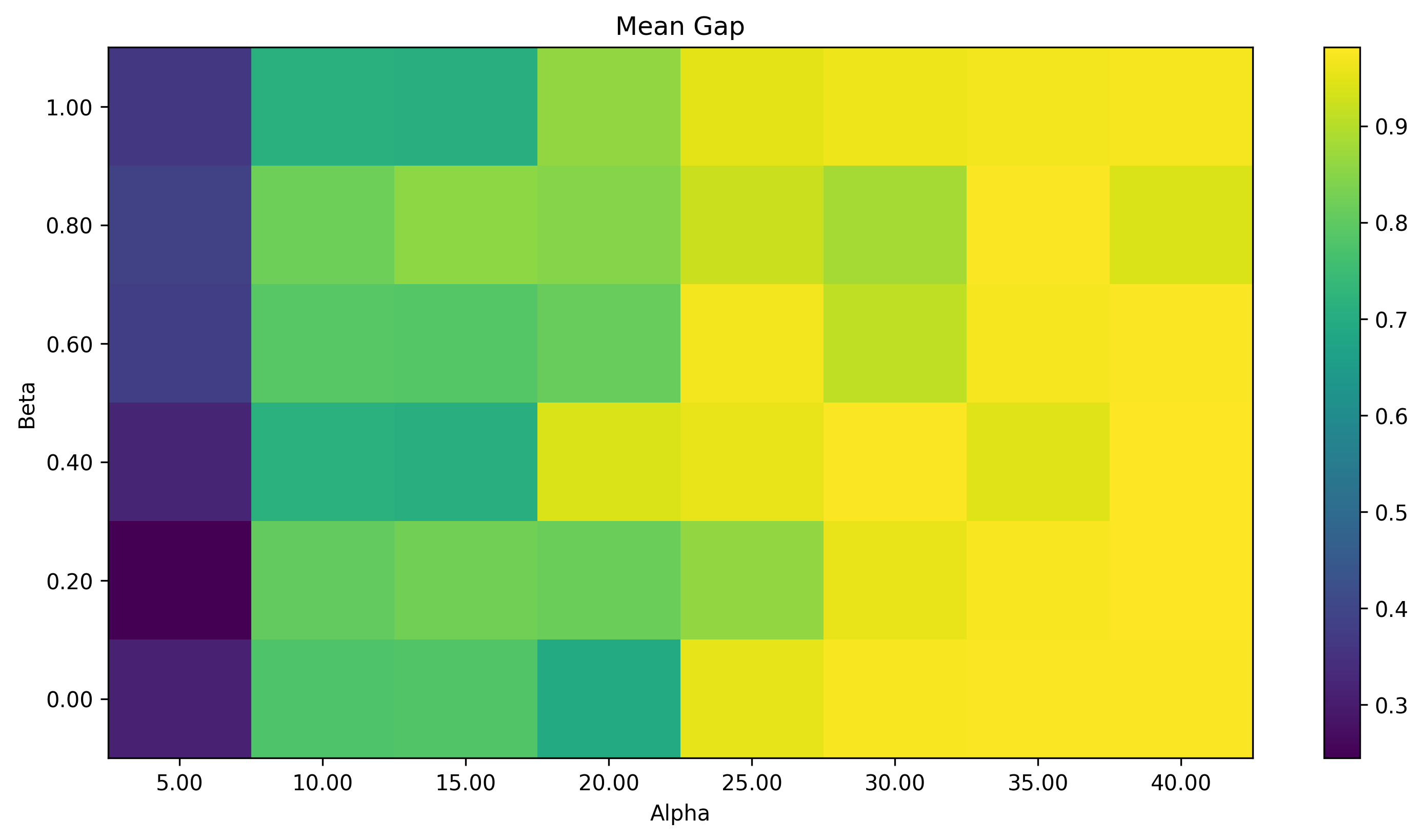}}
    \subfloat[$\texttt{min}_\texttt{gap}$]{\includegraphics[width=0.5\textwidth]{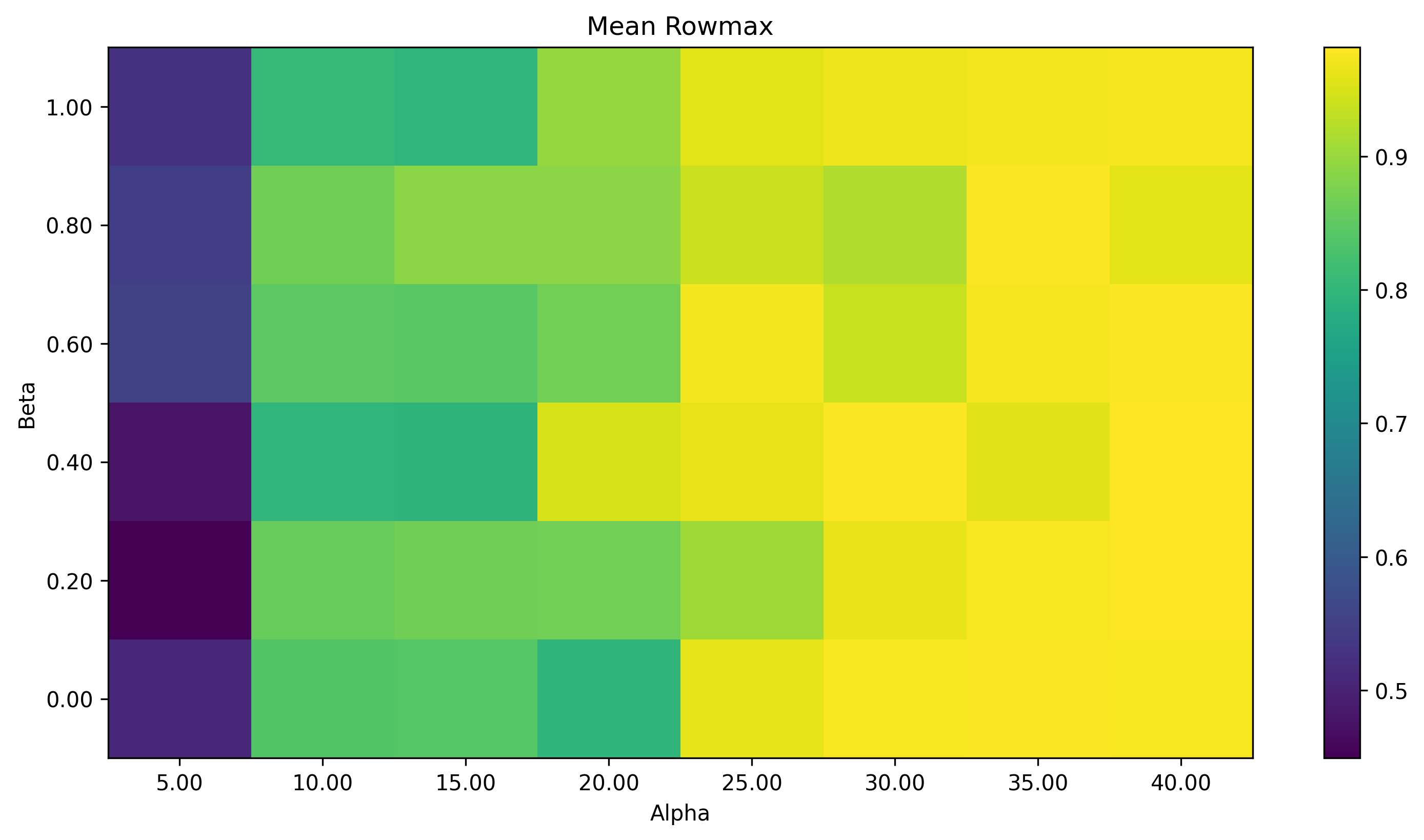}}\\[2mm]

    % ---------------- Row 2 ----------------
    \subfloat[$\texttt{min}_\texttt{gap}$]{\includegraphics[width=0.5\textwidth]{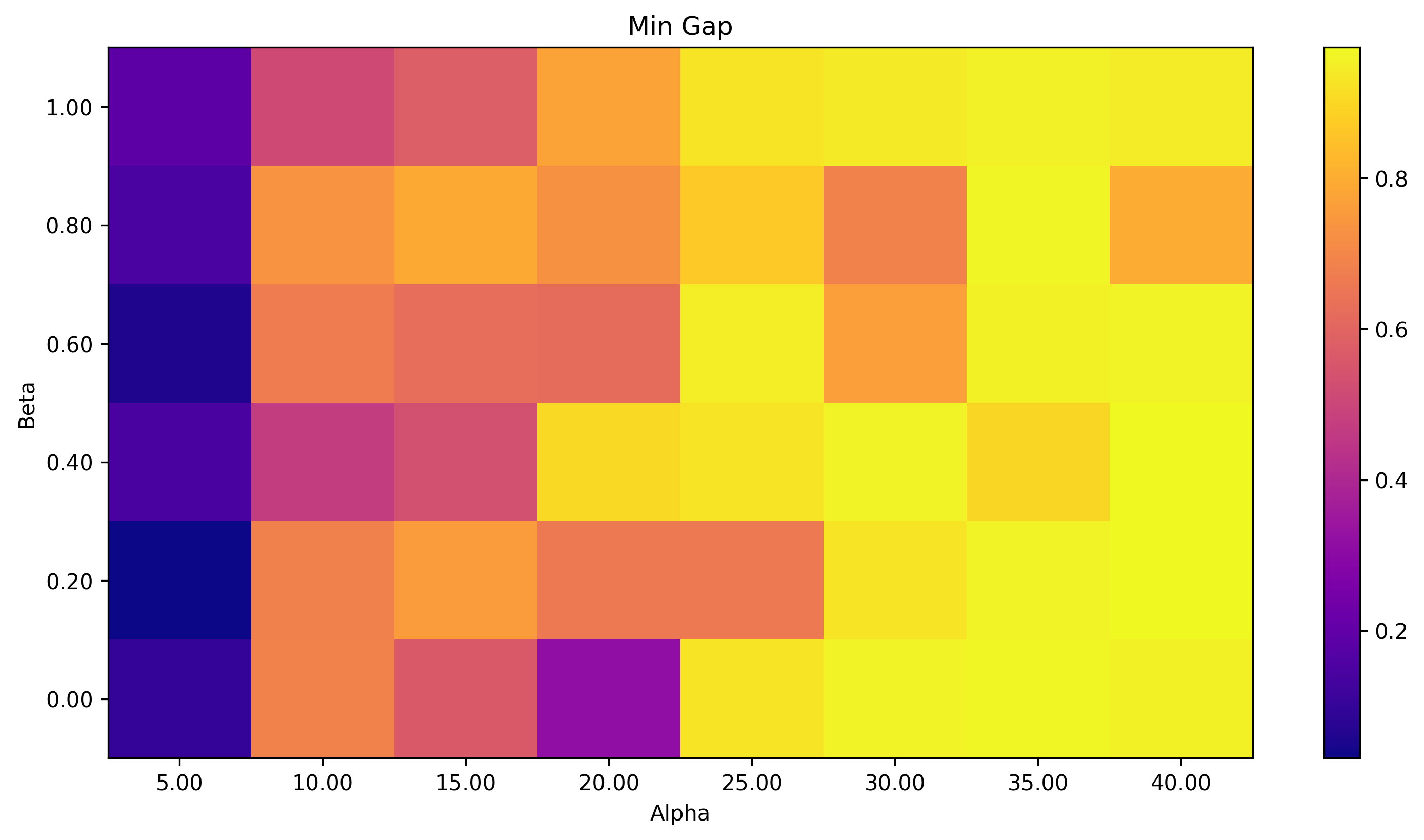}}
    \subfloat[$\texttt{min}_\texttt{rowmax}$]{\includegraphics[width=0.5\textwidth]{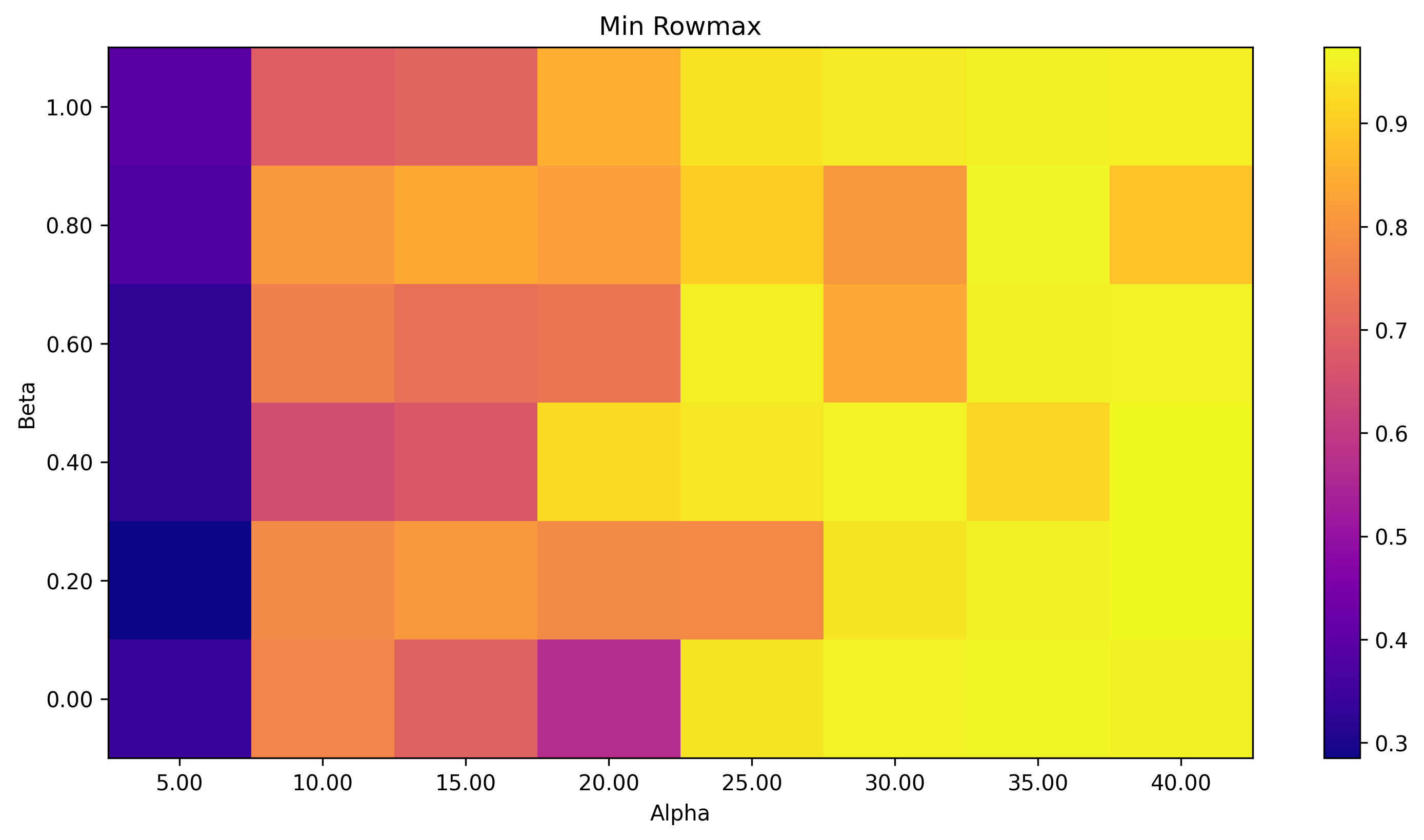}}\\[2mm]

    \caption{Heat map of metrics over the relaxed variables for instance ${\mathcal{N}}=10$ of the TSP and $k=3$, as a function of $\alpha$ and $\beta$, averaged over five random initialisations. The highlighted configuration corresponds to the default penalty setting.}
    \label{fig: Default_Metrics_of_relaxation}
\end{figure}

Figure~\ref{fig: Default_Metrics_of_relaxation} corresponds to the reference configuration that consistently produces feasible solutions. In this case, both $\texttt{mean}_\texttt{gap}$ and $\texttt{min}_\texttt{gap}$ attain large values, reflecting a clear separation between the dominant candidate and its competitors. In particular, for values of $\alpha$ above approximately $20.0$, both metrics remain close to $0.9$, indicating that, on average, there exists a very pronounced difference between the most activated variable in each row and the remaining candidates. This translates into a stable binarisation of the relaxed solution and a practically complete absence of ambiguity during the reconstruction process. Similarly, the large values of $\texttt{mean}_\texttt{rowmax}$ and $\texttt{min}_\texttt{rowmax}$ indicate strong and homogeneous activations across all rows. The combination of both properties gives rise to well-defined solution strings whose discrete reconstruction is considerably more robust.

\begin{figure}[H]
    \centering
    % ---------------- Row 1 ----------------
    \subfloat[$\texttt{mean}_\texttt{gap}$ ]{\includegraphics[width=0.5\textwidth]{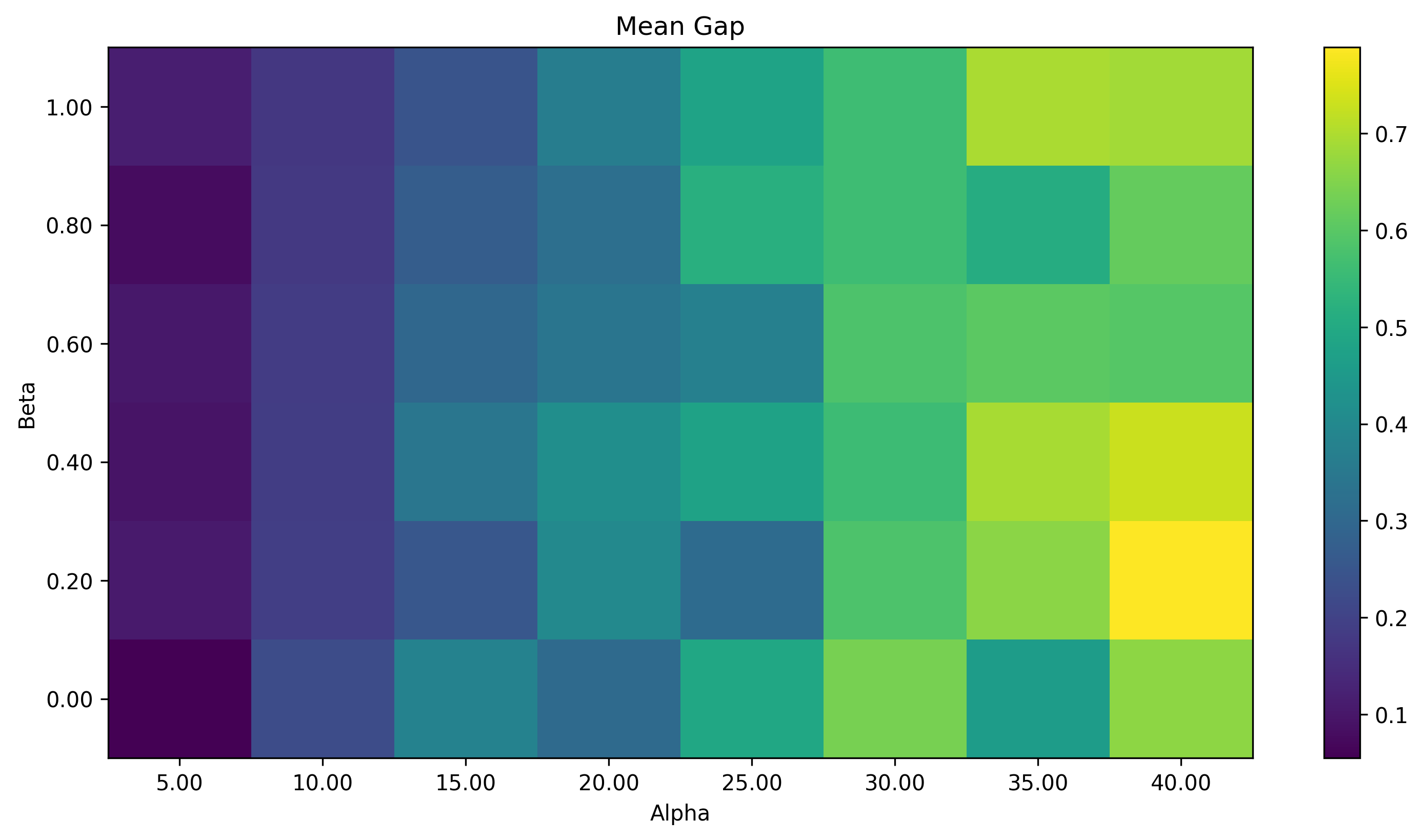}}
    \subfloat[$\texttt{mean}_\texttt{rowmax}$]{\includegraphics[width=0.5\textwidth]{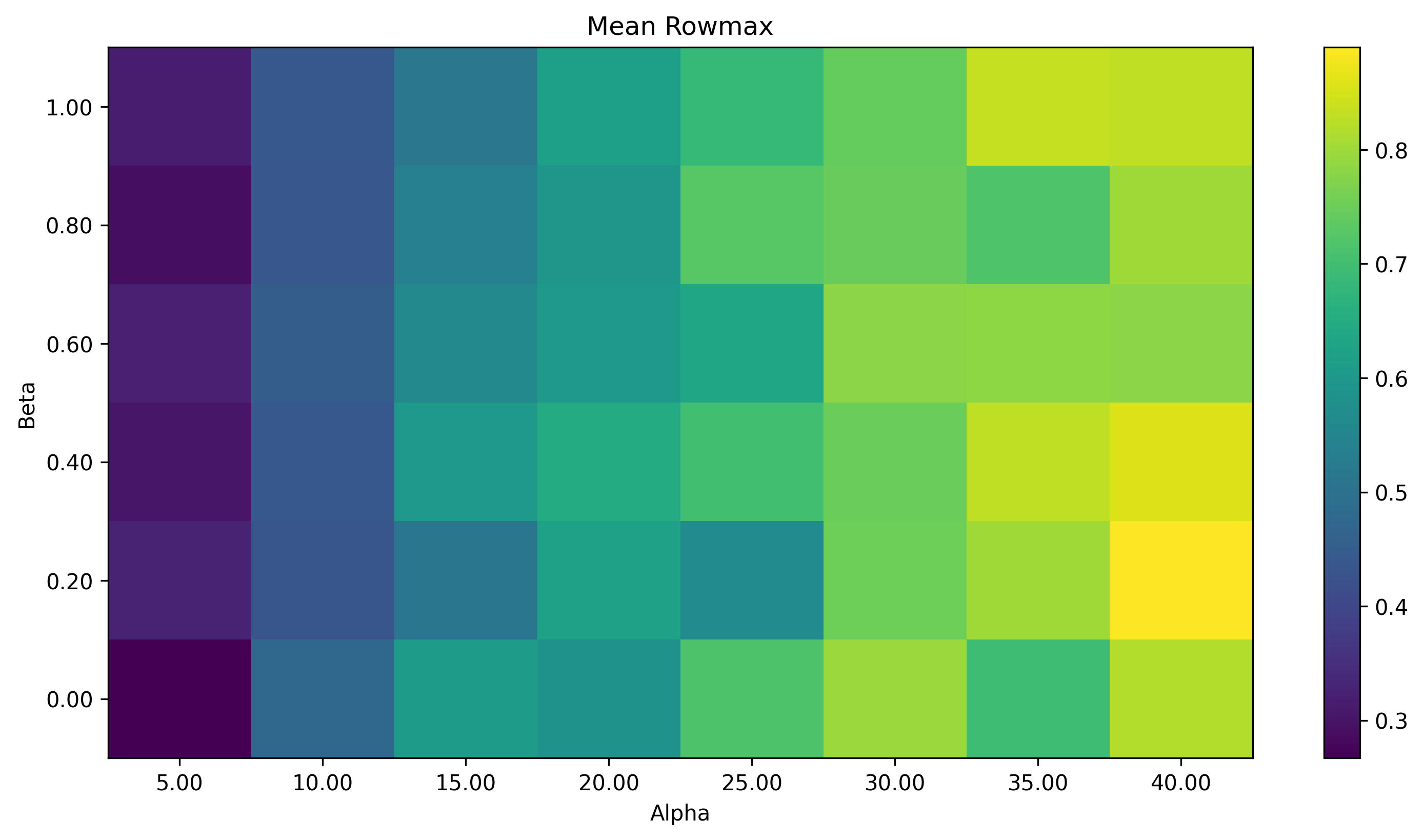}}\\[2mm]

    % ---------------- Row 2 ----------------
    \subfloat[$\texttt{min}_\texttt{gap}$]{\includegraphics[width=0.5\textwidth]{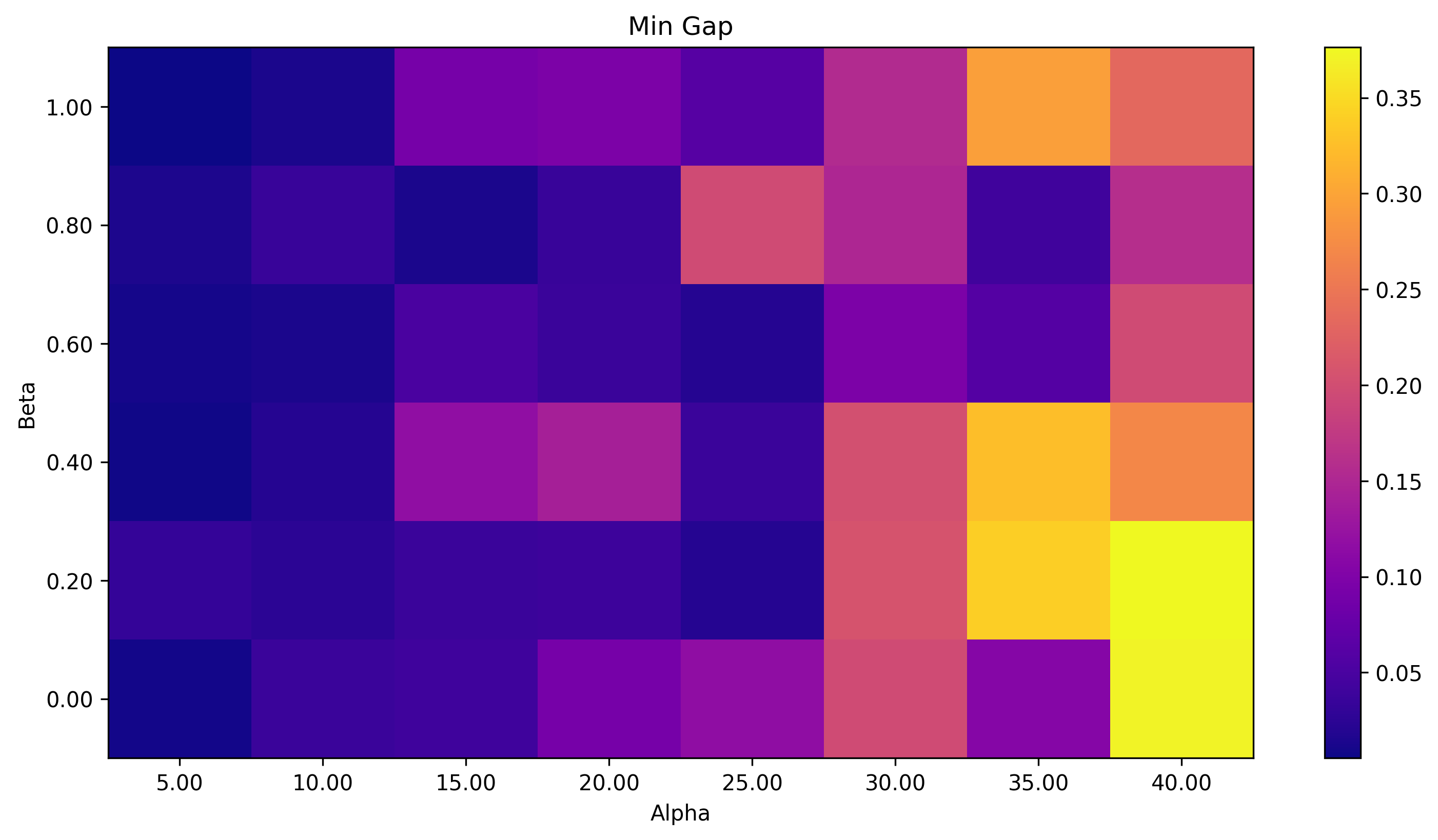}}
    \subfloat[$\texttt{min}_\texttt{rowmax}$]{\includegraphics[width=0.5\textwidth]{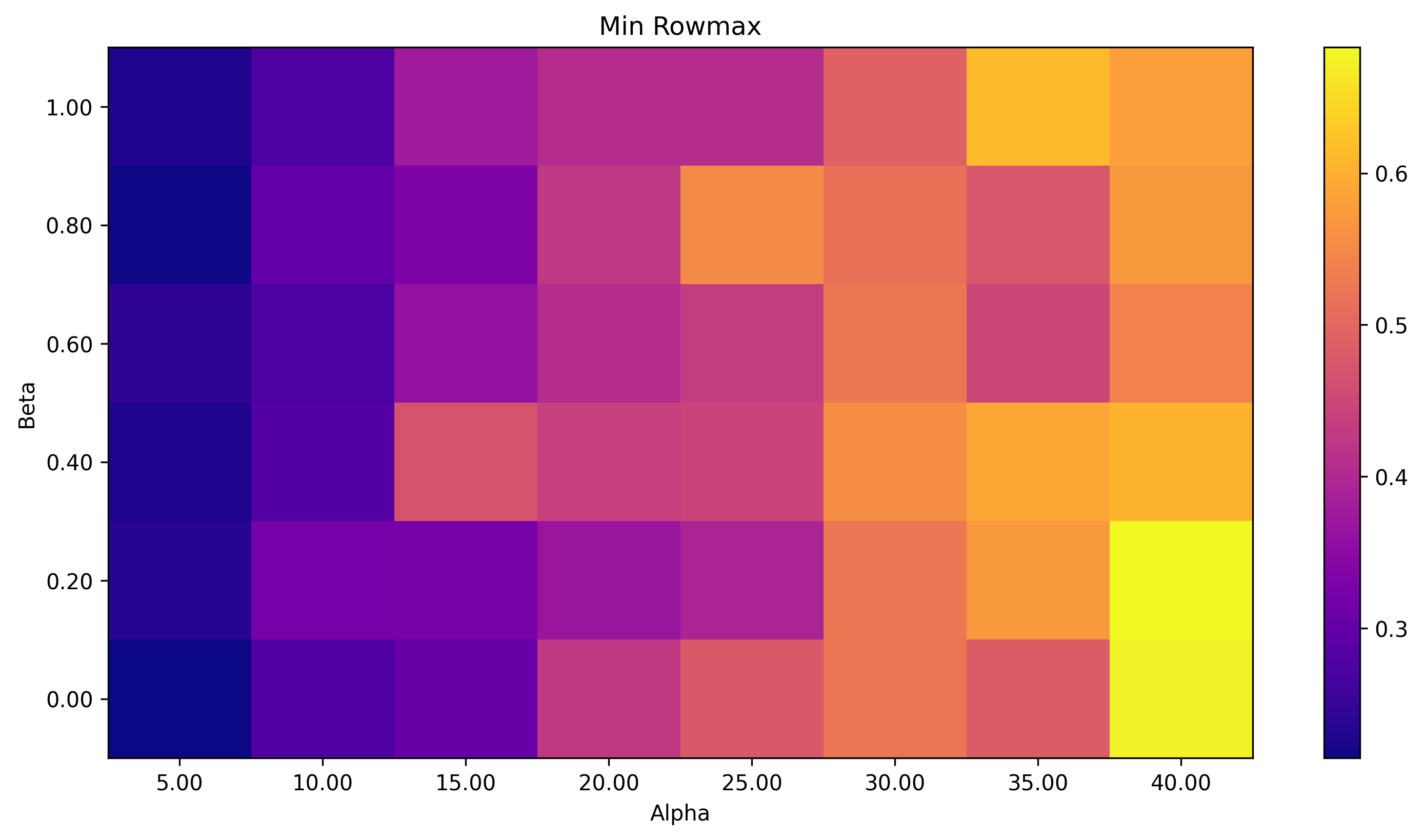}}\\[2mm]

    \caption{Heat map of metrics over the relaxed variables for instance ${\mathcal{N}}=10$ of the TSP and $k=3$, as a function of $\alpha$ and $\beta$, averaged over five random initialisations. The highlighted configuration corresponds to an over-penalised penalty setting.}
    \label{fig: Over_Penalised_Metrics_of_relaxation}
\end{figure}

In contrast, Figure~\ref{fig: Over_Penalised_Metrics_of_relaxation} shows a configuration associated with a low feasibility rate. In this case, both $\texttt{mean}_\texttt{gap}$ and $\texttt{mean}_\texttt{rowmax}$ remain around $0.5$ for values of $\alpha$ above $20.0$, only reaching values close to $0.7$ in isolated cases and for significantly larger values of $\alpha$. These values differ substantially from those observed in the feasible configuration and suggest an insufficient separation between candidates, together with only moderate activation levels of the dominant variables. Likewise, the metrics $\texttt{min}_\texttt{gap}$ and $\texttt{min}_\texttt{rowmax}$ experience a significant reduction, indicating the presence of rows in which either several candidate variables exhibit similar activation levels, generating ambiguity during discretisation, or all activations remain weak, resulting in partially inactive rows.

\begin{figure}[H]
    \centering
    % ---------------- Row 1 ----------------
    \subfloat[Default penalty setting ]{\includegraphics[width=0.5\textwidth]{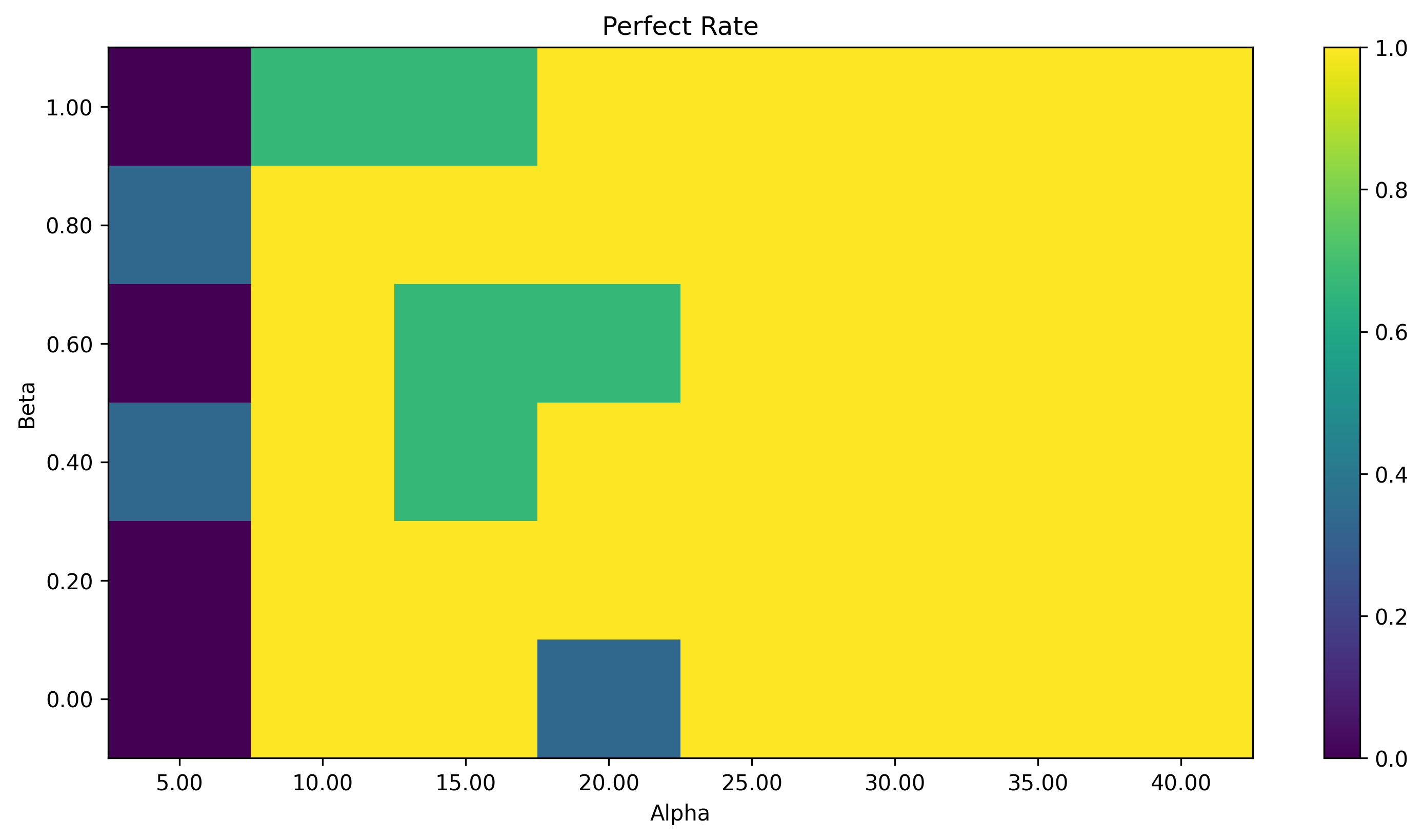}}
    \subfloat[Over-penalised penalty setting]{\includegraphics[width=0.5\textwidth]{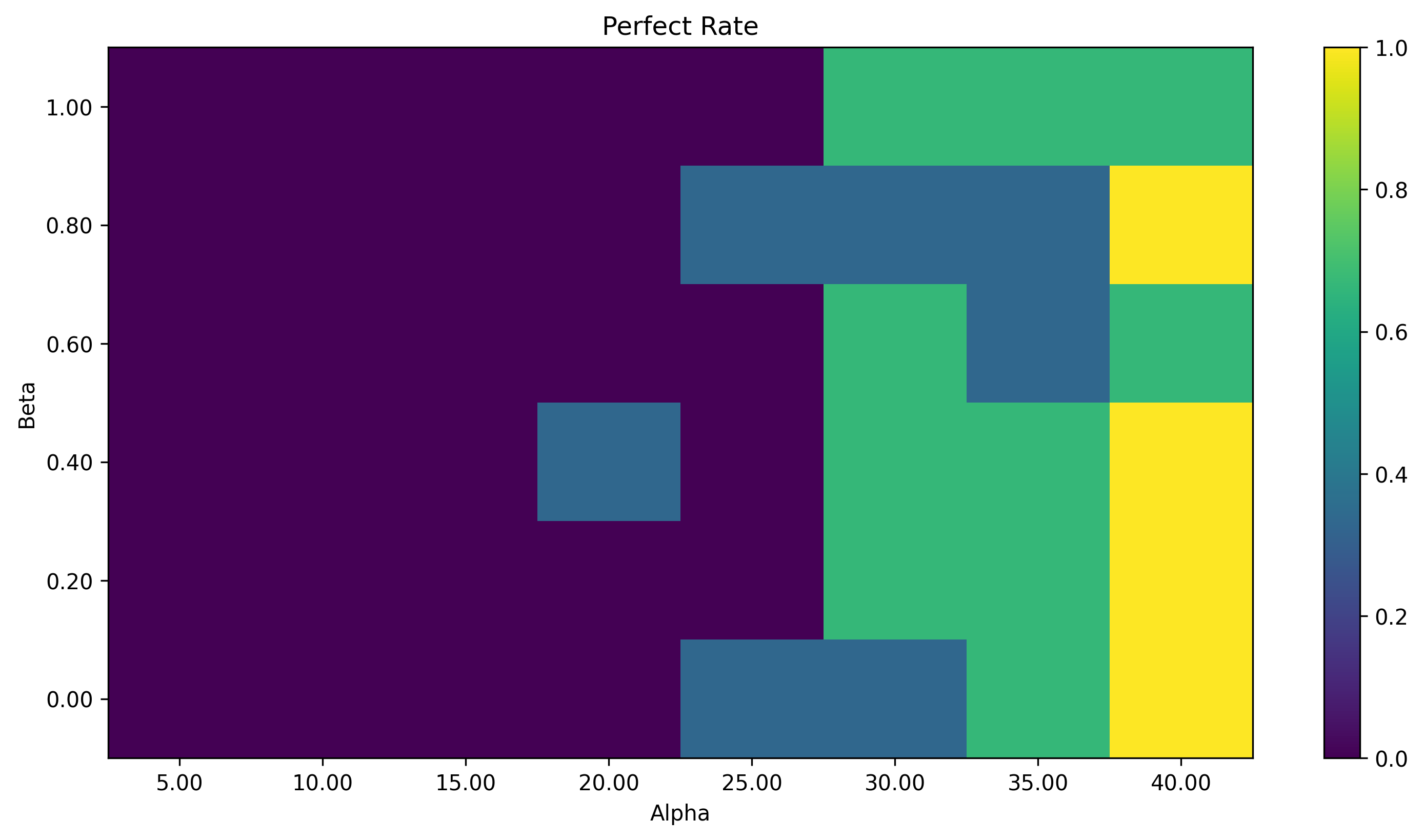}}\\[2mm]

    \caption{Heat map of rate of feasible solutions over the relaxed variables for instance ${\mathcal{N}}=10$ of the TSP and $k=3$, as a function of $\alpha$ and $\beta$, averaged over five random initialisations.}
    \label{fig: Rates_of_relaxation}
\end{figure}

Taken together, these results reveal that the relaxed solution loses the structure required to identify an unambiguous dominant candidate in each row, thereby hindering the reconstruction of a valid discrete solution. A direct validation of this interpretation can be observed in Figure~\ref{fig: Rates_of_relaxation}, which shows the proportion of feasible solutions obtained for both configurations. Consistently with the analysed metrics, the reference configuration exhibits a significantly higher feasibility rate than the over-penalised regime, confirming the close relationship between the quality of the relaxed variables and the ability to reconstruct valid solutions.

These results are consistent with the observations discussed throughout the main analysis of the present work. In particular, regimes associated with large values of $\alpha$ and appropriately calibrated penalty terms tend produce more strongly binarised and better separated relaxed variables, whereas less favourable configurations promote the appearance of weak rows, assignment ambiguities, and fractional solutions. The proposed metrics make it possible to quantify these differences explicitly and provide a direct explanation for the variations observed in the feasibility rate.

\end{document}